\documentclass[12pt]{article}

\usepackage{amsmath}
\usepackage{graphics}
\usepackage{epsfig}
\usepackage{cite}
\usepackage{colordvi}
\usepackage{color}
\input epsf

\titlepage
\author{S.P.~Baranov$^1$, A.V.~Lipatov$^{2,\,3}$}
\title{Are there any challenges in the charmonia production and polarization at the LHC?}

\begin{document}

\maketitle

\begin{center}
{\it $^1$P.N.~Lebedev Institute of Physics, Moscow 119991, Russia}\\
{\it $^2$Skobeltsyn Institute of Nuclear Physics, Lomonosov Moscow State University, Moscow 119991, Russia}\\
{\it $^3$Joint Institute for Nuclear Research, Dubna 141980, Moscow Region, Russia}\\

\end{center}

\vspace{0.5cm}

\begin{center}

{\bf Abstract }

\end{center}

We analyze the inclusive prompt production of $\psi^\prime$, $\chi_c$, $J/\psi$ and $\eta_c$ 
mesons at the LHC using the $k_T$-factorization approach. Our consideration is based on the 
off-shell production amplitudes for hard partonic subprocesses, nonrelativistic QCD formalism
for the formation of bound states and transverse momentum dependent (or unintegrated) gluon 
densities in a proton derived from Catani-Ciafaloni-Fiorani-Marchesini equation.
The nonperturbative color octet transitions are treated in terms of the multipole radiation 
theory. We extract the corresponding long-distance matrix elements from a combined fit to 
transverse momentum distributions measured at various LHC experiments.
We make predictions for the polarization parameters $\lambda_\theta$, $\lambda_\phi$,  
$\lambda_{\theta \phi}$ and the frame-independent parameter $\lambda^*$
and compare them to the available $\psi^\prime$ and $J/\psi$ data. Finally, we present 
a universal set of parameters that provides a reasonable simultaneous 
description for the whole body of the LHC data (on the $p_T$ distributions, relative 
production rates and polarization observables) for the whole charmonium family.

\vspace{1.0cm}

\noindent
PACS number(s): 12.38.-t, 13.20.Gd, 14.40.Pq

\newpage

\section{Introduction} \indent

Since it was first observed, the prompt charmonia production in hadronic collisions 
remains a topic of considerable theoretical and experimental interest. 
A commonly accepted theoretical framework for the description of charmonia production 
and decay is provided by the nonrelativistic QCD (NRQCD) factorization\cite{1,2}. 
In this formalism, the perturbatively calculated cross sections for the short-distance 
production of a $c\bar c$ pair in an intermediate Fock state $^{2S + 1}L_J^{(a)}$ 
with spin $S$, orbital angular momentum $L$, total angular momentum $J$, and color 
representation $a$ are accompanied with long-distance matrix elements (LDMEs), which
describe the transition from that intermediate $c\bar c$ state into a physical meson
via soft gluon radiation. The LDMEs are assumed to be universal (process- and 
energy-independent) and obeying certain hierarchy in powers of the relative charmed 
quarks velocity $v$.

At present, the cross sections of $\psi^\prime$, $\chi_c$, $J/\psi$ and $\eta_c$ 
production in $pp$ collisions are known at the next-to-leading order (NLO) accuracy
\cite{3,4,5,6,7,8,9,10,11,12,13,14,15}. 
The dominant tree-level next-to-next-to-leading order (NNLO$^{*}$) corrections 
to the color-singlet (CS) mechanism have been calculated\cite{16}.
The long-distance matrix elements (LDMEs) are not calculable within the theory and can
only be extracted from fits to the data. Then, with properly adjusted LDMEs values, 
a reasonably good description of the $\psi^\prime$, $\chi_c$ and $J/\psi$ transverse 
momentum distributions measured at the Tevatron and LHC is achieved\cite{3,4,5,6,7,8,9,10}.
However, the extracted LDMEs dramatically depend on the minimal charmonium transverse 
momentum used in the fits and are incompatible with one another when obtained from 
fitting different data sets. Moreover, none of the fits is able to reasonably accommodate
the polarization measurements (the so called "polarization puzzle").

The fits involving low $p_T$ data lead to the conclusion that $\psi^\prime$ and $J/\psi$ 
production at large transverse momenta must be dominated by the $^3S_1^{[8]}$ contributions 
with strong transverse polarization, that contradicts to the unpolarized production seen 
by the CDF Collaboration at the Tevatron\cite{17,18} and CMS\cite{19} and LHCb\cite{20,21} 
Collaborations at the LHC. To obtain an unpolarized $J/\psi$ meson, it is necessary to
assume that its production is dominated by the scalar $^1S_0^{[8]}$ intermediate state
\cite{4}. This comes to an immediate conflict with the recent LHCb data\cite{22} on the 
$\eta_c$ meson production, 
as the respective $\eta_c$ and $J/\psi$ LDMEs are related by one of the basic NRQCD 
principles, the heavy quark spin symmetry (HQSS). 
The impact of the $\eta_c$ data\cite{22} on the general understanding of the charmonia 
production and polarization phenomena was investigated in\cite{12}.
The overall situation is found difficult and has been even called `challenging'\cite{13}.
At present, the conventional NRQCD is yet far from understanding the data
(see also discussions\cite{23,24,25}). 
So, the further theoretical studies are still an urgent task.

Recently, a solution to the polarization puzzle has been proposed\cite{26} 
in the framework of a model that interprets the soft final state gluon radiation 
as a series of color-electric dipole transitions. In this way the LDMEs are represented 
in a form ispired by the classical multipole radiation theory, so that the spin structure 
of the transition amplitudes is explicitly specified. 
The calculations made in this approach lead to weak final $J/\psi$ polarization,
either because of the cancellation between the $^3P_1^{[8]}$ and $^3P_2^{[8]}$ 
contributions, or as a result of two successive color-electric (E1) dipole 
transitions in the chain $^3S_1^{[8]}\to {^3P}_{J}^{[8]}\to \, ^3S_1^{[1]}$ with 
$J = 0, 1, 2$. This solves completely the polarization puzzle for $J/\psi$ mesons
and, also, the production puzzle for $\eta_c$ mesons (see\cite{27,28,29,30}).
Since we no longer need the polarization-diluting $^1S_0^{[8]}$ contribution to $J/\psi$,
we neither need its HQSS counterpart process, the $^3S_1^{[8]}$ contribution to $\eta_c$,
while the production of $\eta_c$ is saturated by the color singlet mechanism alone.

We follow this approach\cite{26} in the present paper and carry out a global study 
of the production and polarization phenomena for the entire charmonium family 
($\psi^\prime$, $\chi_c$, $J/\psi$ and $\eta_c$) at the LHC.
To describe the perturbative production of a $c\bar c$ pair in hard scattering 
subprocess we employ the $k_T$-factorization formalism\cite{31,32}.
This formalism is based on the Balitsky-Fadin-Kuraev-Lipatov (BFKL)\cite{33} or 
Catani-Ciafaloni-Fiorani-Marchesini (CCFM)\cite{34} evolution equations and has
certain technical advantages in the ease of including higher-order radiative corrections
(namely, a part of NLO + NNLO +... terms corresponding to the initial-state real gluon
emissions) in the form of transverse momentum dependent (TMD) gluon 
densities\footnote{See reviews\cite{35,36} for more information.}.
Then we perform a simultaneous fit for charmonia LDMEs using the latest LHC data 
collected by the ATLAS\cite{37,38,39}, CMS\cite{40,41,42} and 
LHCb\cite{43,44,45,46,47,48,49} Collaborations at $\sqrt s = 7$, $8$, and $13$~TeV%
\footnote{In our previous studies\cite{27,28,29} such fits were performed using the 
LHC data at $\sqrt s = 7$~TeV only.}.
We also pay attention to the relative production rates, for example,
$\sigma(\chi_{c2})/\sigma(\chi_{c1})$, as the latter are sensitive
to the color siglet and color octet production mechanisms.
A clear understanding of $\chi_c$ (and, of course, $\psi^\prime$)
production is an important component of any general description of $J/\psi$
production and polarization since the feed-down contributions from radiative
decays constitute about 30\% of the visible $J/\psi$ cross section at the LHC.
Using the fitted LDMEs, we make predictions for the polarization parameters $\lambda_\theta$, 
$\lambda_\phi$ and $\lambda_{\theta \phi}$ 
and compare them to the currently available data on $\psi^\prime$ and $J/\psi$ mesons.
Our main goal is to show that the consistently used approach\cite{26} meets no troubles 
with the available charmonia data (including transverse momentum distributions, relative 
production rates and polarization observables). We end up with presenting a universal set
of parameters that provides a reasonable simultaneous description of everything.

The outline of our paper is the following. In Section 2 we describe the basic steps of our 
calculations. In Section 3 we perform a numerical fit and extract the charmonia LDMEs 
from the latest LHC data\cite{37,38,39,40,41,42,43,44,45,46,47,48,49} on the transverse 
momentum distributions. Later in this section we check the compatibility of the extracted 
LDMEs with the available data\cite{50} on charmonia polarization. The comparison is followed 
by a discussion. Our final conclusions are collected in Section 4.

\section{Theoretical framework} \indent

We start with recalling the essential calculation steps.
Our consideration is based on the following leading-order off-shell gluon-gluon fusion
subprocesses for $\psi^\prime$ and $J/\psi$ mesons:
\begin{equation}
  g^*(k_1) + g^*(k_2) \to c\bar c\left[ ^3S_1^{[1]} \right](p) + g(k),
\end{equation}
\begin{equation}
  g^*(k_1) + g^*(k_2) \to c\bar c\left[ ^1S_0^{[8]}, \, ^3S_1^{[8]}, \, ^3P_J^{[8]} \right](p),
\end{equation}
for $\chi_c$ mesons (with $J = 0$, $1$, $2$):
\noindent 
\begin{equation}
  g^*(k_1) + g^*(k_2) \to c\bar c\left[ ^3P_J^{[1]}, \, ^3S_1^{[8]} \right](p),
\end{equation}
and for $\eta_c$ mesons:
\noindent
\begin{equation}
  g^*(k_1) + g^*(k_2) \to c\bar c\left[ ^1S_0^{[1]}, \, ^1S_0^{[8]}, \, ^3S_1^{[8]}, \, ^1P_1^{[8]} \right] (p)
\end{equation}
\noindent
Additionally, we took into account the feed-down contribution to $\eta_c$
production from the decays $h_c \to \eta_c \gamma$. The leading contributions to $h_c$
come from the off-shell partonic subprocesses
\begin{equation}
  g^*(k_1) + g^*(k_2) \to c\bar c\left[ ^1P_1^{[1]} \right](p) + g(k),
\end{equation}
\begin{equation}
  g^*(k_1) + g^*(k_2) \to c\bar c\left[ ^1S_0^{[8]} \right](p)
\end{equation}
\noindent 
where the four-momenta of all particles are indicated in the parentheses.
The corresponding production amplitudes contain spin projection operators
which discriminate the spin-singlet and spin-triplet $c\bar c$ states\cite{51}:
\begin{equation}
\Pi_0 = {1\over (2m_c)^{3/2} } \, (\hat p_{\bar c} - m_c) \gamma_5 (\hat p_c + m_c),\\
\end{equation}
\begin{equation}
\Pi_1 = {1\over (2m_c)^{3/2} } \, (\hat p_{\bar c} - m_c) \hat \epsilon_{\psi}(S_z) (\hat p_c + m_c),
\end{equation}
\noindent
where $m_c$ is the charmed quark mass. States with various projections of 
the spin momentum onto the $z$ axis are represented by the polarization 
four-vector $\epsilon(S_z)$. Here $p_c$ and $p_{\bar c}$ are the four-momenta of
the charmed quark and anti-quark:
\begin{equation}
  p_c = {1 \over 2} p + q, \quad p_{\bar c} = {1 \over 2} p - q.
\end{equation}
\noindent
The relative momentum $q$ of the quarks in a bound state is associated with the 
orbital angular momentum $L$. According to the general formalism\cite{52,53},
the terms showing no dependence on $q$ are identified with the contributions to the 
$L = 0$ states while the terms linear (quadratic) in $q^\mu$ are related to the 
$L = 1$ ($L = 2$) states with the proper polarization vector $\epsilon^\mu(L_z)$ 
(resp., polarization tensor $\epsilon^{\mu\nu}(L_z)$).

The hard scattering amplitude ${\cal A}(q)$ has to be multiplied by the bound state 
wave finction $\Psi^{[a]}(q)$ and integrated over $q$. The integration is done after 
expanding the amplitude around $q = 0$:
\begin{equation}
  {\cal A}(q) = {\cal A}|_{q = 0} + q^\mu(\partial{\cal A}/\partial q^\mu)|_{q = 0} + ...
\end{equation}
\noindent
A term-by-term integration of this series employs the identities\cite{51}
\begin{equation}
  \int {d^3 q \over (2 \pi)^3} \Psi^{[a]}(q) = {1\over \sqrt{4\pi}} {\cal R}^{[a]}(0),
\end{equation}
\begin{equation}
  \int {d^3 q \over (2 \pi)^3} q^\mu \Psi^{[a]}(q) = - i \epsilon^\mu {\sqrt 3 \over \sqrt{4\pi}} {\cal R}^{\prime [a]}(0),
\end{equation}

\noindent
where ${\cal R}^{[a]}(x)$ is the radial wave function in the coordinate representation.
The first term in (10) only contributes to $S$-waves, but vanishes for $P$-waves. On the 
contrary, the second term only contributes only to $P$-waves, but vanishes for $S$-waves.
The corresponding LDMEs are related to the wave functions ${\cal R}^{[a]}(x)$ and their 
derivatives\cite{1,2} as
\begin{equation}
  \left\langle {\cal O}^S \left[^{2S + 1}L_J^{[a]}\right] \right\rangle = 2 N_c (2J + 1) |{\cal R}^{[a]}(0)|^2 / 4\pi
\end{equation}
\noindent
for $S$-waves and
\begin{equation}
  \left\langle {\cal O}^P \left[^{2S + 1}L_J^{[a]}\right] \right\rangle = 6 N_c (2J + 1) |{\cal R}^{\prime [a]}(0)|^2 / 4\pi
\end{equation}

\noindent
for $P$-waves. 
All algebraic calculations are straightforward and have been done in our previous 
papers\cite{27,28,29,30}. 
The resulting expressions have been tested for gauge invariance by substituting 
the gluon momenta for corresponding polarization vectors. We have observed gauge 
invariance even with off-shell initial gluons.

Now, let us turn to non-perturbative ingredients of the theory.
As it is motivated by the HQSS relations, the LDMEs should 
be identical for transitions in both directions (from vectors to scalars and vice versa)
and can only differ by an overall normalizing factor representing the averaging over 
spin degrees of freedom. Thus, we strictly have from this property\cite{1,2}:
\begin{eqnarray}
  \left\langle {\cal O}^{\eta_c} \left[ ^1S_0^{[1]} \right] \right\rangle 
   &=& {1\over 3} \left\langle {\cal O}^{J/\psi} \left[ ^3S_1^{[1]} \right] \right\rangle,\\
  \left\langle {\cal O}^{\eta_c} \left[ ^1S_0^{[8]} \right] \right\rangle 
   &=& {1\over 3} \left\langle {\cal O}^{J/\psi} \left[ ^3S_1^{[8]} \right] \right\rangle,\\
  \left\langle {\cal O}^{\eta_c} \left[ ^3S_1^{[8]} \right] \right\rangle 
   &=& \left\langle {\cal O}^{J/\psi} \left[ ^1S_0^{[8]} \right] \right\rangle,\\
  \left\langle {\cal O}^{\eta_c} \left[ ^1P_1^{[8]} \right] \right\rangle 
   &=& 3 \left\langle {\cal O}^{J/\psi} \left[ ^3P_0^{[8]} \right] \right\rangle,\\
  \left\langle {\cal O}^{h_c} \left[ ^1P_1^{[1]} \right] \right\rangle 
   &=& 3 \left\langle {\cal O}^{\chi_{c0}} \left[ ^3P_0^{[1]} \right] \right\rangle,\\
  \left\langle {\cal O}^{h_c} \left[ ^1S_0^{[8]} \right] \right\rangle 
   &=& 3 \left\langle {\cal O}^{\chi_{c0}} \left[ ^3S_1^{[8]} \right] \right\rangle,\\
  \left\langle {\cal O}^{\cal Q} \left[ ^3P_J^{[a]} \right] \right\rangle 
   &=& (2J + 1) \left\langle {\cal O}^{\cal Q} \left[ ^3P_0^{[a]} \right] \right\rangle
\end{eqnarray}
for all $S$- and $P$-wave bound states $\cal Q$ and color states $a$.
The relations between the different LDMEs require that the fit be done simultaneously for
the entire charmonium family.

Following the ideas of\cite{26}, we employ the classical multipole radiation theory
to describe nonperturbative transformations of the color-octet $c\bar c$ pairs produced 
in hard subprocesses into observed final state charmonia.
Only a single E1 transition is needed to transform a $P$-wave state into an
$S$-wave state, and the structure of the respective ${^3P_J^{[8]}}\to {^3S_1^{[1]}}+g$
amplitudes is taken as\cite{54}
\begin{equation}
  {\cal A}(^3P_0^{[8]} \to \, ^3S_1^{[1]} + g) \sim k_\mu \, p^\mu \, \epsilon_\nu (l) \epsilon^\nu(k),
\end{equation}
\begin{equation}
  {\cal A}(^3P_1^{[8]} \to \, ^3S_1^{[1]} + g) \sim e^{\mu \nu \alpha \beta} k_\mu \, \epsilon_\nu(p) \, \epsilon_\alpha (l) \epsilon_\beta(k),
\end{equation}
\begin{equation}
  {\cal A}(^3P_2^{[8]} \to \, ^3S_1^{[1]} + g) \sim p^\mu \, \epsilon^{\alpha \beta}(p) \, \epsilon_\alpha (l) \left[ k_\mu \epsilon_\beta(k) - k_\beta \epsilon_\mu(k) \right], 
\end{equation}

\noindent
where $k$ and $l = p - k$ are the four-momenta of the emitted gluon and the produced meson, 
$\epsilon^\mu(k)$, $\epsilon^\mu(l)$, $\epsilon^\mu(p)$ and $\epsilon^{\mu\nu}(p)$ are 
the polarization vectors (tensor) of the respective particles and $e^{\mu \nu \alpha \beta}$
is the fully antisymmetric Levi-Civita tensor.
The transformation of an $S$-wave state into another $S$-wave state 
(such as $J/\psi$ meson) is treated as two successive E1 transitions
${^3S_1}^{[8]}\to ~{^3P_J}^{[8]}+g$, ${^3P_J}^{[8]}\to {^3S_1}^{[1]}+g$
proceeding via either of the three intermediate states:
${^3P_0}^{[8]}$, ${^3P_1}^{[8]}$, or ${^3P_2}^{[8]}$. For each of the two transitions 
we exploit the same effective coupling vertices (23) --- (25).
Note that the expressions describing E1 transitions are the same for gluons and photons
(up to an overall color factor) and therefore can also be used to calculate the polarization 
variables in radiative decays in feed-down processes
$\psi'\to\chi_{cJ}+\gamma$ and $\chi_{cJ}\to J/\psi+\gamma$.
The polarization of the outgoing mesons can then be calculated without any ambiguity.

The production cross section for a charmonium ${\cal Q}$ is calculated as a convolution 
of the off-shell partonic cross section and the TMD gluon densities in a proton. 
We have for the $2\to 1$ and $2\to 2$ subprocesses, respectively:
\begin{equation}
  \displaystyle \sigma(pp \to {\cal Q} + X) = \int {2\pi \over x_1 x_2 s F} \, f_g(x_1, {\mathbf k}_{1T}^2), \mu^2) f_g(x_2, {\mathbf k}_{2T}^2), \mu^2) \, \times \atop {
  \displaystyle \times \, |{\cal \bar A}(g^* + g^* \to {\cal Q})|^2 d{\mathbf k}_{1T}^2 d{\mathbf k}_{2T}^2 dy {d\phi_1 \over 2\pi} {d\phi_2 \over 2\pi} },
\end{equation}
\begin{equation}
  \displaystyle \sigma(pp \to {\cal Q} + X) = \int {1 \over 16 \pi (x_1 x_2 s)^2 } \, f_g(x_1, {\mathbf k}_{1T}^2), \mu^2) f_g(x_2, {\mathbf k}_{2T}^2), \mu^2) \, \times \atop {
  \displaystyle \times \, |{\cal \bar A}(g^* + g^* \to {\cal Q} + g)|^2 d{\mathbf p}_{T}^2 d{\mathbf k}_{1T}^2 d{\mathbf k}_{2T}^2 dy dy_g {d\phi_1 \over 2\pi} {d\phi_2 \over 2\pi} },
\end{equation}

\noindent
where $f_g(x, {\mathbf k}_{T}^2), \mu^2)$ is the transverse momentum dependent (TMD, or unintegrated) gluon
density in a proton, ${\mathbf p}_{T}$ and $y$ are the transverse momentum and
rapidity of produced charmonium ${\cal Q}$, $y_g$ is the rapidity of outgoing gluon and
$\sqrt s$ is the $pp$ center-of-mass energy.
The initial off-shell gluons have fractions $x_1$ and $x_2$
of the parent protons longitudinal momenta, non-zero transverse momenta
${\mathbf k}_{1T}$ and ${\mathbf k}_{2T}$
and azimuthal angles $\phi_1$ and $\phi_2$.
In accordance with the general definition\cite{55}, the off-shell gluon 
flux factor in (26) is taken as $ F = 2\lambda^{1/2}(\hat{s},k_1^2,k_2^2)$,
where $\hat{s}=(k_1 + k_2)^2$. 

In the numerical analysis below, we have tried a few sets of 
TMD gluon densities in a proton, referred to as A0\cite{56}, JH'2013 set 1 and JH'2013 set 2\cite{57}.
These gluon densities were obtained from CCFM evolution equation where the input
parametrization (used as boundary conditions) was fitted to the proton structure 
function $F_2(x,Q^2)$ and, in the case of JH'2013 set 2, to $F^c_2(x,Q^2)$ also. 
The CCFM equation provides a suitable tool for our phenomenologycal study 
since it smoothly interpolates between the
small-$x$ BFKL gluon dynamics and high-$x$ DGLAP one. 
The renormalization and factorization scales 
were set to $\mu_R^2 = m_{\cal Q}^2 + {\mathbf p}_T^2$ and $\mu_F^2 = \hat{s} + {\mathbf Q}_T^2$, 
where $m_{\cal Q}$ and ${\mathbf Q}_T$
are the produced charmonium ${\cal Q}$ mass
and the transverse momentum of the initial off-shell gluon pair, respectively.
The choice of $\mu_R$ is rather standard for charmonia production, while
the unusual choice of $\mu_F$ is connected with the CCFM evolution (see\cite{56,57} for more details).
The multidimensional phase space integration has been performed by means of
the Monte-Carlo technique using the routine \textsc{vegas}\cite{58}.

\section{Numerical results} \indent

In the numerical analysis below we set $m_{\psi^\prime} = 3.686097$~GeV, $m_{\chi_{c1}} = 3.51066$~GeV,
$m_{\chi_{c2}} = 3.5562$~GeV, $m_{J/\psi} = 3.096916$~GeV, $m_{\eta_c} = 2.9839$~GeV and 
$m_{h_c} = 3.52538$~GeV, the branching fractions 
$B(\psi^\prime \to \mu^+ \mu^-) = 0.0079$, $B(J/\psi \to \mu^+ \mu^-) = 0.05961$,
$B(\chi_{c1} \to J/\psi \gamma) = 0.339$, $B(\chi_{c2} \to J/\psi \gamma) = 0.192$,
$B(\psi^\prime \to J/\psi + X) = 0.614$ and $B(h_c \to \eta_c \gamma) = 0.51$\cite{59}.
As for the CS LDMEs, we take them from the known
$\psi^\prime \to \mu^+\mu^-$ and $J/\psi \to \mu^+\mu^-$ decay widths\footnote{In our previous 
paper\cite{29} the CS LDMEs for $J/\psi$ meson were extracted from the LHC data. The fitted values 
were found to be close to commonly used conventional ones.}:
$\left\langle {\cal O}^{\psi^\prime} \left[ ^3S_1^{[1]} \right] \right\rangle = 0.7038$~GeV$^3$
and $\left\langle {\cal O}^{J/\psi} \left[ ^3S_1^{[1]} \right] \right\rangle = 1.16$~GeV$^3$\cite{3,4,5,6,7}.

\begin{table}[h!] \footnotesize
\caption{Charmonia LDMEs as determined from the different fits} \bigskip
\begin{tabular}{lcccc}
\hline \hline \\[0.1mm]
  & A0 & JH'2013 set 1 & JH'2013 set 2 & NLO NRQCD fits\\[3mm]
\hline \hline
\\
$\left\langle {\cal O}^{\psi^\prime} \big[ \, ^3S_1^{[1]} \big] \right\rangle$/GeV$^3$
  & 0.7038 & 0.7038 & 0.7038 & 0.529\cite{7}  \\
\\
$\left\langle {\cal O}^{\psi^\prime} \bigl[ \, ^1S_0^{[8]} \bigr] \right\rangle$/GeV$^3$
  & $(1.7 \pm 0.4) \cdot 10^{-2}$ & $(1.2 \pm 0.7) \cdot 10^{-2}$ & $(5.0 \pm 5.0) \cdot 10^{-3}$ & $-1.2 \cdot 10^{-4}$\cite{7} \\
\\
$\left\langle {\cal O}^{\psi^\prime} \bigl[ \, ^3S_1^{[8]} \bigr] \right\rangle$/GeV$^3$
  & $(2.3 \pm 0.1) \cdot 10^{-3}$ & $(6.9 \pm 0.6) \cdot 10^{-4}$ & $(1.6 \pm 0.1) \cdot 10^{-3}$ & $3.4 \cdot 10^{-3}$\cite{7} \\
\\  
$\left\langle {\cal O}^{\psi^\prime} \bigl[ \, ^3P_0^{[8]} \bigr] \right\rangle$/GeV$^5$
  & $(2.0 \pm 1.0) \cdot 10^{-3}$ & $(1.4 \pm 0.3) \cdot 10^{-2}$ & $(1.6 \pm 0.2) \cdot 10^{-2}$ & $9.45 \cdot 10^{-3}$\cite{7} \\
\\  
\hline
\\
$|{\cal R}^{\prime [1] \chi_{c1}}(0)|^2$/GeV$^5$
  & $0.13 \pm 0.01$ & $0.24 \pm 0.03$ & $0.25 \pm 0.04$ & $7.5 \cdot 10^{-2}$\cite{10}  \\
\\  
$|{\cal R}^{\prime [1] \chi_{c2}}(0)|^2$/GeV$^5$
  & $(4.8 \pm 3.0) \cdot 10^{-2}$ & $(1.0 \pm 0.1) \cdot 10^{-1}$ & $(9.0 \pm 1.0) \cdot 10^{-2}$ & $7.5 \cdot 10^{-2}$\cite{10} \\
\\  
$\left\langle {\cal O}^{\chi_c} \big[ \, ^3S_1^{[8]} \big] \right\rangle$/GeV$^3$
  & $(5.0 \pm 3.0) \cdot 10^{-4}$ & $(2.0 \pm 1.0) \cdot 10^{-4}$ & $(5.0 \pm 3.0) \cdot 10^{-4}$ & $2.01 \cdot 10^{-3}$\cite{10}  \\
\\  
\hline
\\
$\left\langle {\cal O}^{J/\psi} \big[ \, ^3S_1^{[1]} \big] \right\rangle$/GeV$^3$
  & 1.16 & 1.16 & 1.16 & 1.16\cite{7}  \\
\\
$\left\langle {\cal O}^{J/\psi} \bigl[ \, ^1S_0^{[8]} \bigr] \right\rangle$/GeV$^3$
  & 0.0 & 0.0 & 0.0 & $ 9.7 \cdot 10^{-2}$\cite{7} \\
\\
$\left\langle {\cal O}^{J/\psi} \bigl[ \, ^3S_1^{[8]} \bigr] \right\rangle$/GeV$^3$
  & $(2.5 \pm 0.3) \cdot 10^{-3}$ & $(4.2 \pm 0.9) \cdot 10^{-4}$ & $(1.6 \pm 0.2) \cdot 10^{-3}$ & $-4.6 \cdot 10^{-3}$\cite{7} \\
\\
$\left\langle {\cal O}^{J/\psi} \bigl[ \, ^3P_0^{[8]} \bigr] \right\rangle$/GeV$^5$
  & $(1.3 \pm 0.2) \cdot 10^{-2}$ & $(2.3 \pm 0.2) \cdot 10^{-2}$ & $(2.4 \pm 0.2) \cdot 10^{-2}$ & $-2.14 \cdot 10^{-2}$\cite{7} \\
\\  
\hline
\\
$\left\langle {\cal O}^{\eta_c} \bigl[ \, ^1S_0^{[1]} \bigr] \right\rangle$/GeV$^3$
  & 0.39 & 0.39 & 0.39 & 0.39\cite{7} \\
\\  
$\left\langle {\cal O}^{\eta_c} \bigl[ \, ^1S_0^{[8]} \bigr] \right\rangle$/GeV$^3$
  & $(8.3 \pm 0.1) \cdot 10^{-4}$ & $(1.4 \pm 0.3) \cdot 10^{-4}$ & $(5.3 \pm 0.7) \cdot 10^{-4}$ & $-1.53 \cdot 10^{-3}$\cite{7} \\
\\  
$\left\langle {\cal O}^{\eta_c} \bigl[ \, ^3S_1^{[8]} \bigr] \right\rangle$/GeV$^3$
  &  0.0 & 0.0 & 0.0 & 0.097\cite{7} \\
\\  
$\left\langle {\cal O}^{\eta_c} \bigl[ \, ^1P_1^{[8]} \bigr] \right\rangle$/GeV$^5$
  & $(3.9 \pm 0.6) \cdot 10^{-2}$  & $(6.9 \pm 0.6) \cdot 10^{-2}$  & $(7.2 \pm 0.6) \cdot 10^{-2}$ & $-6.42 \cdot 10^{-2}$\cite{7} \\
\\  
\hline
\\
$\left\langle {\cal O}^{h_c} \bigl[ \, ^1P_1^{[1]} \bigr] \right\rangle$/GeV$^5$
  & $0.2 \pm 0.1$ &  $0.43 \pm 0.04$ & $0.39 \pm 0.04$ &  $0.32$\cite{10}  \\
\\  
$\left\langle {\cal O}^{h_c} \bigl[ \, ^1S_0^{[8]} \bigr] \right\rangle$/GeV$^3$
  & $(1.5 \pm 0.9) \cdot 10^{-3}$ & $(6.0 \pm 3.0) \cdot 10^{-4}$ & $(1.5 \pm 0.9) \cdot 10^{-3}$ & $6.03 \cdot 10^{-3}$\cite{10} \\
\\  
\hline \hline

\end{tabular}
\end{table}

\subsection{Global fit of charmonia LDMEs based on the LHC data} \indent

We have performed a global fit to the charmonium production data at the LHC for the 
entire $c\bar{c}$ family and determined the corresponding LDMEs.
Specifically, for $\psi^\prime$ mesons, we included in the fitting procedure 
the transverse momentum distributions measured by ATLAS\cite{38,39} and CMS\cite{41,42} 
Collaborations at moderate and large transverse momenta $8 < p_T < 130$~GeV 
at $\sqrt s = 7$, $8$ and $13$~TeV, where the NRQCD formalism is believed to be most 
reliable.
We have excluded from our fit the LHCb data\cite{45} since they mainly lie in 
the low $p_T$ region, where a more accurate treatment of large logarithms
$\ln m_{\psi^\prime}^2/p_T^2$ and other nonperturbative effects becomes 
necessary\footnote{Large terms proportional to $\ln m_{\psi^\prime}^2/p_T^2$ could be 
resummed using Collins-Soper-Sterman approach\cite{60} and absorbed into the TMD gluon 
density. However, this point is out of our consideration.}.
In the case of $\chi_c$ mesons, we considered the $\chi_{c1}$ and $\chi_{c2}$ 
transverse momentum distributions measured by ATLAS Collaboration\cite{37} 
at $\sqrt s = 7$~TeV and also include in the fitting procedure the ratio 
of the production rates $\sigma(\chi_{c2})/\sigma(\chi_{c1})$ measured by 
CMS\cite{40}, ATLAS\cite{37} and LHCb\cite{43,44} Collaborations at the same energy.
Note that most of the theoretical uncertainties cancel out in the ratio;
in particular, the uncertainties related to the behavior of the TMD gluon densities 
in the low-$p_T$ region.

Following the suggestion\cite{61}, we consider the CS wave functions of $\chi_{c1}$ and 
$\chi_{c2}$ mesons as independent (not necessarily identical) parameters.
The reasoning for such a suggestion is that treating charmed quarks as spinless particles
(as in the potential models\cite{62,63,64,65}) might be an oversimplification, and that
radiative corrections to the wave functions may be large\footnote{The same scenario 
was applied in our previous paper\cite{28}.}.

To determine the LDMEs of $J/\psi$ mesons (as well as their $\eta_c$ counterparts)
we performed a simultaneous fit of $J/\psi$ and $\eta_c$ transverse momentum 
distributions using the latest CMS\cite{41,42}, ATLAS\cite{38} and LHCb\cite{22} data 
taken at $7$, $8$ and $13$ TeV. Here, the NRQCD factorization
principle seems to be on solid theoretical grounds again because of not too low $p_T$
values for both $J/\psi$ and $\eta_c$ (at least, $p_T > 8$~GeV for $\eta_c$ mesons).
Of course, we took into account the feed-down contributions to $J/\psi$ and $\eta_c$
production from radiative decays of $\chi_c$, $\psi^\prime$ and $h_c$ mesons
using the corresponding branching fractions as listed above.

Nowhere we impose any kinematic restrictions but the experimental acceptance.
The fitting procedure was separately done in each of the rapidity subdivisions
(using the fitting algorithm as implemented in the \textsc{gnuplot} package\cite{66})
under the requirement that the LDMEs be strictly positive, and then the 
mean-square average of the fitted values was taken.
The relevant uncertainties are estimated in the conventional way
using Student's t-distribution at the confidence level $P = 95$\%.

To estimate the TMD scale uncertainties, the variations in the scale 
$\mu_R\to 2\mu_R$ or $\mu_R\to\mu_R/2$
were introduced through replacing the gluon distributions A0 and JH'2013 (sets 1 and 2) 
with A+ and JH'2013+, or with A- and JH'2013-, respectively. This was done to preserve
the intrinsic correspondence between the TMD set and the scale used in the evolution
equation (see\cite{56,57}).

The results of the LDME fits for $\psi^\prime$, $\chi_c$, $J/\psi$ and $\eta_c$ mesons 
are collected in Table~1. For comparison, we also present there the LDME values
obtained in the NLO NRQCD by other authors\cite{5,9}.
For the reader's convenience, the LDMEs for $\eta_c$ and $h_c$ mesons are translated
from the $J/\psi$ and $\chi_c$ ones using the HQSS relations (15) --- (20).
All the data used in the fits are compared with our predictions in Figs.~1 --- 9.
The shaded bands represent the theoretical uncertainties of our calculations, 
which include both the scale uncertainties and the uncertainties coming from the 
LDME fitting procedure.

We observe in Figs.~1 --- 9 quite a nice agreement between our calculations and the LHC 
data for the entire charmonium family at different energies and in a wide $p_T$ range 
for all of the considered TMD gluons (with the LDMEs values shown in Table~1). 
In particular, we have achieved good simultaneous description of the prompt $\eta_c$
and $J/\psi$ production, see Figs.~6 --- 9. Such an agreement turned out to be impossible
in the traditional NRQCD scheme, where the calcullated cross sections for $\eta_c$ are 
either at odds with the measurements or at odds with theoretical principles\cite{11,13}.
Further on, we have achieved a good agreement with the LHCb data\cite{45,46,47,48,49}, 
originally not included in the fitting procedure (see Figs.~3 and 9).
The extracted LDMEs values strongly depend on the TMD gluon density (see Table~1),
that reflects the different $x$ and ${\mathbf k}_T$ behavior of the latter 
(see discussion\cite{57}).
The estimated theoretical uncertainties of our predictions are rather small 
and comparable with the uncertainties of the NLO NRQCD calculations.

Our fits show unequal values for the $\chi_{c1}$ and $\chi_{c2}$ wave functions 
at the origin $|{\cal R}^{\prime [1]}(0)|^2$. We present these values in Table~1.
The difference in the values of the wave functions mainly follows from the prompt measurements
of the ratio $\sigma(\chi_{c1})/\sigma(\chi_{c2})$.
%
For each of the considered gluon densities, our extracted values of 
$|{\cal R}^{\prime [1]\chi_{c2}}(0)|^2$ (but not $|{\cal R}^{\prime [1]\chi_{c1}}(0)|^2$) 
are close to the estimations based on the potential models\cite{62,63,64,65} and two-photon 
decay width\cite{59}; namely, $|{\cal R}^{\prime [1]}(0)|^2 = 7.5 \cdot 10^{-2}$~GeV$^5$ 
(that is a widely adopted choice). However, it differs significantly from 
$|{\cal R}^{\prime [1]}(0)|^2 = 3.5 \cdot 10^{-1}$~GeV$^5$ obtained from a combined 
fit\cite{9} to the Tevatron and LHC data. Note that the fit\cite{9} was performed under 
the assumption of equal $\chi_{c1}$ and $\chi_{c2}$ wave functions.
We interpret the available LHC data\cite{37,40,43,44} as supporting their unequal values,
that qualitatively agrees with the previous results\cite{28,61}.
In such an interpretation, the data on the $\sigma(\chi_{c2})/\sigma(\chi_{c1})$ ratio
lie almost inside the theoretical uncertainty bands, as one can see in Fig.~5.
Moreover, the ratio $|{\cal R}^{\prime [1]\chi_{c2}}(0)|^2/|{\cal R}^{\prime [1]\chi_{c1}}(0)|^2 \simeq 2.5$ 
is practically independent on the TMD gluon density.
Finally, we find that $\chi_c$ production is dominated by the CS contributions in the 
considered $p_T$ range, that agrees with some earlier conclusions\cite{9}.
The obtained LDMEs for $\psi^\prime$ and $\chi_c$ mesons
were further used to calculate the feed-down contributions to $J/\psi$ production.
The results of our fits for $J/\psi$ and $\psi^\prime$ polarization parameters
are discussed in the next Section.

\subsection{$J/\psi$ and $\psi^\prime$ polarization} \indent

It is known that the polarization of $\psi^\prime$ or $J/\psi$ mesons
can be described with three parameters $\lambda_\theta$, $\lambda_\phi$ and 
$\lambda_{\theta\phi}$, which determine the spin density matrix of a charmonium decaying 
into a lepton pair. 
In general, the double differential angular distribution of the decay leptons 
in the charmonium rest frame can be written as 
\begin{equation}
 {d\sigma\over d\cos\theta^*\,d\phi^*}\sim{1\over 3+\lambda_\theta}
 \left(1+\lambda_\theta\cos^2\theta^*+\lambda_\phi\sin^2\theta^*\cos 2\phi^*
 +\lambda_{\theta\phi}\sin2\theta^*\cos\phi^* \right),
\end{equation}
\noindent
where $\theta^*$ and $\phi^*$ are the polar and azimuthal angles of the decay lepton.
So, the angular parameters $\lambda_\theta$, $\lambda_\phi$ and 
$\lambda_{\theta\phi}$ can be 
measured experimentally. The case of $(\lambda_\theta, \lambda_\phi, \lambda_{\theta \phi}) = (0,0,0)$
corresponds to unpolarized state, while $(\lambda_\theta, \lambda_\phi, \lambda_{\theta \phi}) = (1,0,0)$
and $(\lambda_\theta, \lambda_\phi, \lambda_{\theta \phi}) = (-1,0,0)$
refer to fully transverse and longitudinal polarizations.
The CMS Collaboration has measured\cite{50} all these
parameters as functions of $J/\psi$ and $\psi^\prime$
transverse momentum in the complementary frames: the Collins-Soper, helicity and perpendicular helicity 
ones\footnote{The LHCb Collaboration has also measured $J/\psi$ and $\psi^\prime$ polarization\cite{67,68}. However, 
these data were obtained at rather low $p_T < 14$~GeV and, therefore, we will not analyze these data in the present paper.}. In the Collins-Soper frame the polarization axis $z$ bisects
the two beam directions whereas the polarization axis in the helicity frame 
coincides with the direction of the charmonium momentum in the laboratory frame. 
In the perpendicular helicity frame the $z$ axis is orthogonal to that in the 
Collins-Soper frame and lies in the plane spanned by
the two beam ($P_1$ and $P_2$) momenta.
In all cases, the $y$ axis is taken to be in the direction of the vector product of the 
two beam directions in the charmonium rest frame, $\vec P_1 \times \vec P_2$ and
$\vec P_2 \times \vec P_1$ for positive and negative rapidities, respectively. 
Additionally, the frame-independent polarization parameter
$\lambda^* = (\lambda_\theta + 3\lambda_\phi)/(1 - \lambda_\phi)$ 
was investigated\cite{50}.

To estimate the polarization parameters $\lambda_\theta$, $\lambda_\phi$, 
$\lambda_{\theta\phi}$ and $\lambda^*$ we generally follow the experimental procedure. 
We collect the simulated events in the kinematical region defined by
the CMS measurement\cite{50}, generate the decay lepton angular
distributions according to the production and decay matrix elements 
and then apply a three-parametric fit based on~(27).
Of course, in the case of $J/\psi$ production we took into account
the polarization of $J/\psi$ mesons originated from radiative $\chi_c$ and $\psi^\prime$ 
decays, that is in full agreement with the experimental case.
Since the $\psi^\prime \to J/\psi + X$ decay matrix elements are unknown, 
these events were generated according to the phase space.
In Figs.~10 --- 12 we confront our predictions for all polarization parameters with the
CMS data\cite{50}.
For both $J/\psi$ and $\psi^\prime$ mesons we find only weak polarization 
($\lambda_\theta \simeq - 0.2$) at $p_T \sim 15$~GeV in the Collins-Soper and helicity 
frames and practically zero polarization ($\lambda_\theta \simeq - 0.1$ or even close to 
zero) at large transverse momenta $p_T \sim 50$~GeV.
In the perpendicular helicity frame our simulation shows practically unpolarized
$J/\psi$ and $\psi^\prime$ production with $\lambda_\theta \sim 0$ in the whole $p_T$ range.
The $\lambda_\phi$ and $\lambda_{\theta\phi}$ parameters are close to zero everywhere, as one can see in Figs.~10 --- 12.
Moreover, these results are practically independent of the $J/\psi$ and/or $\psi^\prime$ rapidity.
Thus, we demonstrate that treating the soft gluon emission within the NRQCD as a series of explicit color-electric dipole 
transitions leads to unpolarized charmonia production, that is in 
agreement with available LHC data. 
The absense of strong polarization is not connected with parameter tuning, 
but seems to be a natural and rather general feature of the scenario\cite{26}.
We would like to point out here that the conventional NLO CS calculations 
predict large longitudinal charmonia polarization at high transverse momenta, 
while the NLO NRQCD predicts large transverse polarization. None of these predictions 
is supported by the LHC measurements.

The obtained unpolarized $J/\psi$ and $\psi^\prime$ production at the LHC is our main result.
The qualitative predictions for the 
$\lambda_\theta$, $\lambda_\phi$, $\lambda_{\theta\phi}$ and $\lambda^*$
are stable with respect to variations in the model parameters. In fact,
there is no dependence on the strong coupling constant and TMD gluon densities,
i.e. two of important sources of theoretical uncertainties cancel out.
Despite large experimental uncertainties (especially for $\lambda^*$ parameter), 
the agreement between our predictions and the data is rather satisfactory 
and shows no fundamental problems in describing the data.
So, the proposed way, in our opinion,
can provide an easy and natural solution to the long-standing polarization puzzle.

\section{Conclusion} \indent

We have considered the inclusive prompt production of $\psi^\prime$, $\chi_c$, $J/\psi$ and $\eta_c$ mesons 
at the LHC in the framework of $k_T$-factorization approach. 
Our consideration was based on the off-shell production amplitudes
for hard partonic subprocesses (including both color-singlet and color-octet contributions) and 
NRQCD formalism for the formation of bound states.
Treating the nonperturbative color octet transitions in terms of multipole radiation theory
and applying the TMD gluon densities in a proton derived from the CCFM evolution equation,
we extracted charmonia LDMEs in a combined fit to 
transverse momentum distributions measured on various LHC experiments at $\sqrt s = 7$, $8$ and $13$~TeV.
Then, using the extracted LDMEs, we estimated polarization parameters $\lambda_\theta$, $\lambda_\phi$,  
$\lambda_{\theta \phi}$ and frame-independent parameter $\lambda^*$
which determine the charmonia spin density matrix.
We have demonstrated that treating the soft gluon emission as a series of explicit color-electric 
dipole transitions within the NRQCD leads to unpolarized charmonia
production at moderate and large transverse momenta, that is in agreement with the recent
LHC data on $\psi^\prime$ and $J/\psi$ mesons.
Thus, we achieved a reasonable simultaneous description 
for all of the available data (transverse momentum distributions, relative production rates
and polarization observables) on the entire charmonia family at the LHC.

\section*{Acknowledgements} \indent

The authors thank H.~Jung for his interest, very useful discussions and important remarks.
This work was supported by the DESY Directorate in the framework of
Cooperation Agreement between MSU and DESY on phenomenology of the LHC processes
and TMD parton densities.

\newpage

\begin{figure}
\begin{center}
\includegraphics[width=8.1cm]{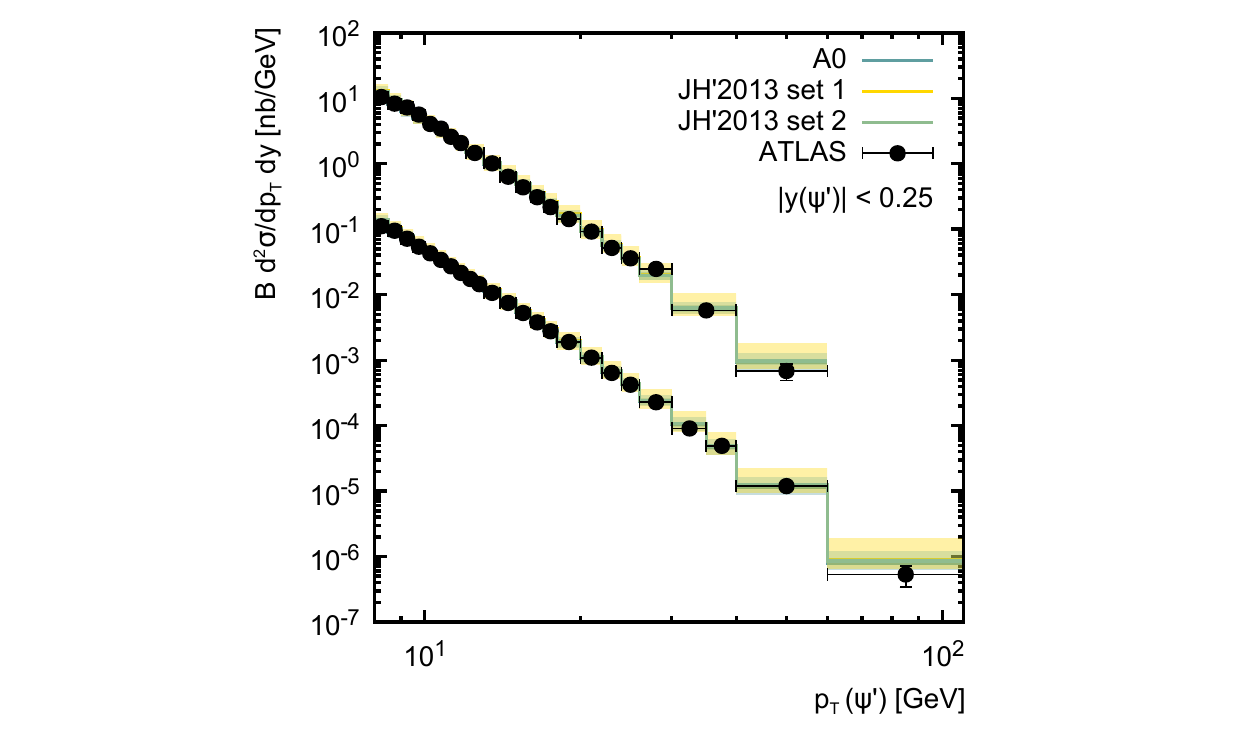}
\includegraphics[width=8.1cm]{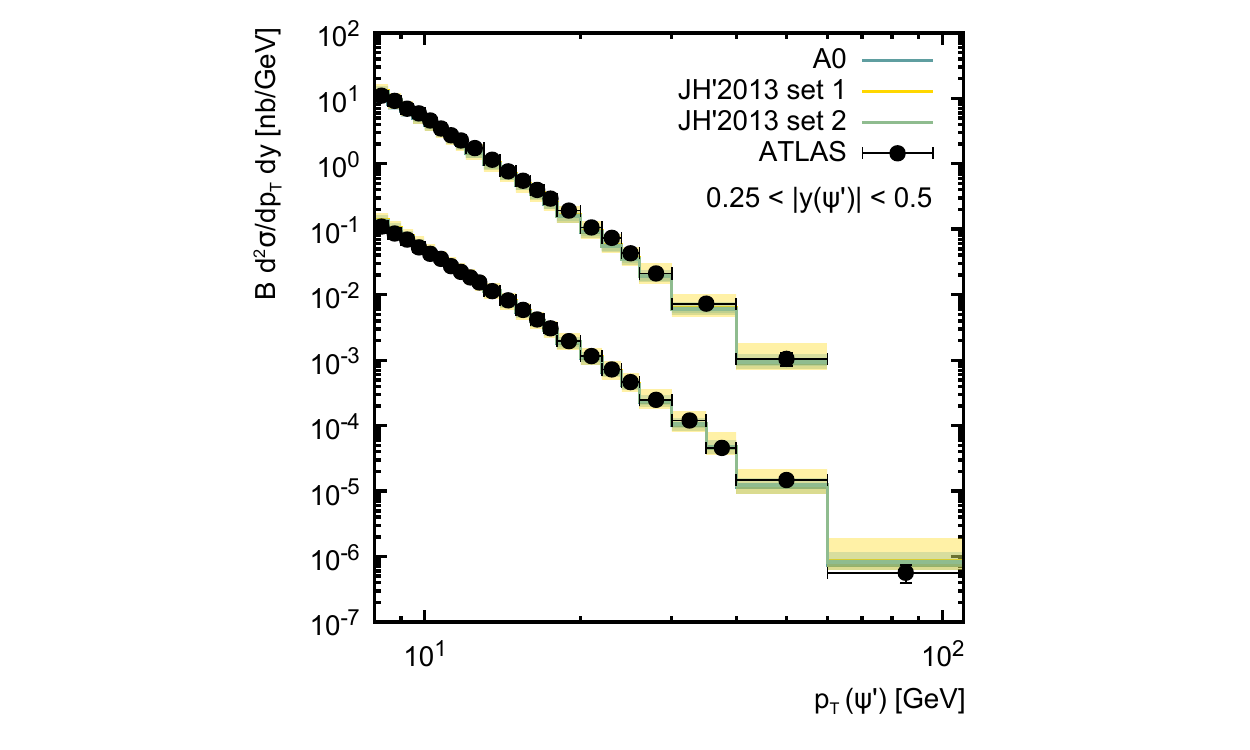}
\includegraphics[width=8.1cm]{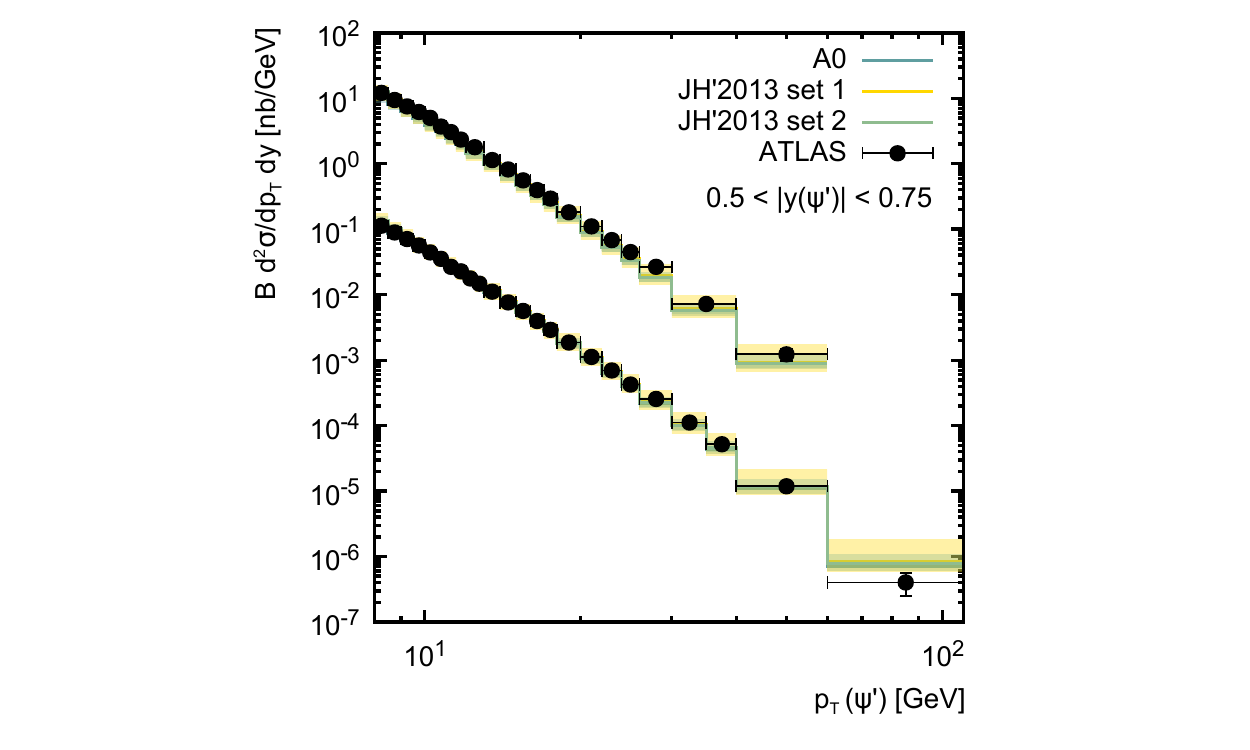}
\includegraphics[width=8.1cm]{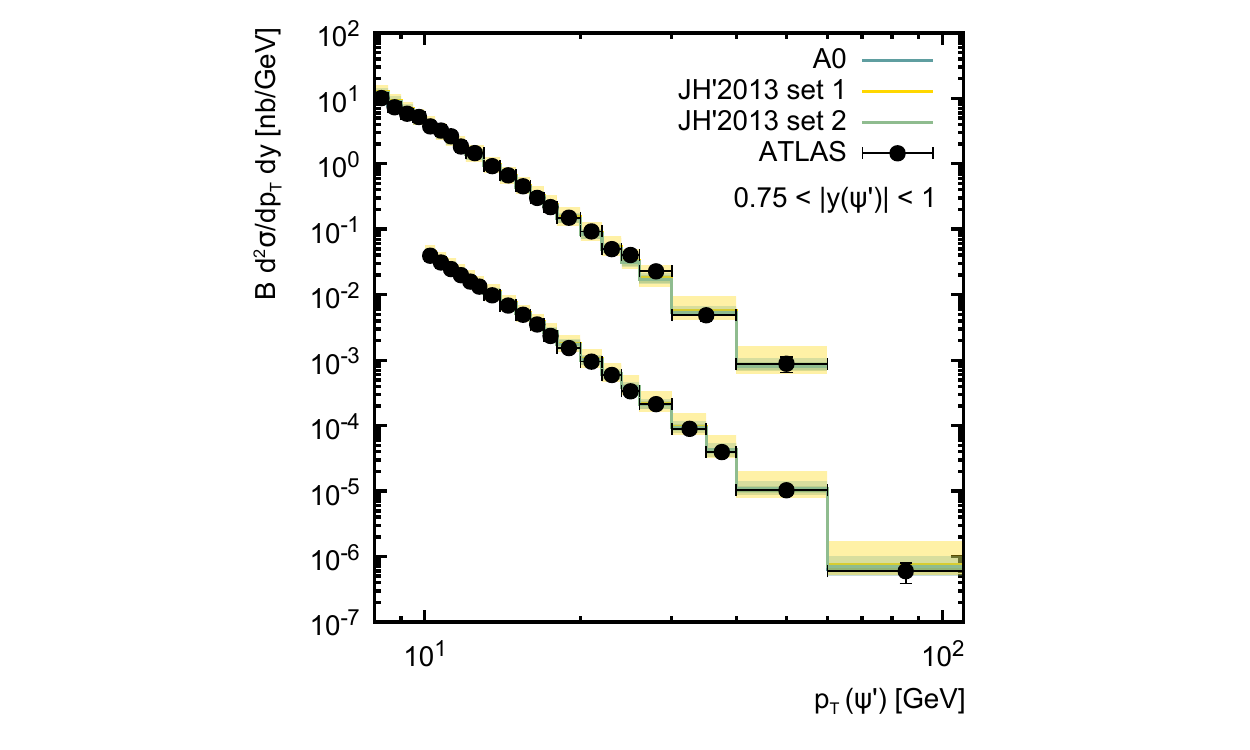}
\includegraphics[width=8.1cm]{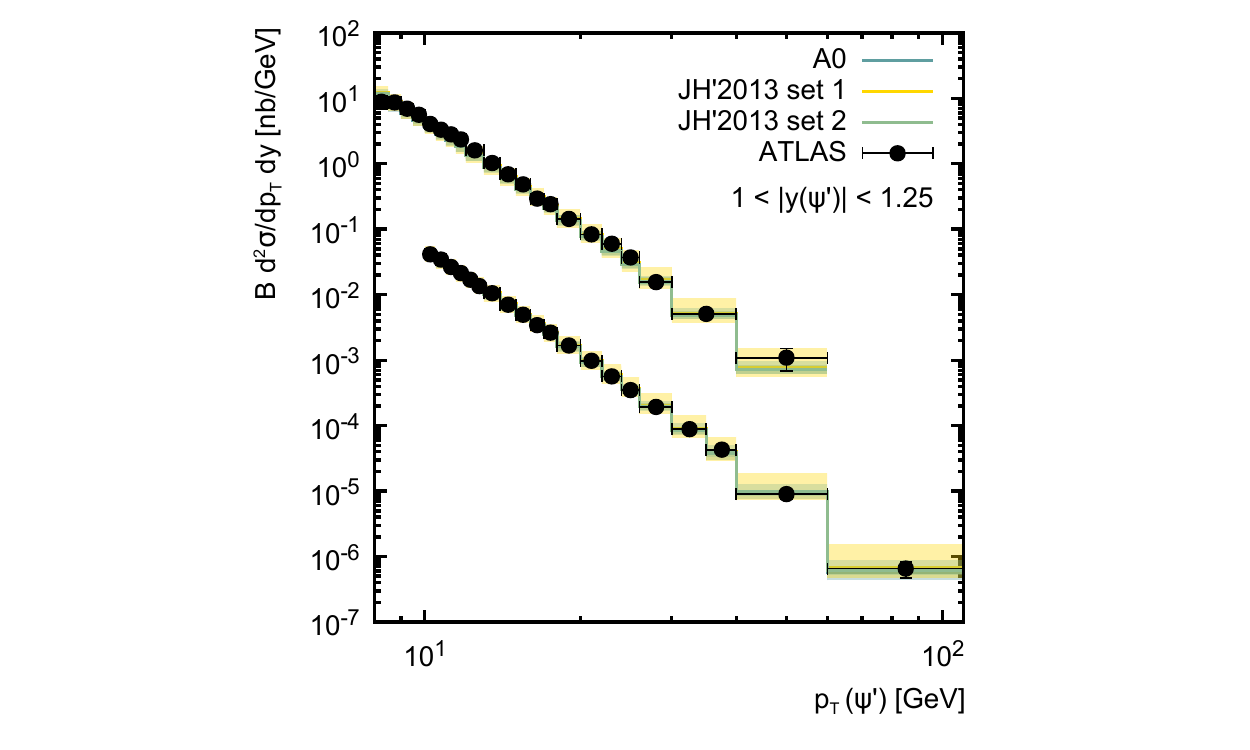}
\includegraphics[width=8.1cm]{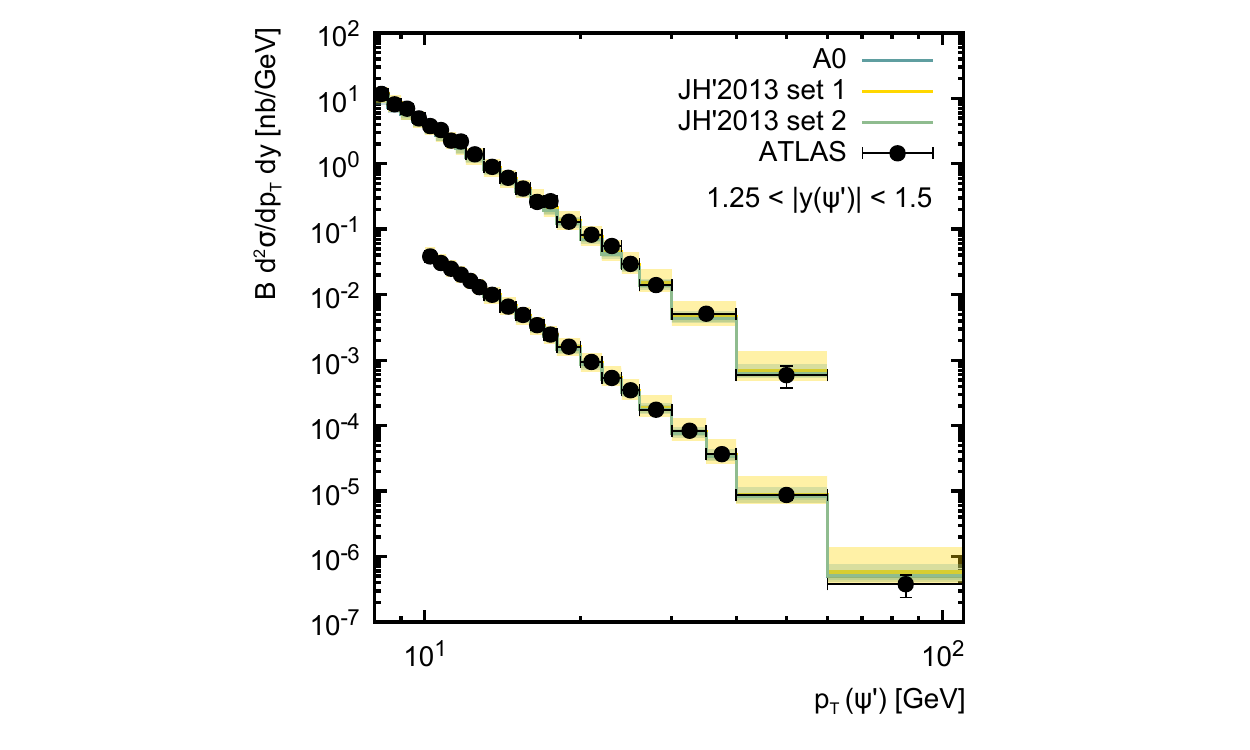}
\includegraphics[width=8.1cm]{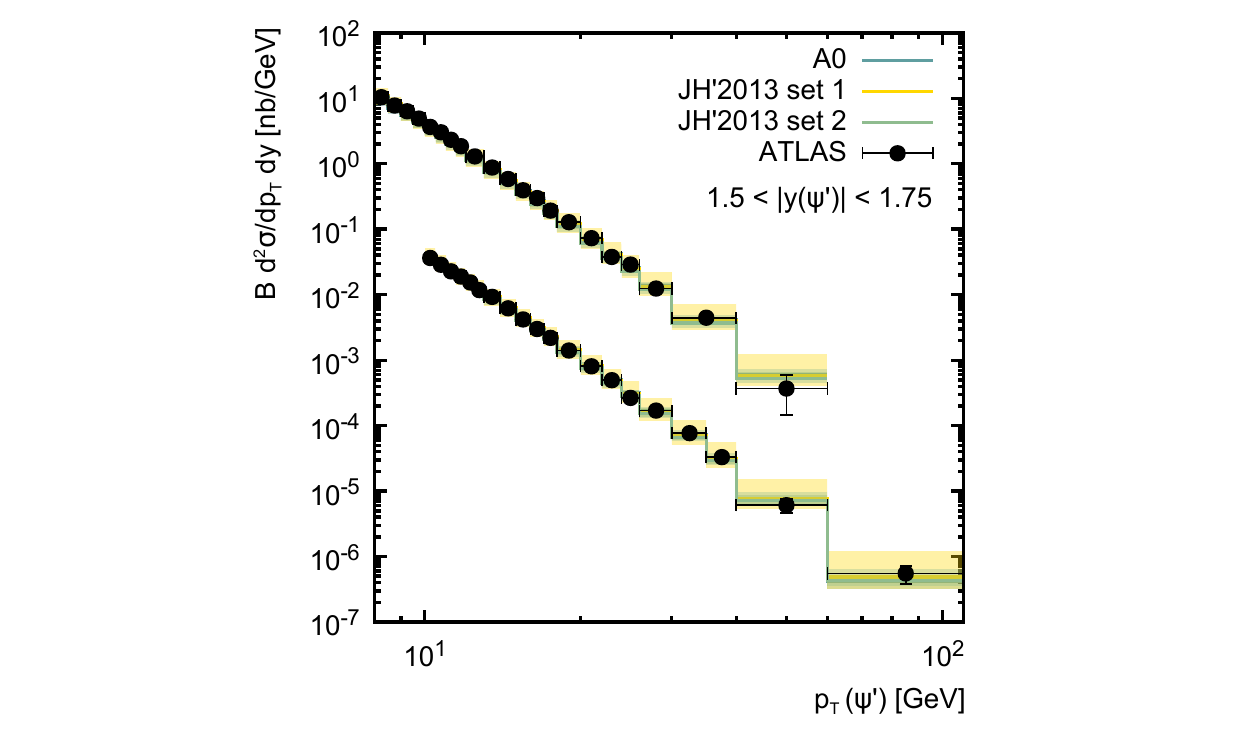}
\includegraphics[width=8.1cm]{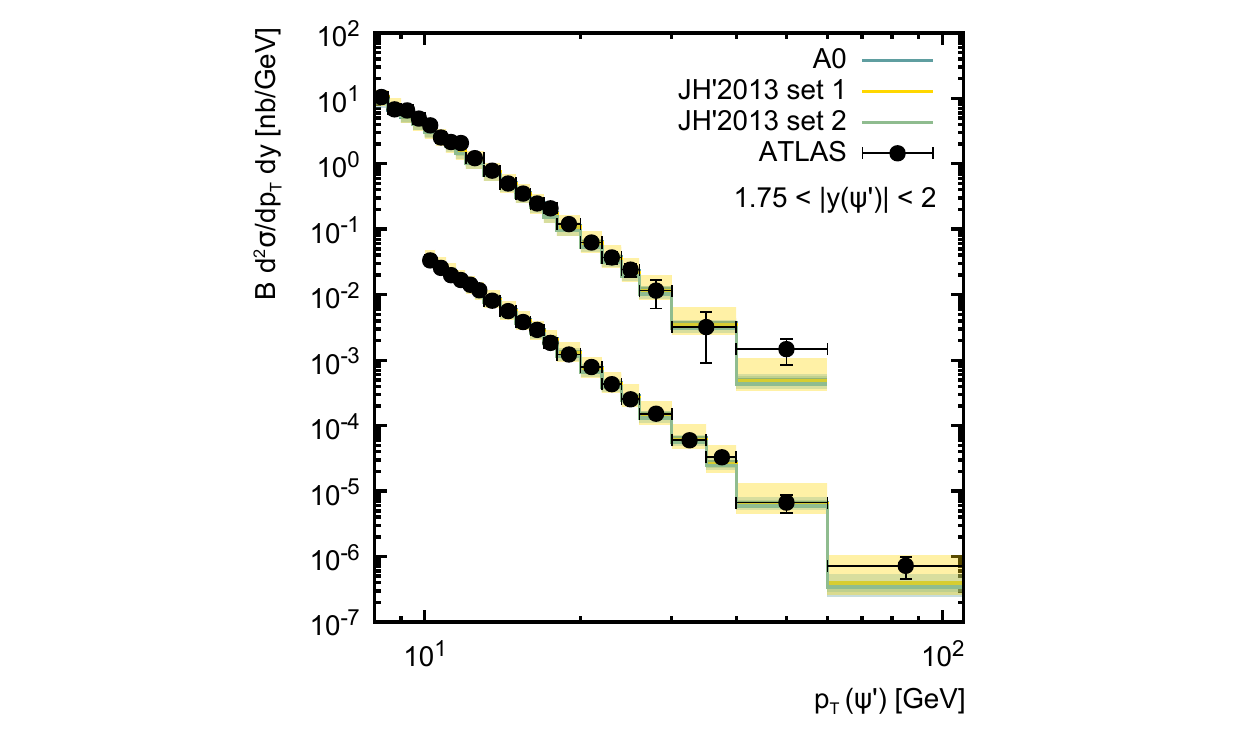}
\caption{Transverse momentum distribution of prompt $\psi^\prime$ mesons 
produced in $pp$ collisions at $\sqrt s = 7$~TeV (upper histograms, multiplied by $100$) and
$\sqrt s = 8$~TeV (lower histograms) at different rapidities. Shaded bands represent the total uncertainties of our 
calculations (scale uncertainties and the uncertainties coming from LDMEs fit, summed in quadrature). 
The experimental data are from ATLAS\cite{38}.}
\label{fig1}
\end{center}
\end{figure}

\begin{figure}
\begin{center}
\includegraphics[width=8.1cm]{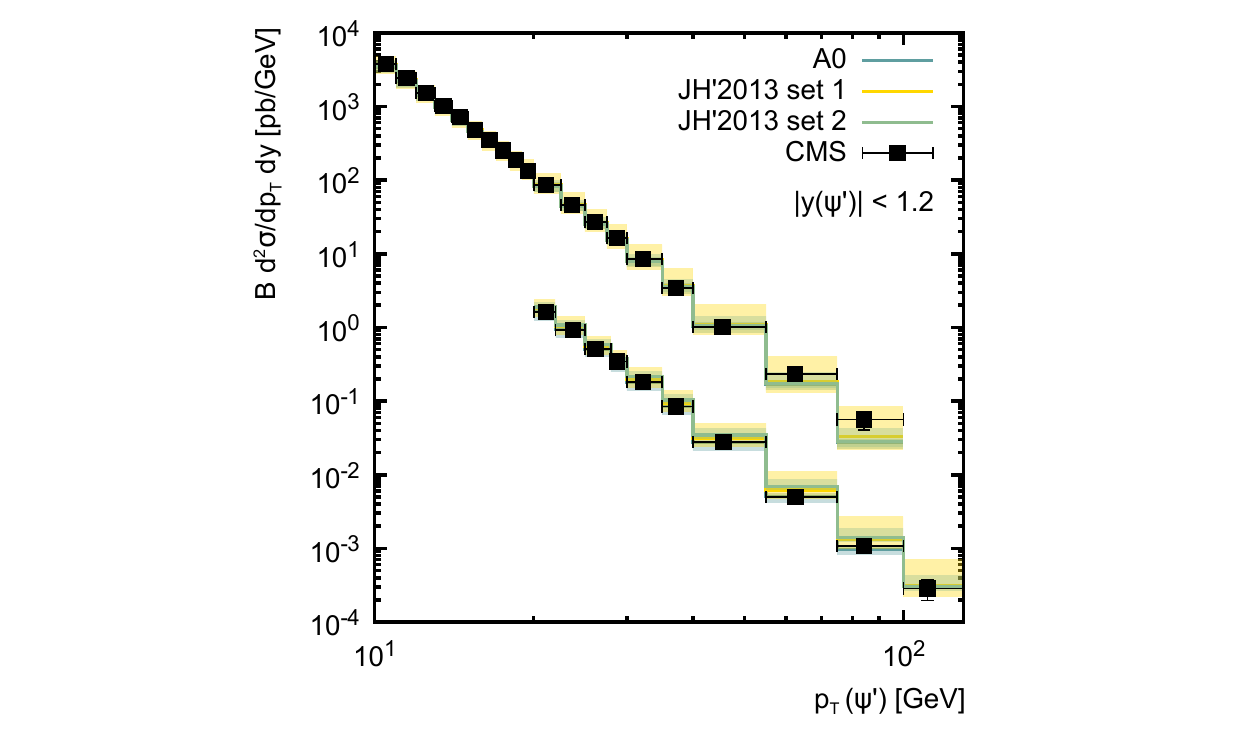}
\includegraphics[width=8.1cm]{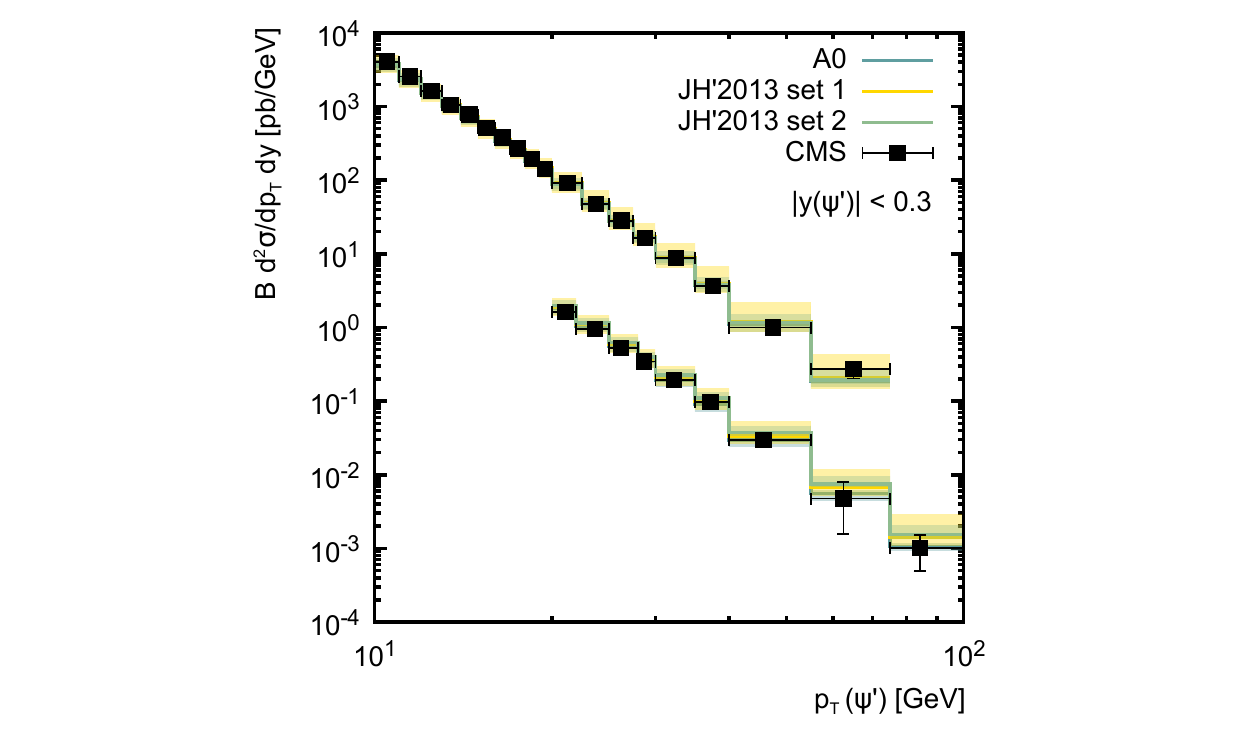}
\includegraphics[width=8.1cm]{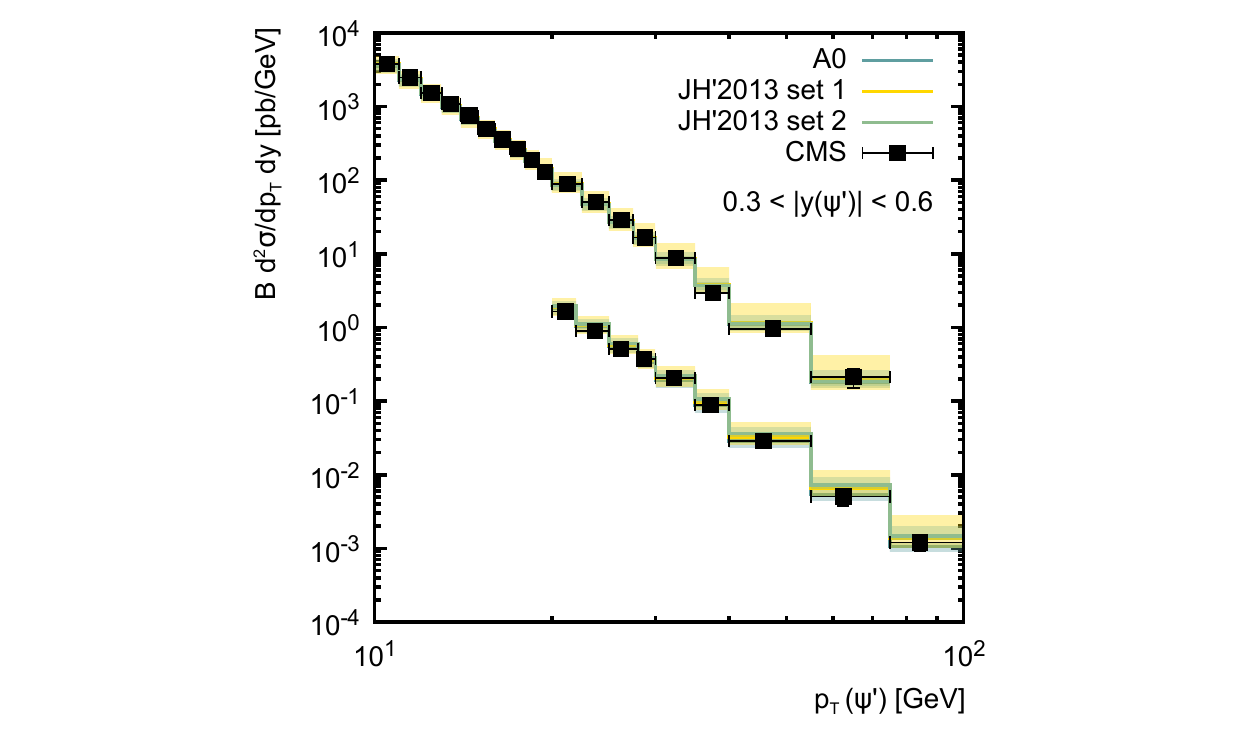}
\includegraphics[width=8.1cm]{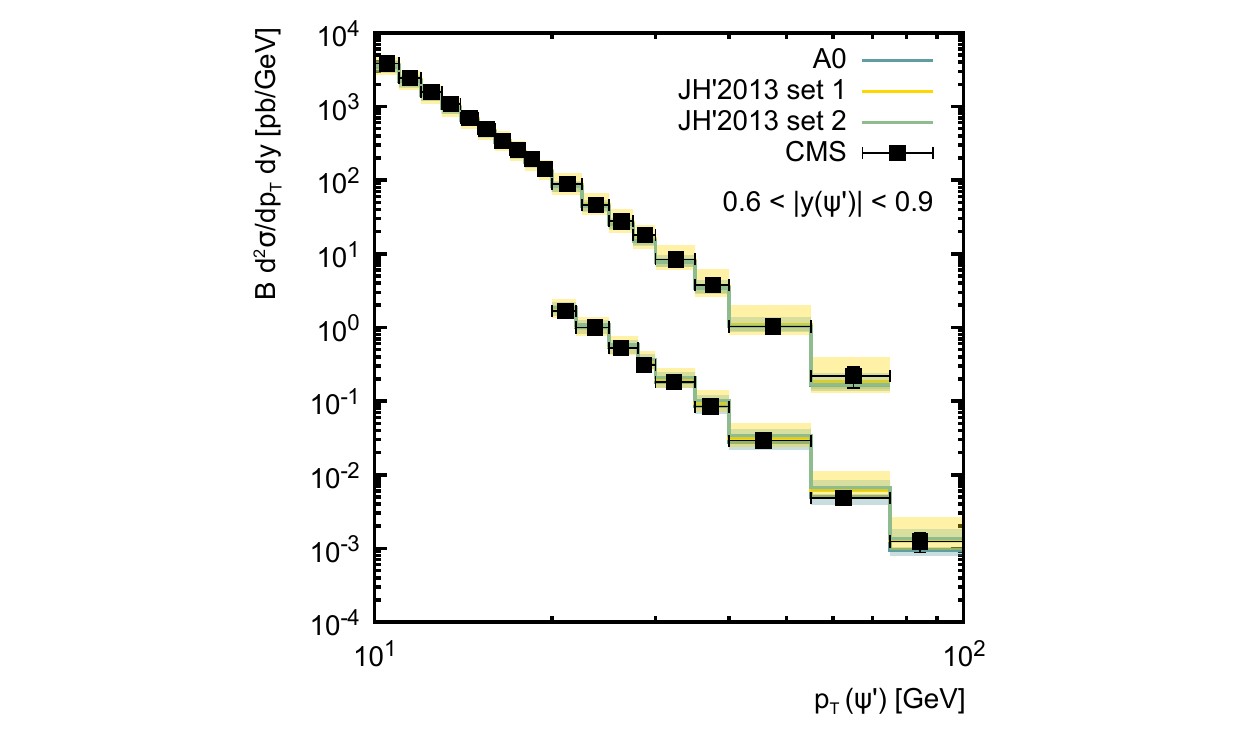}
\includegraphics[width=8.1cm]{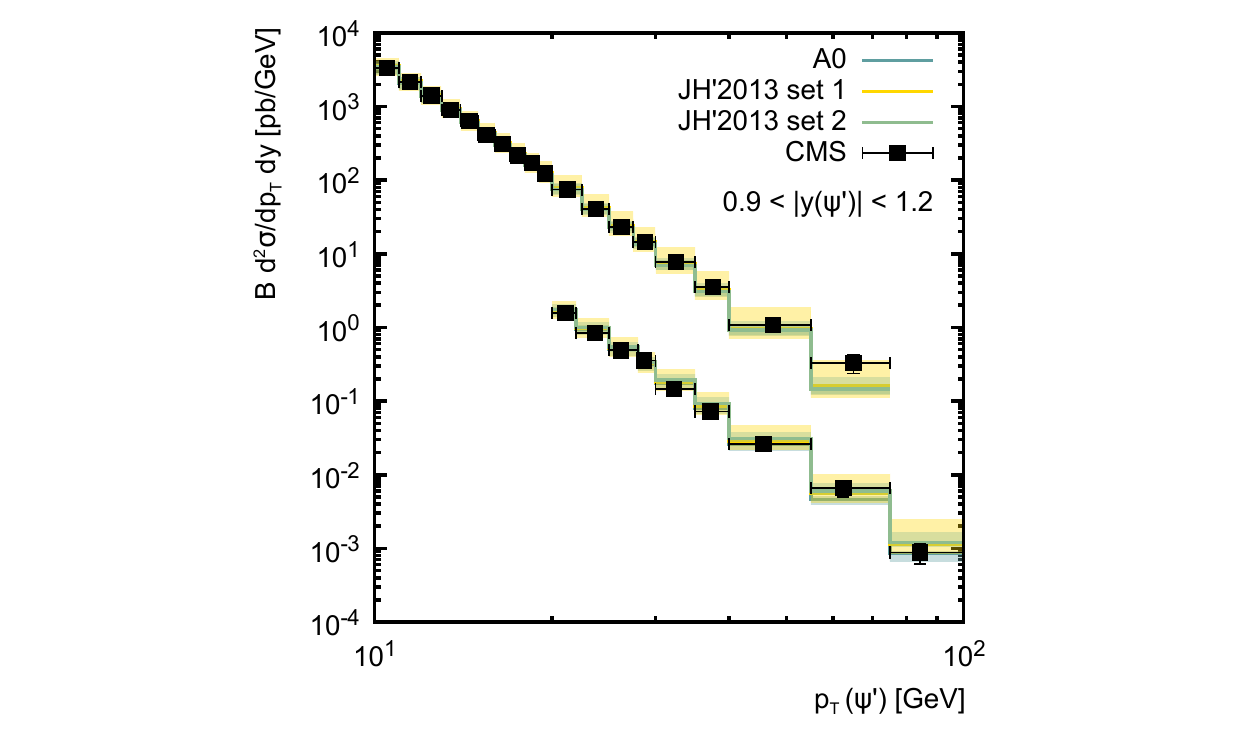}
\caption{Transverse momentum distribution of prompt $\psi^\prime$ mesons 
produced in $pp$ collisions at $\sqrt s = 7$~TeV (upper histograms, multiplied by $100$) and
$\sqrt s = 13$~TeV (lower histograms) at different rapidities. Notation of all histograms 
is the same as in Fig.~1. The experimental data are from CMS\cite{41,42} and LHCb\cite{45}.}
\label{fig2}
\end{center}
\end{figure}

\begin{figure}
\begin{center}
\includegraphics[width=8.1cm]{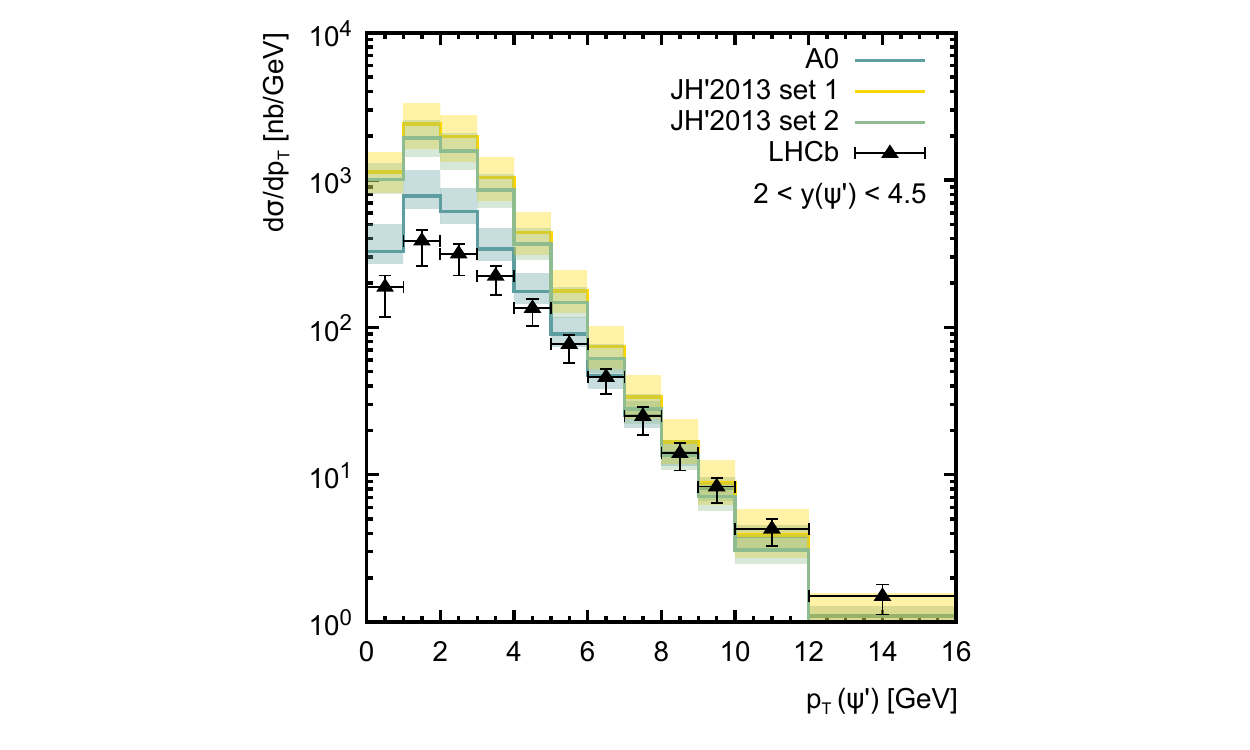}
\includegraphics[width=8.1cm]{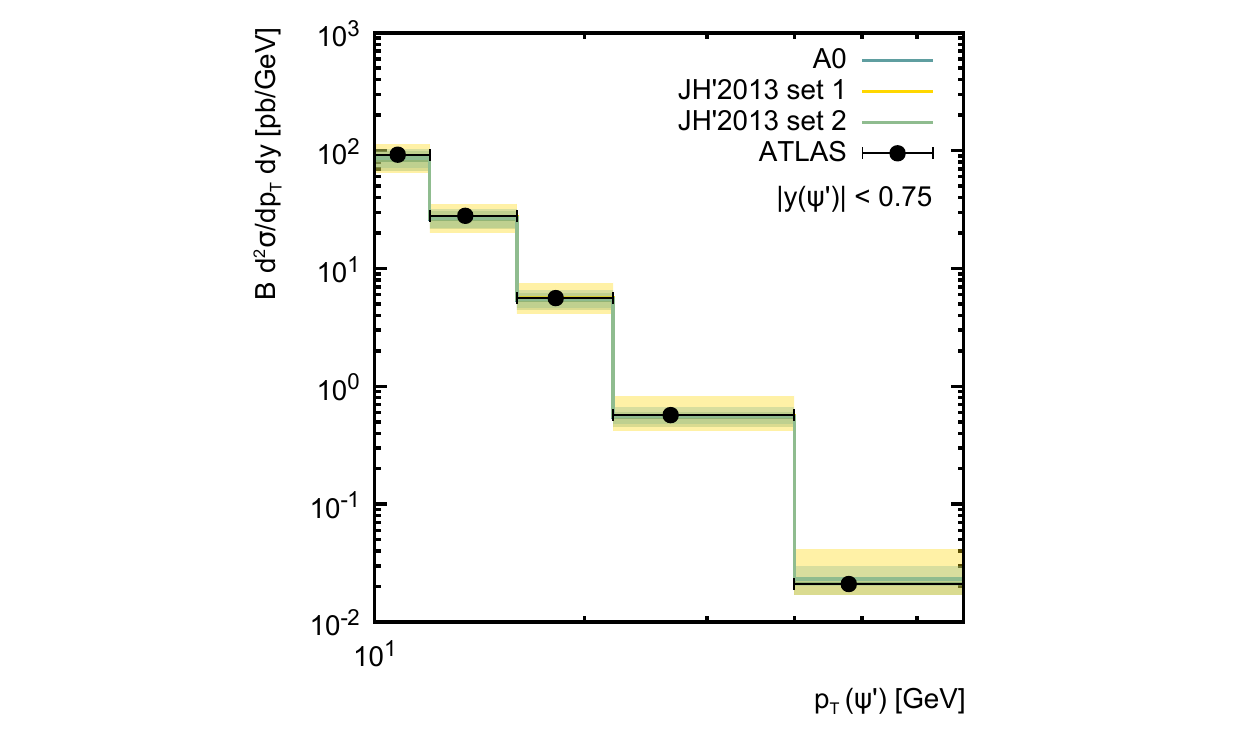}
\caption{Transverse momentum distribution of prompt $\psi^\prime$ mesons 
produced in $pp$ collisions at $\sqrt s = 7$~TeV (left panel) and
$\sqrt s = 8$~TeV (right panel) at different rapidities. Notation of all histograms 
is the same as in Fig.~1. The experimental data are from LHCb\cite{45} and ATLAS\cite{39}.}
\label{fig3}
\end{center}
\end{figure}

\begin{figure}
\begin{center}
\includegraphics[width=8.1cm]{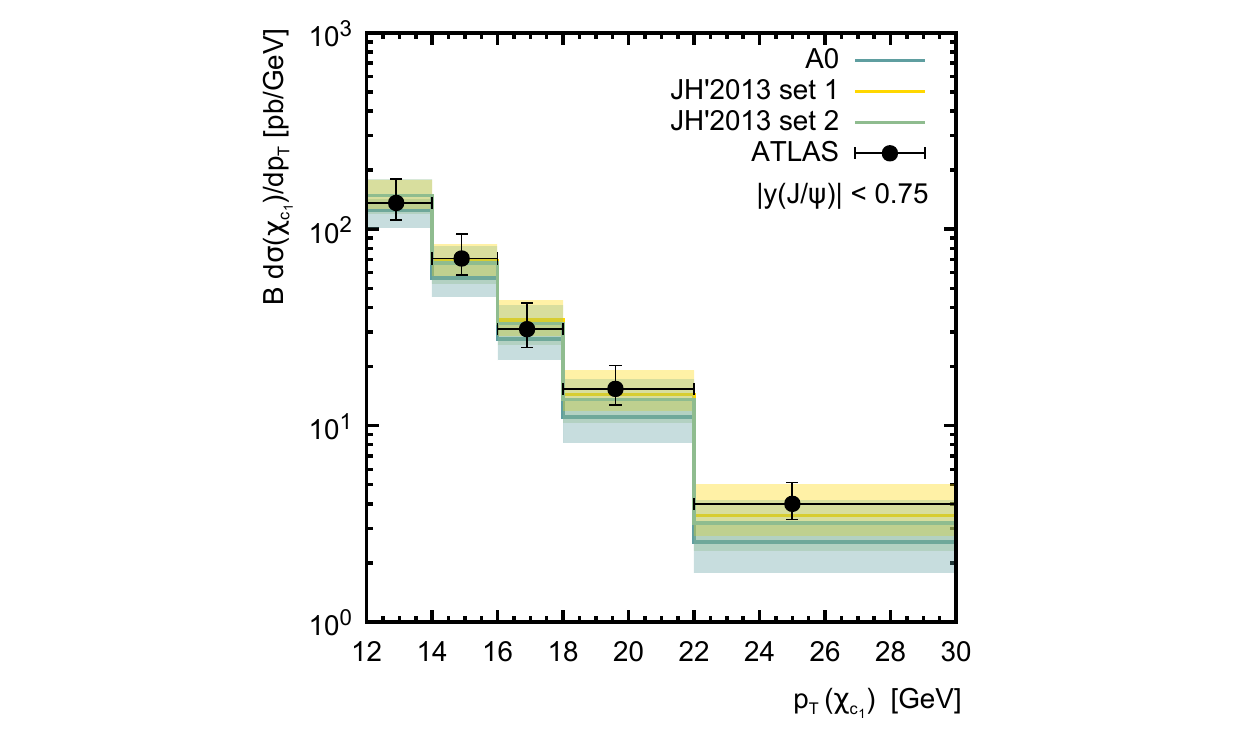}
\includegraphics[width=8.1cm]{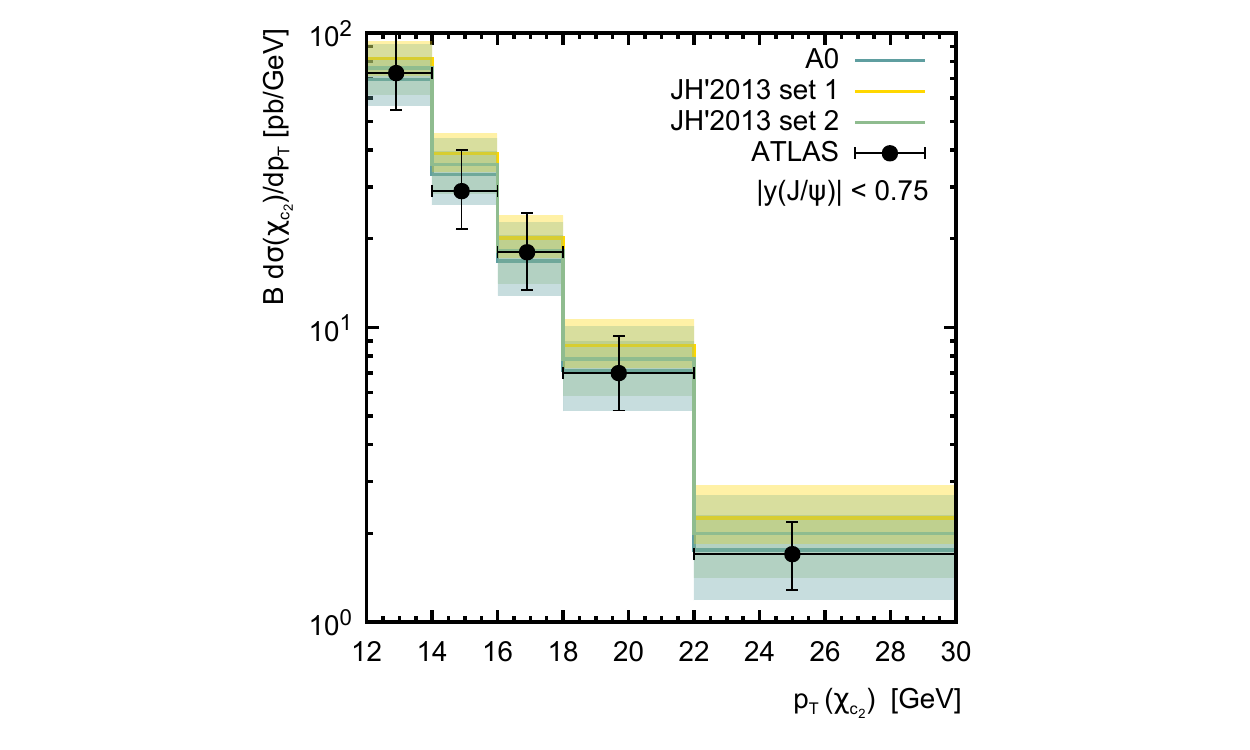}
\includegraphics[width=8.1cm]{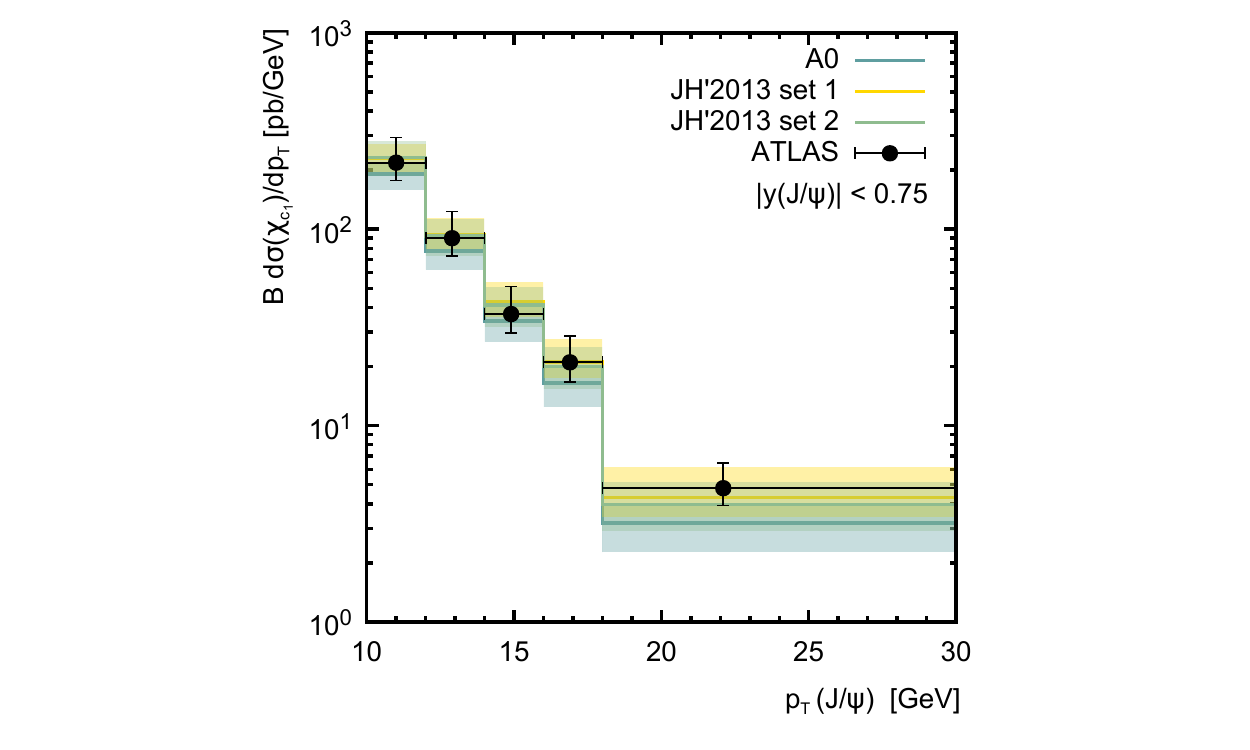}
\includegraphics[width=8.1cm]{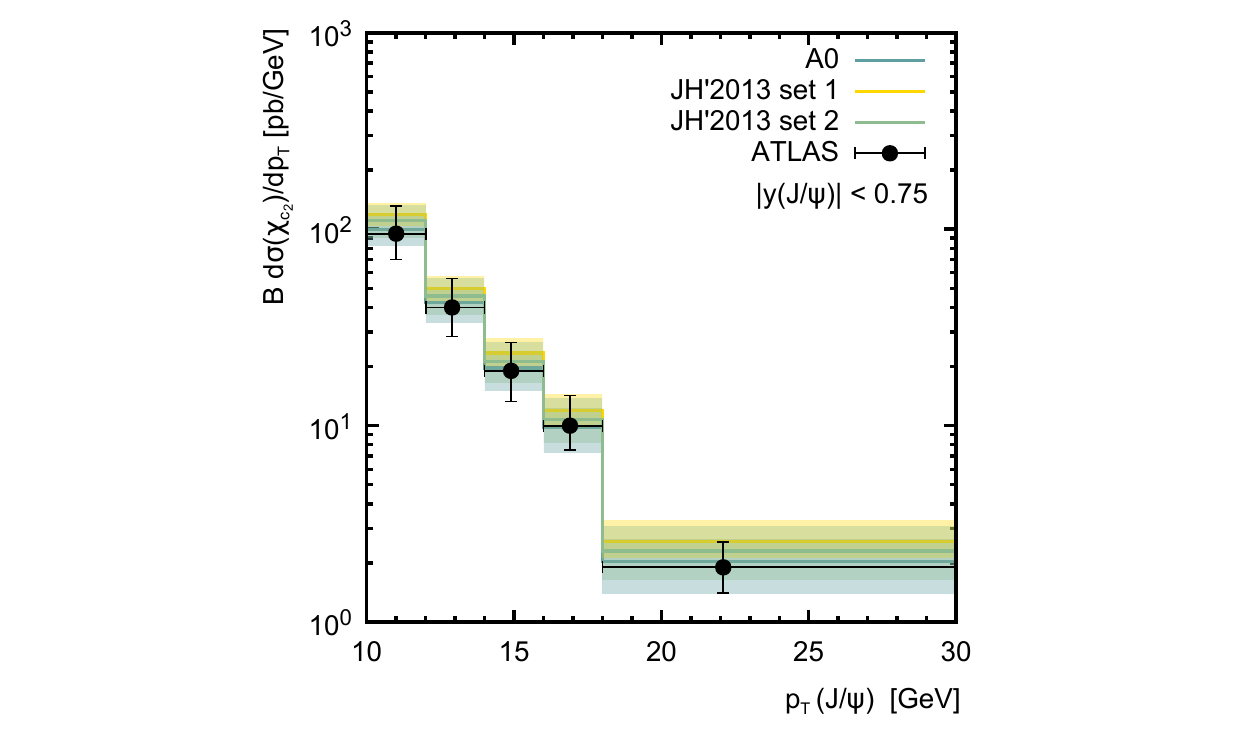}
\caption{The prompt $\chi_{c1}$ and $\chi_{c2}$ meson production 
in $pp$ collisions at $\sqrt s = 7$~TeV as a function of $\chi_c$ (upper panels) and 
decay $J/\psi$ (lower panels) transverse momenta. 
Notation of all histograms 
is the same as in Fig.~1. The experimental data are from ATLAS\cite{37}.}
\label{fig4}
\end{center}
\end{figure}

\begin{figure}
\begin{center}
\includegraphics[width=8.1cm]{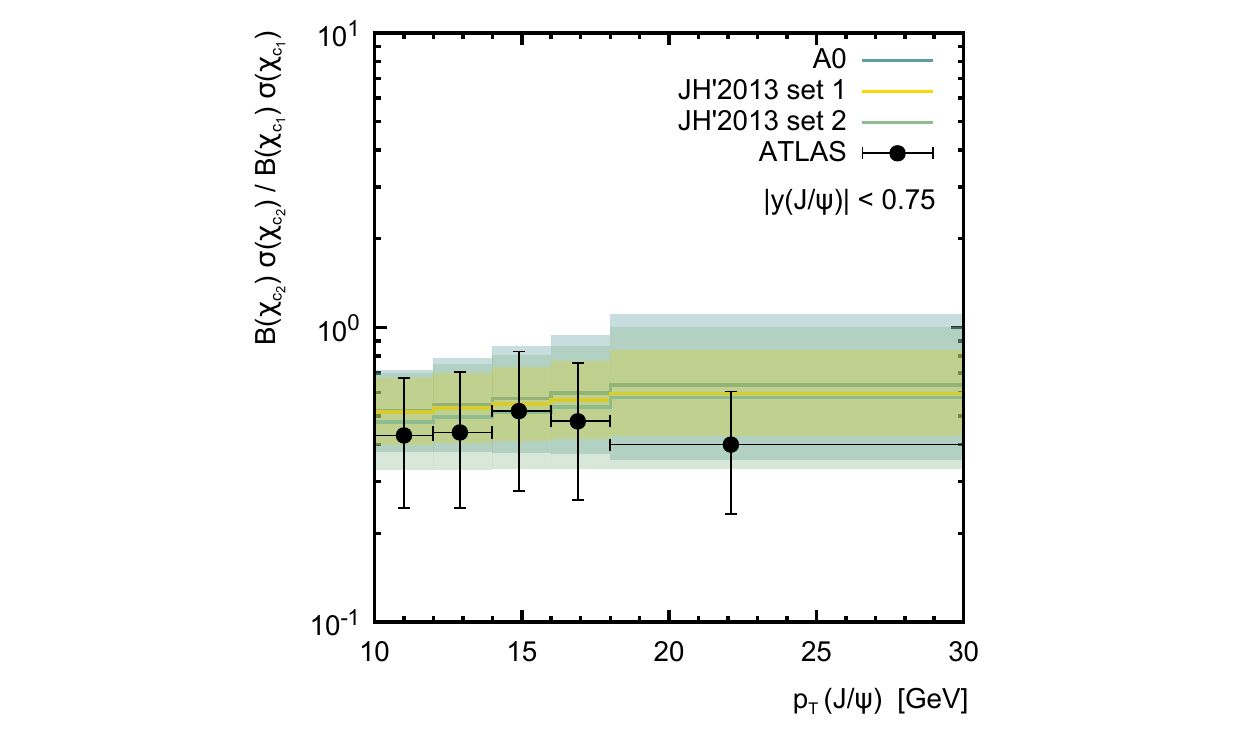}
\includegraphics[width=8.1cm]{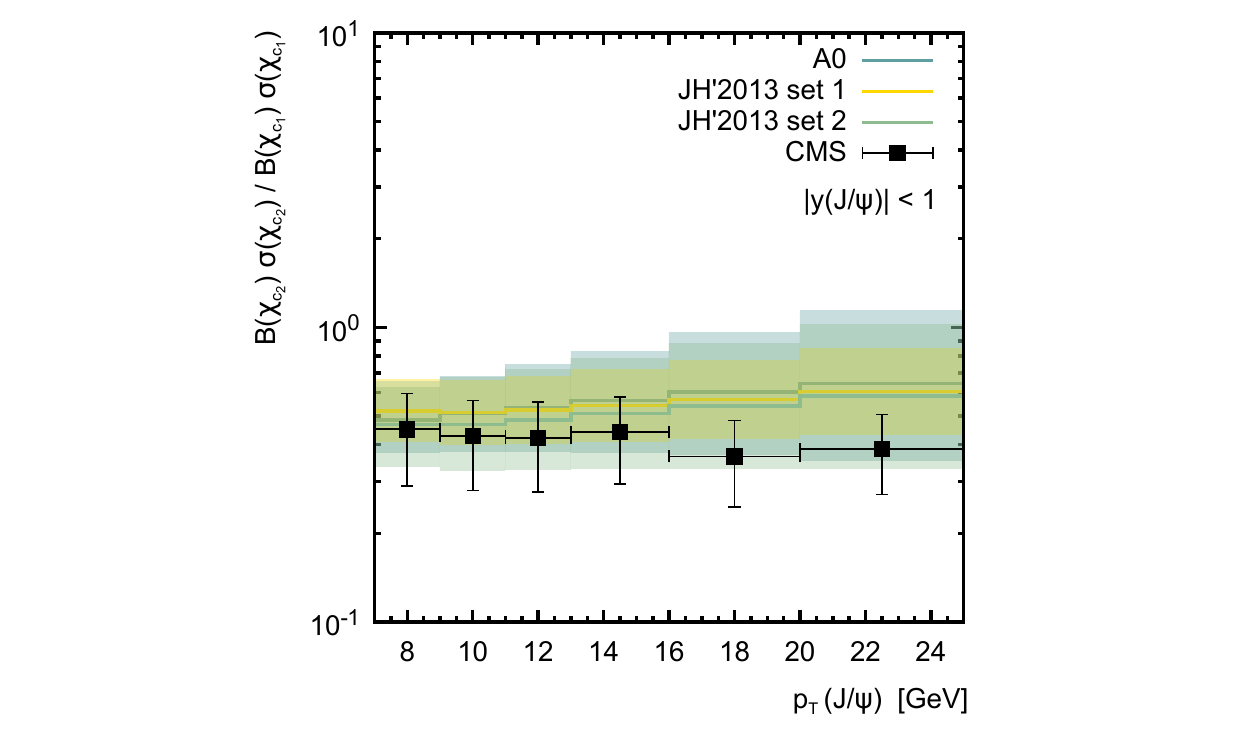} 
\includegraphics[width=8.1cm]{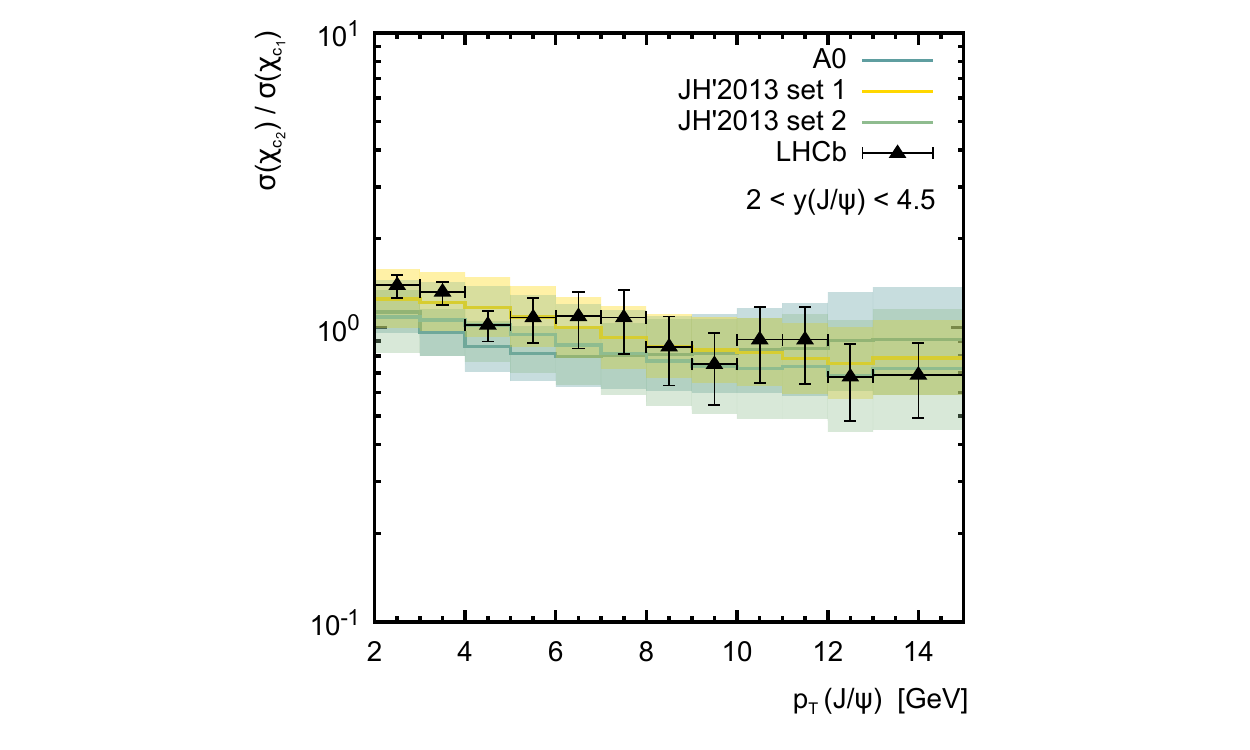} 
\includegraphics[width=8.1cm]{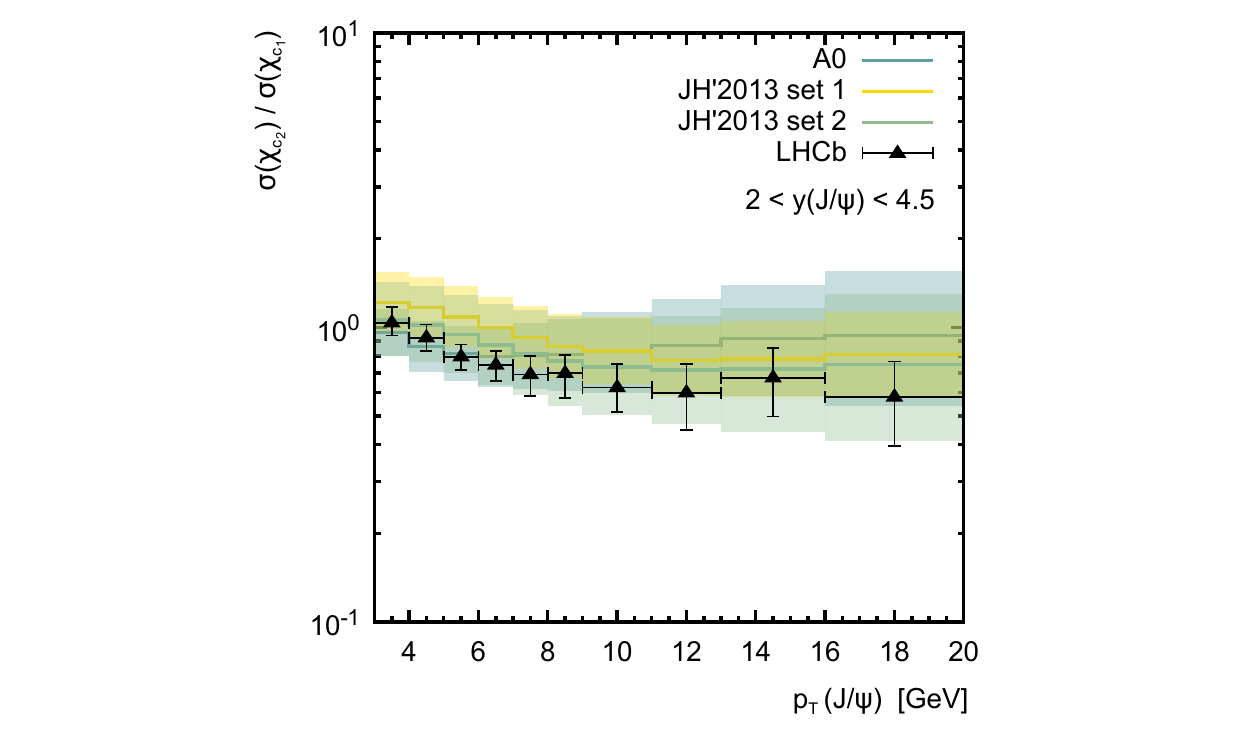} 
\caption{The relative production rate $\sigma(\chi_{c2})/\sigma(\chi_{c1})$
calculated as a function of decay $J/\psi$ meson transverse momenta at $\sqrt s = 7$~TeV. 
Notation of all curves is the same as in Fig.~1. The
experimental data are from ATLAS\cite{37}, CMS\cite{40} and LHCb\cite{43,44}.}
\label{fig5}
\end{center}
\end{figure}

\begin{figure}
\begin{center}
\includegraphics[width=8.1cm]{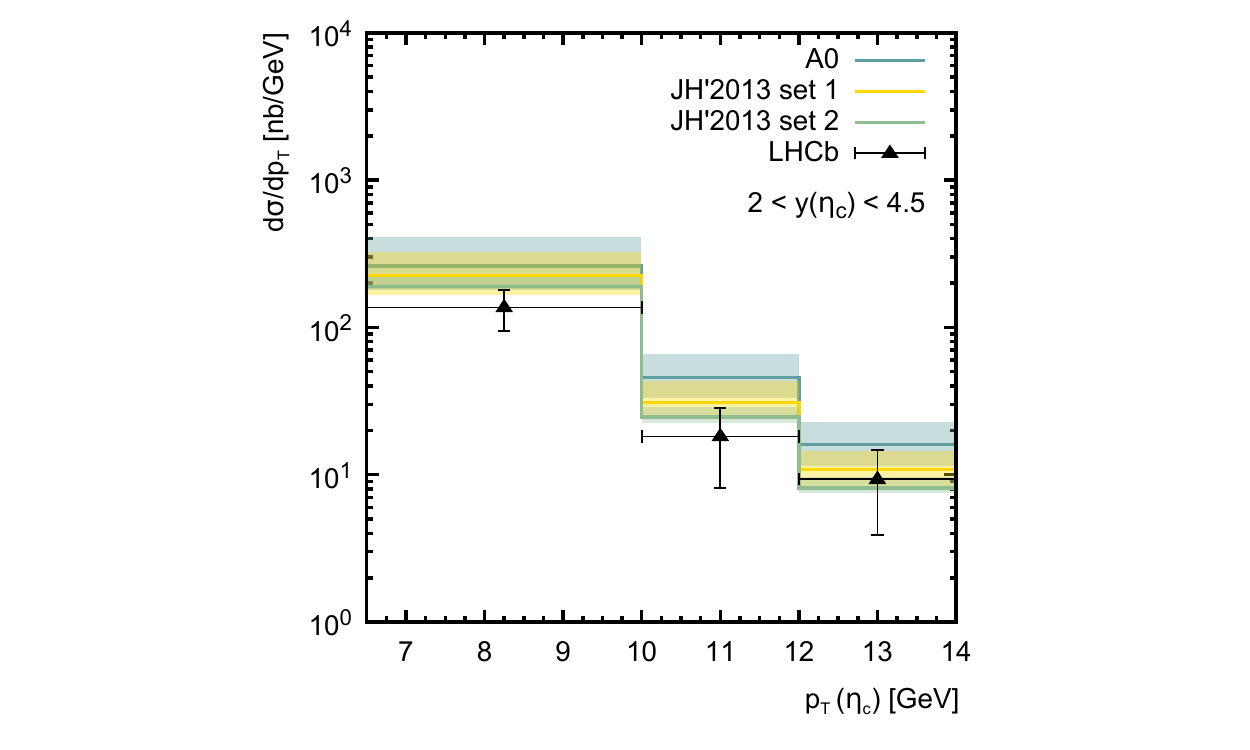}
\includegraphics[width=8.1cm]{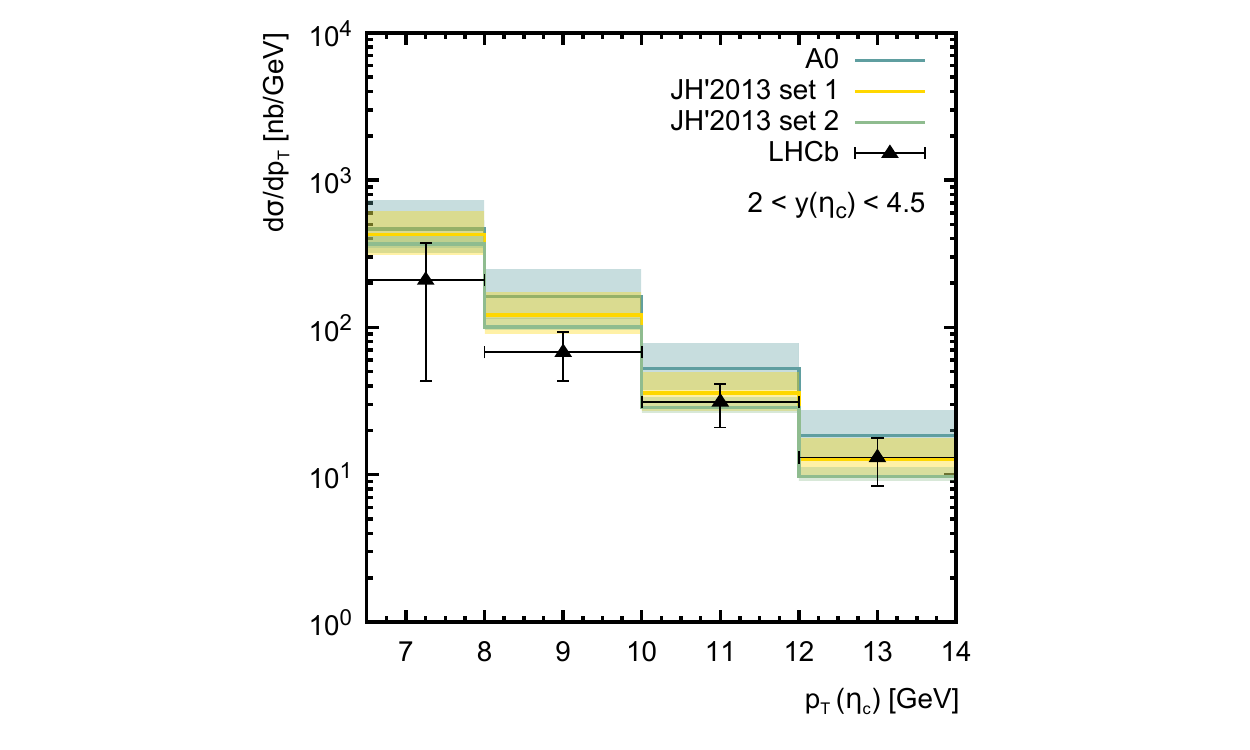}
\caption{Transverse momentum distribution of prompt $\eta_c$ mesons 
produced in $pp$ collisions at $\sqrt s = 7$~TeV (left panel) and $\sqrt s = 8$~TeV (right panel). 
Notation of all histograms is the same as in Fig.~1. The experimental data are from LHCb\cite{22}.}
\label{fig6}
\end{center}
\end{figure}

\begin{figure}
\begin{center}
\includegraphics[width=8.1cm]{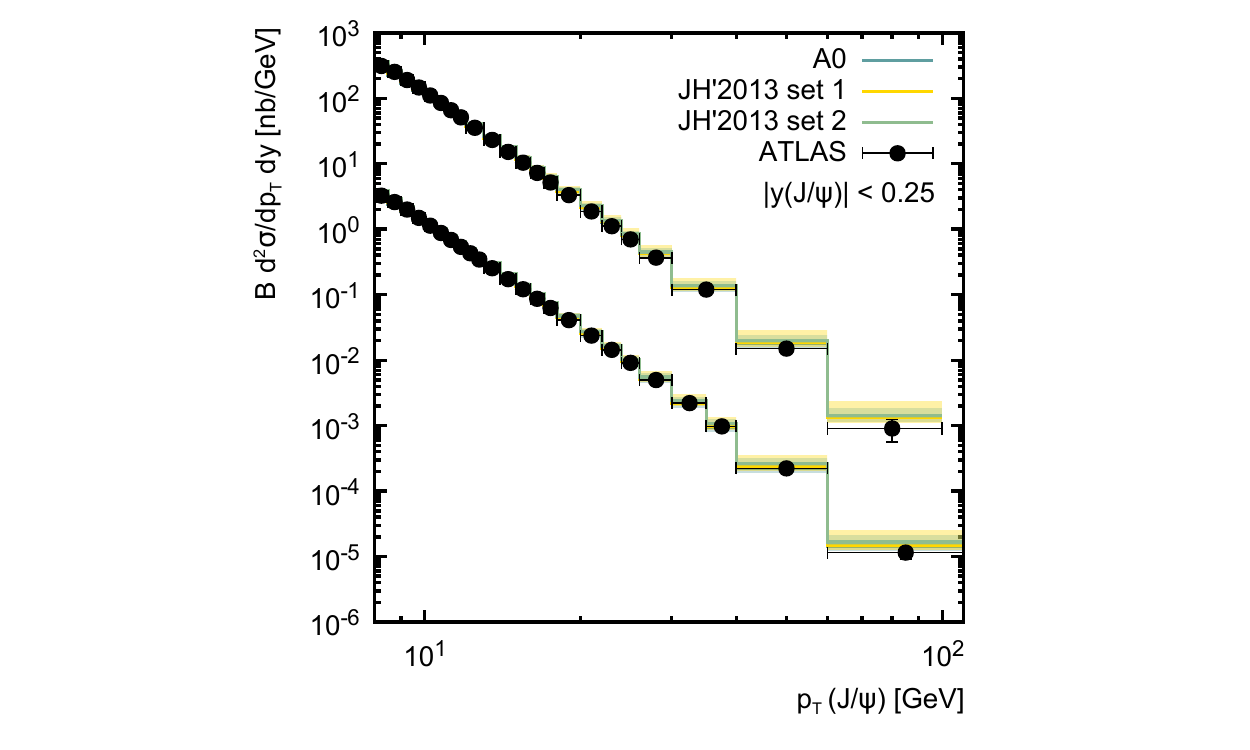}
\includegraphics[width=8.1cm]{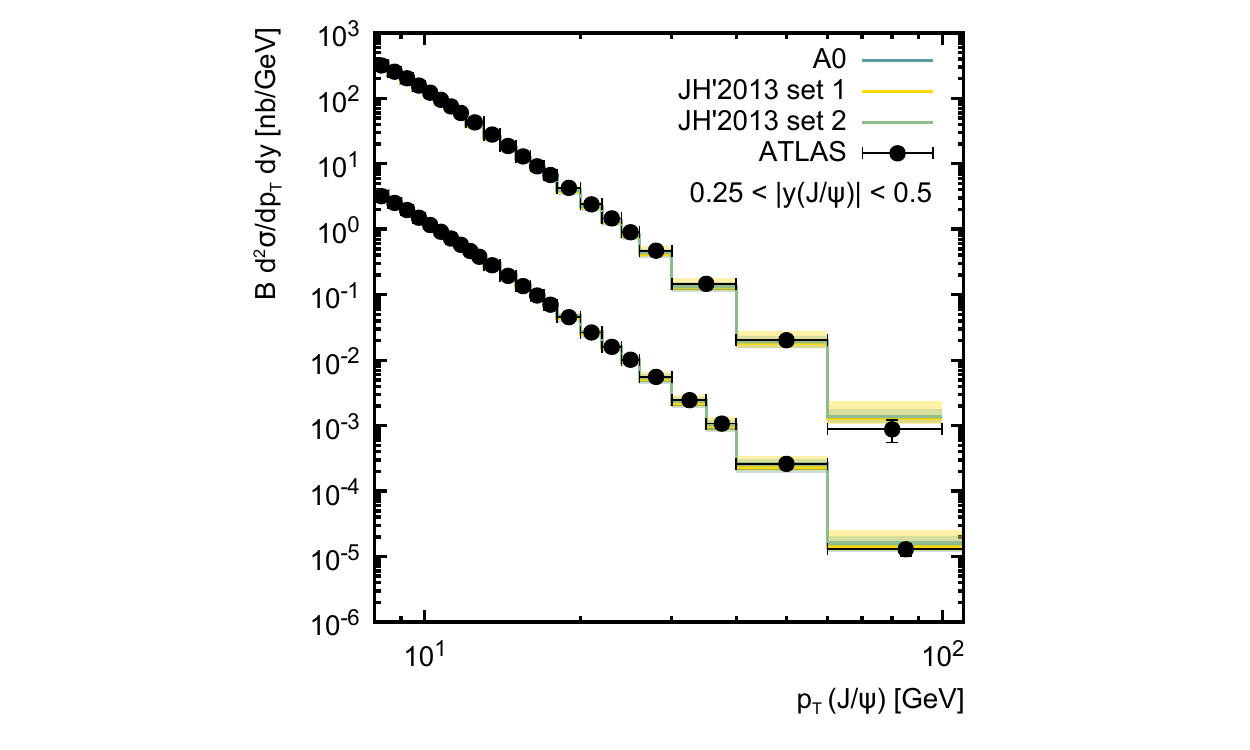}
\includegraphics[width=8.1cm]{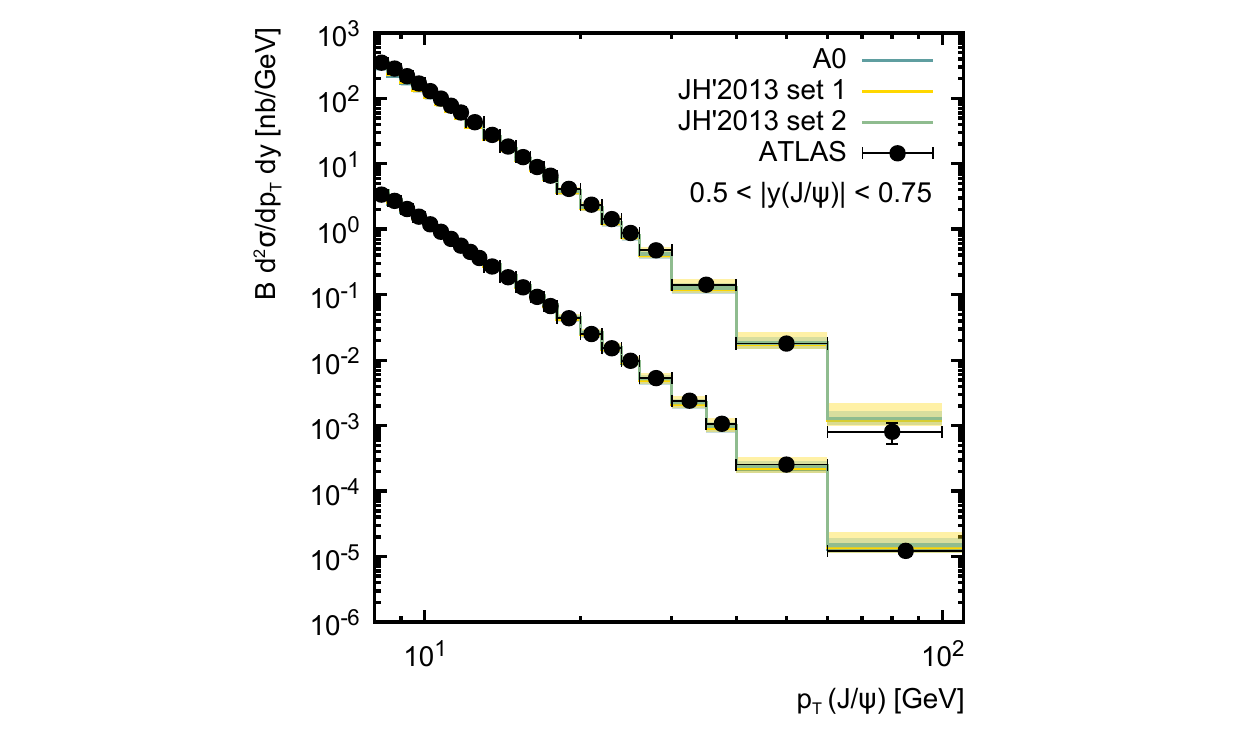}
\includegraphics[width=8.1cm]{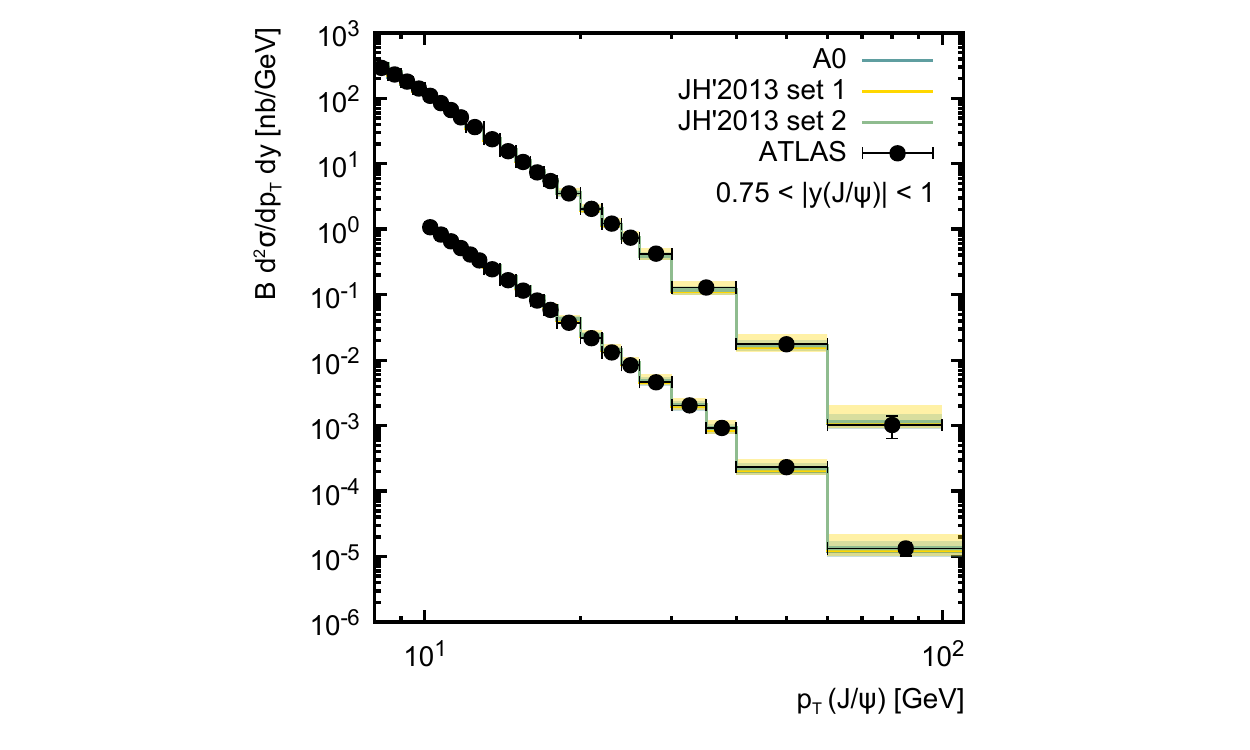}
\includegraphics[width=8.1cm]{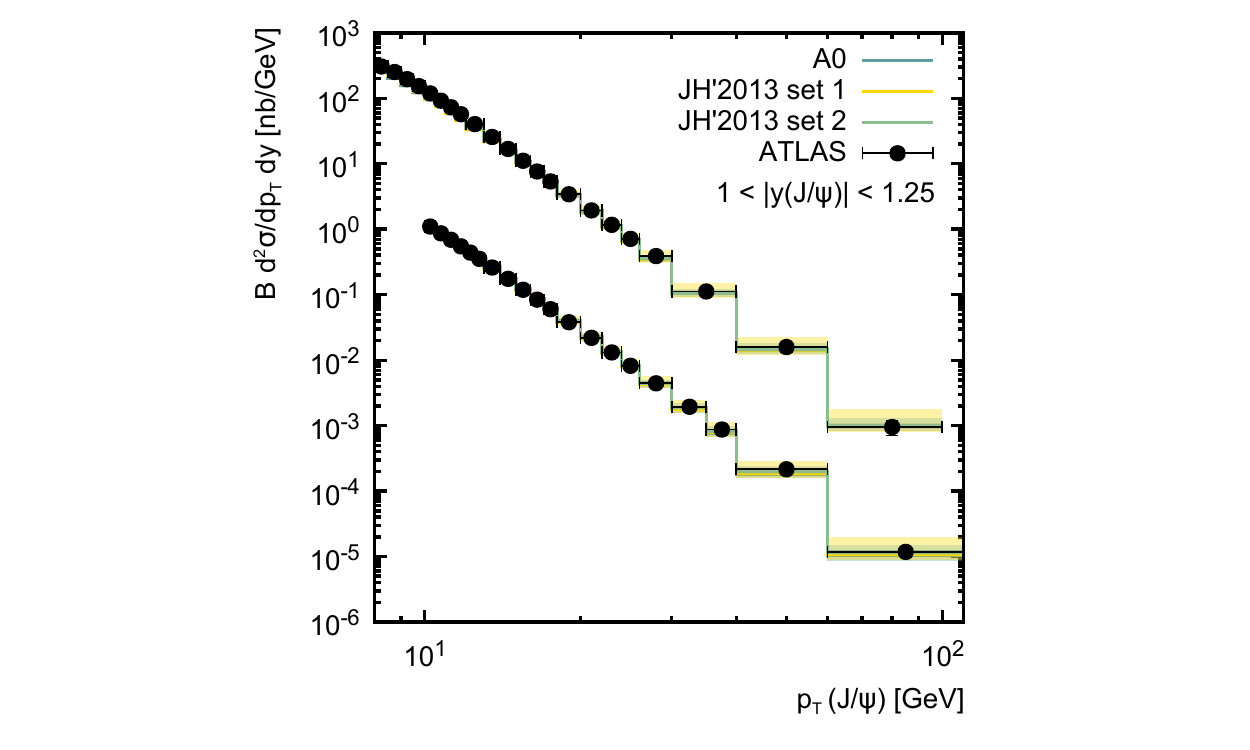}
\includegraphics[width=8.1cm]{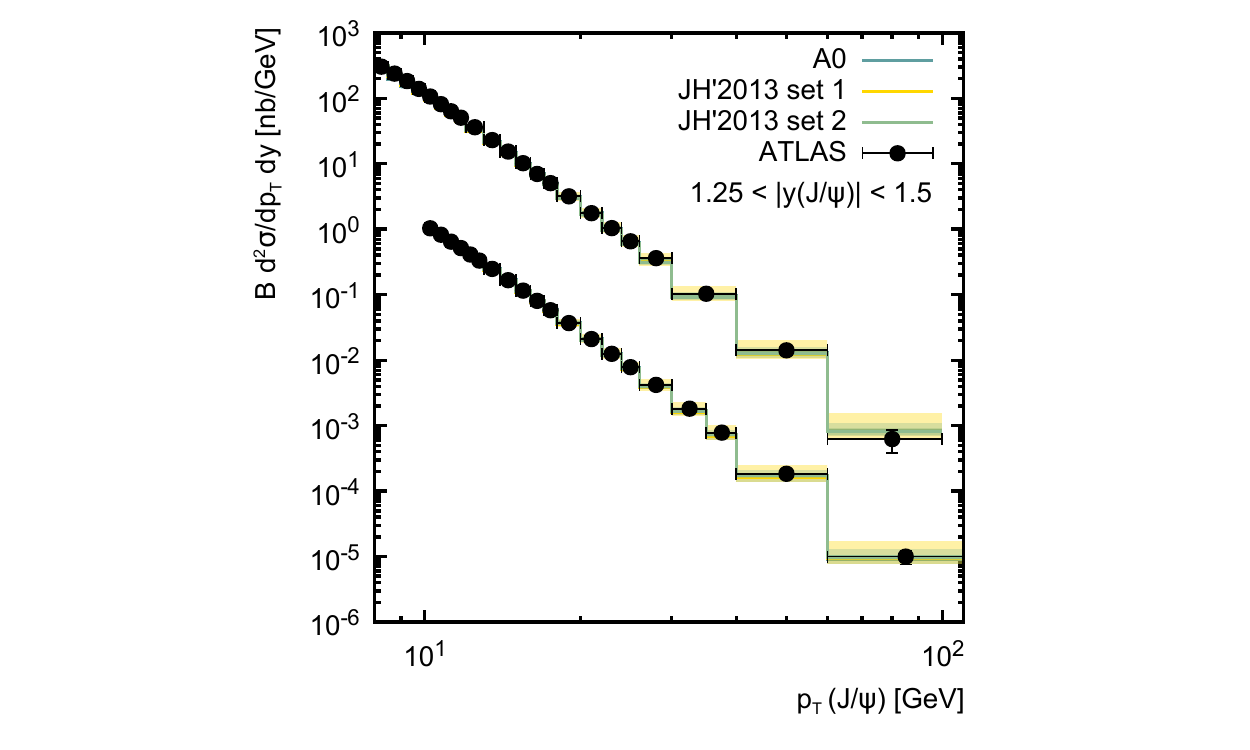}
\includegraphics[width=8.1cm]{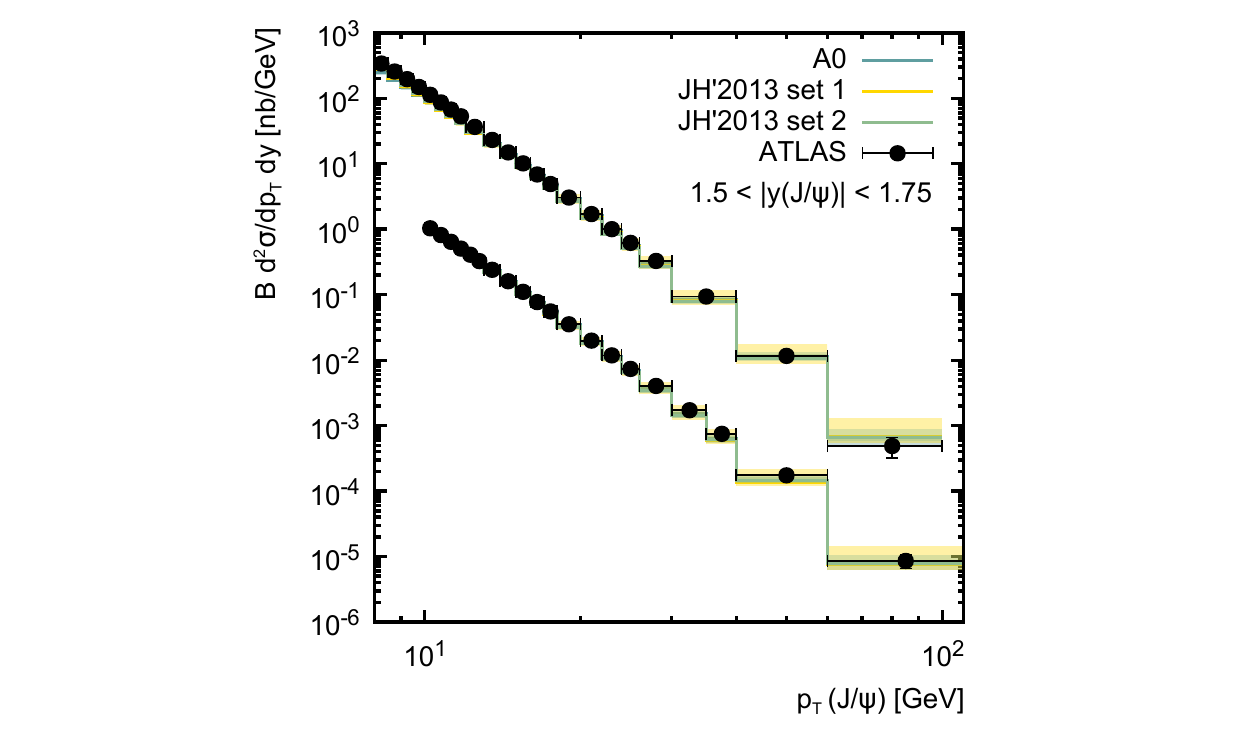}
\includegraphics[width=8.1cm]{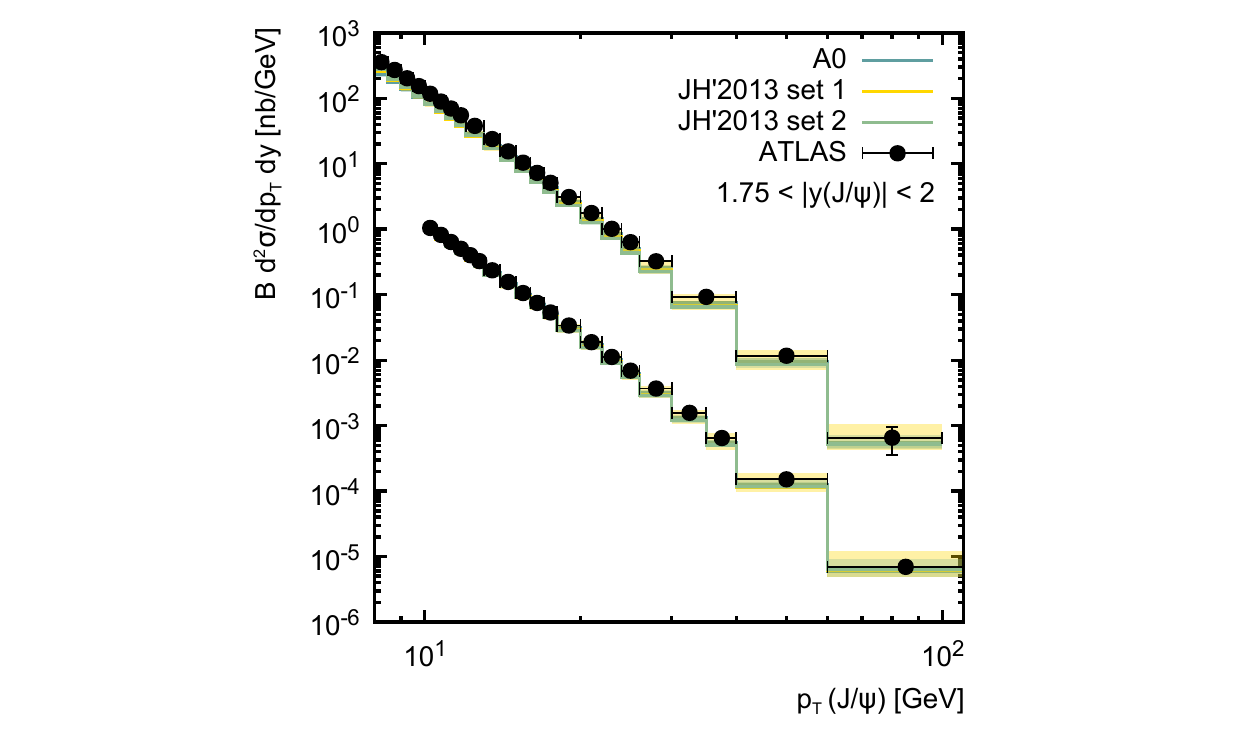}
\caption{Transverse momentum distribution of prompt $J/\psi$ mesons 
produced in $pp$ collisions at $\sqrt s = 7$~TeV (upper histograms, multiplied by $100$) and
$\sqrt s = 8$~TeV (lower histograms) at different rapidities. Notation of all curves is the same as in Fig.~1.
The experimental data are from ATLAS\cite{38}.}
\label{fig7}
\end{center}
\end{figure}

\begin{figure}
\begin{center}
\includegraphics[width=8.1cm]{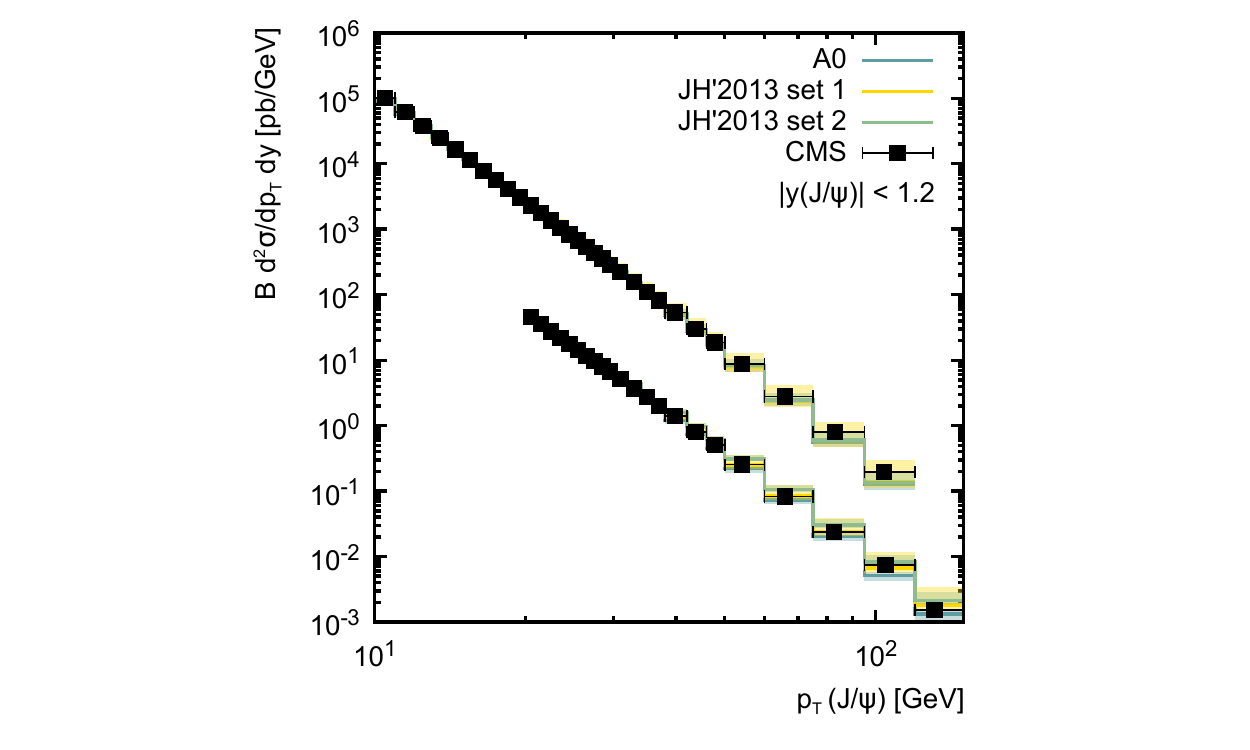}
\includegraphics[width=8.1cm]{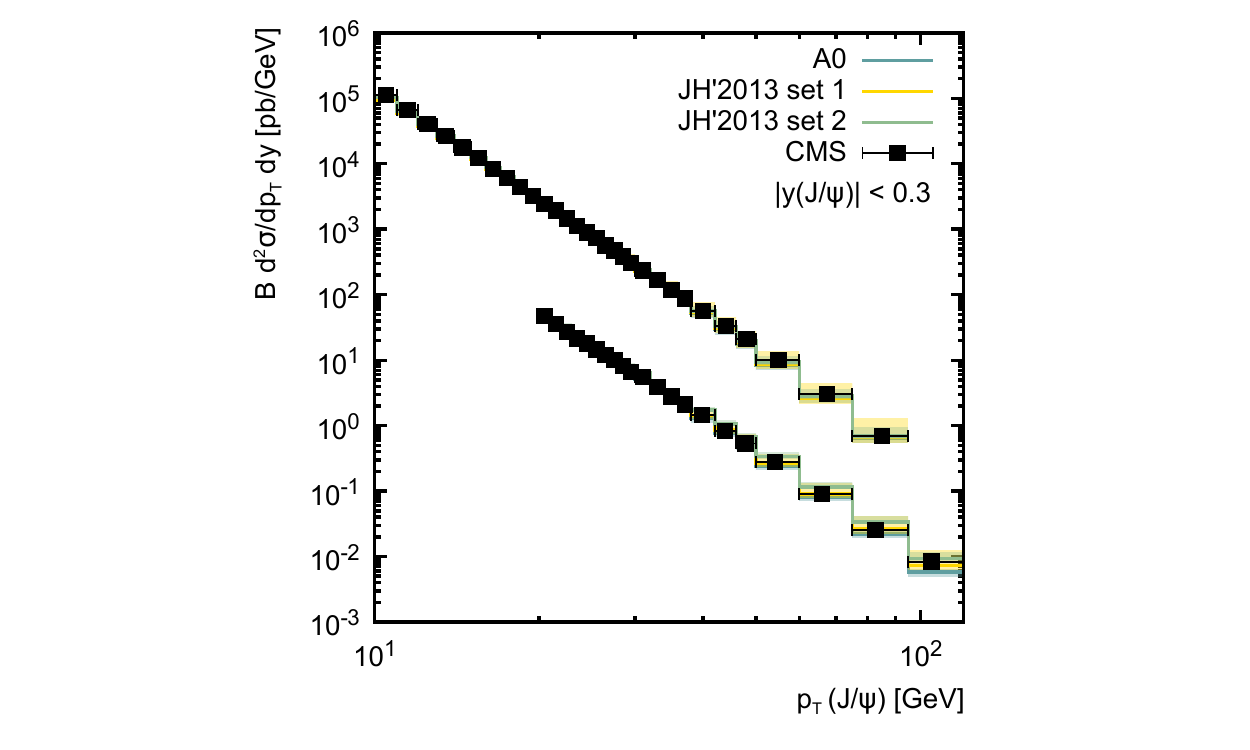}
\includegraphics[width=8.1cm]{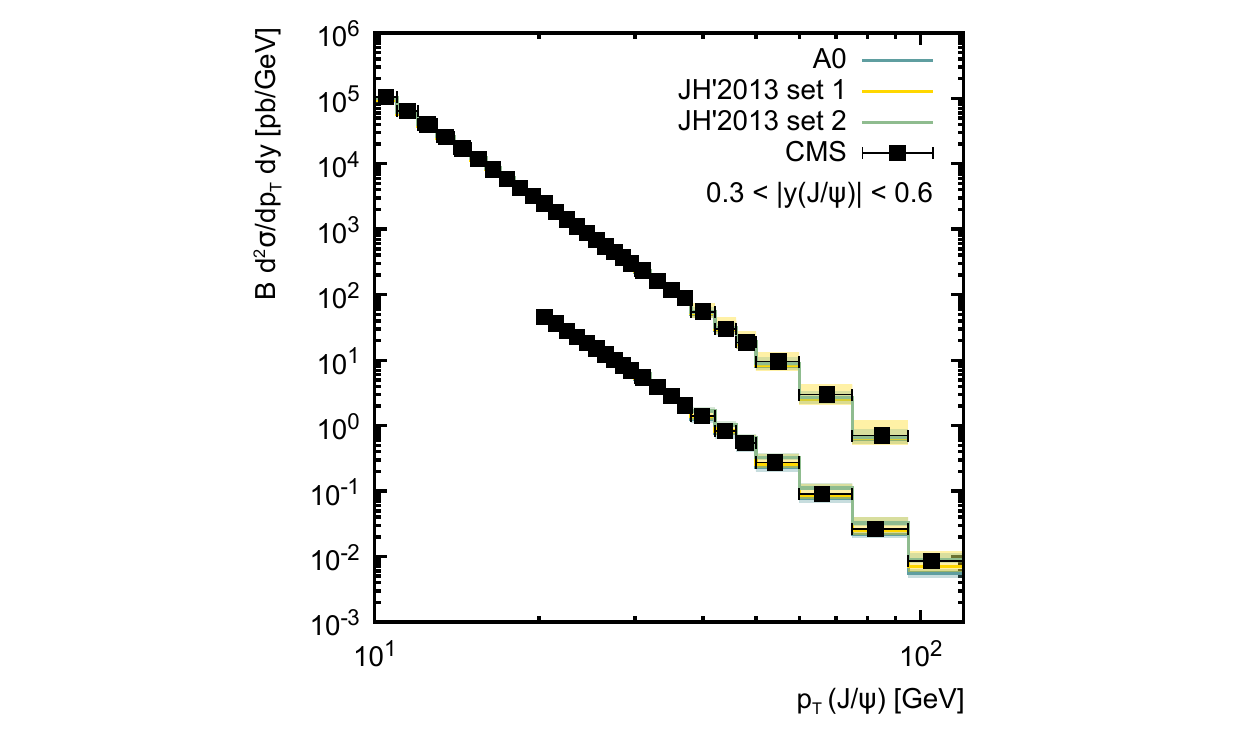}
\includegraphics[width=8.1cm]{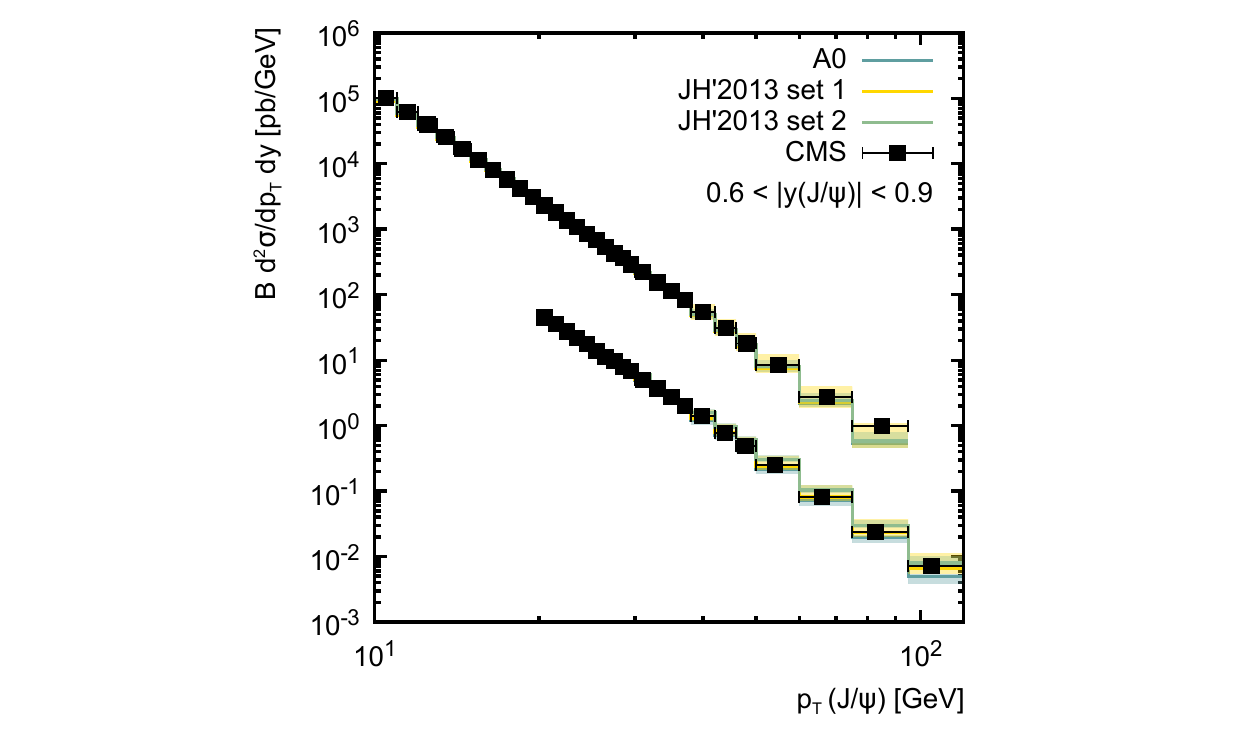}
\includegraphics[width=8.1cm]{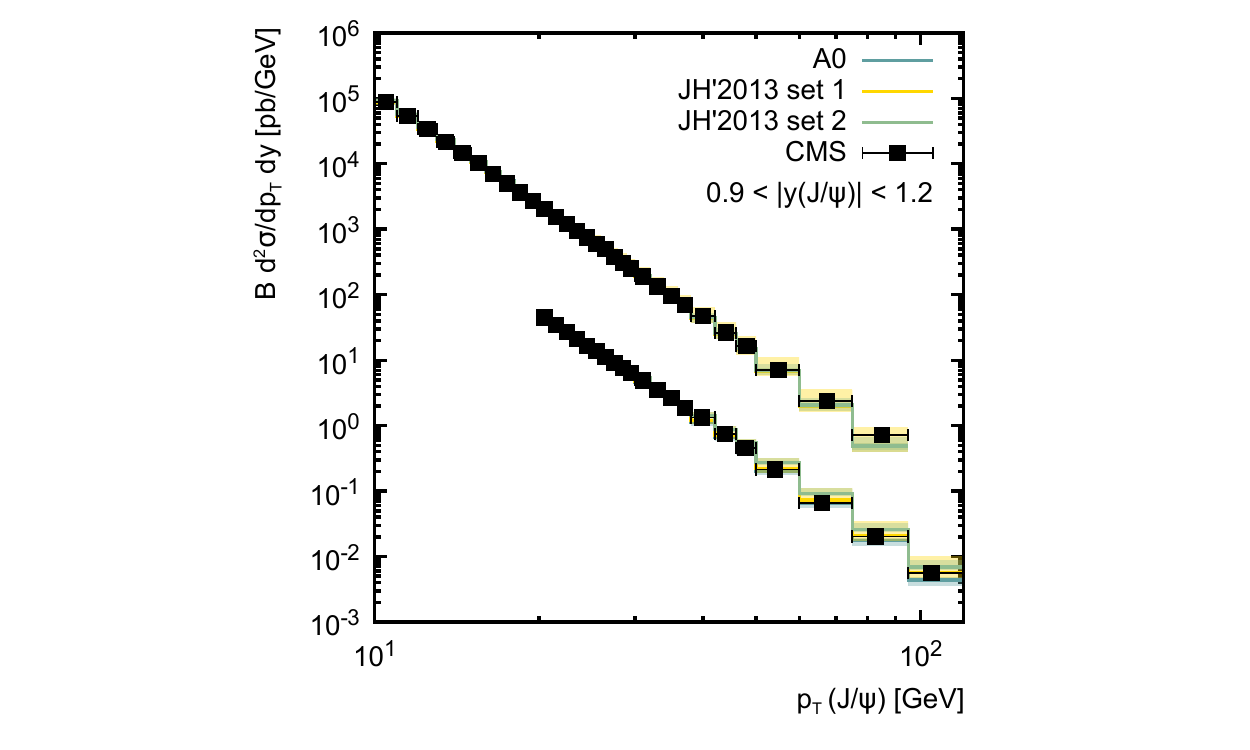}
\caption{Transverse momentum distribution of prompt $J/\psi$ mesons 
produced in $pp$ collisions at $\sqrt s = 7$~TeV (upper histograms, multiplied by $100$) and
$\sqrt s = 13$~TeV (lower histograms) at different rapidities. Notation of all histograms 
is the same as in Fig.~1. The experimental data are from CMS\cite{41,42}.}
\label{fig8}
\end{center}
\end{figure}

\begin{figure}
\begin{center}
\includegraphics[width=8.1cm]{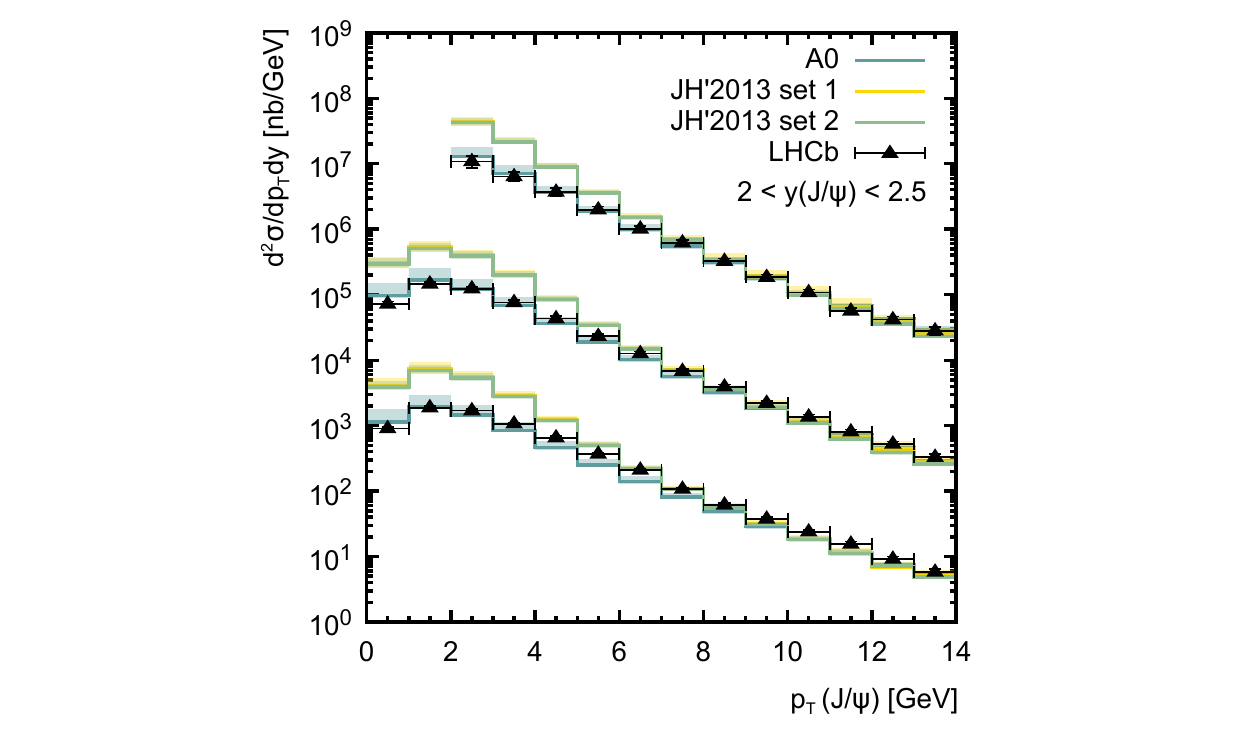}
\includegraphics[width=8.1cm]{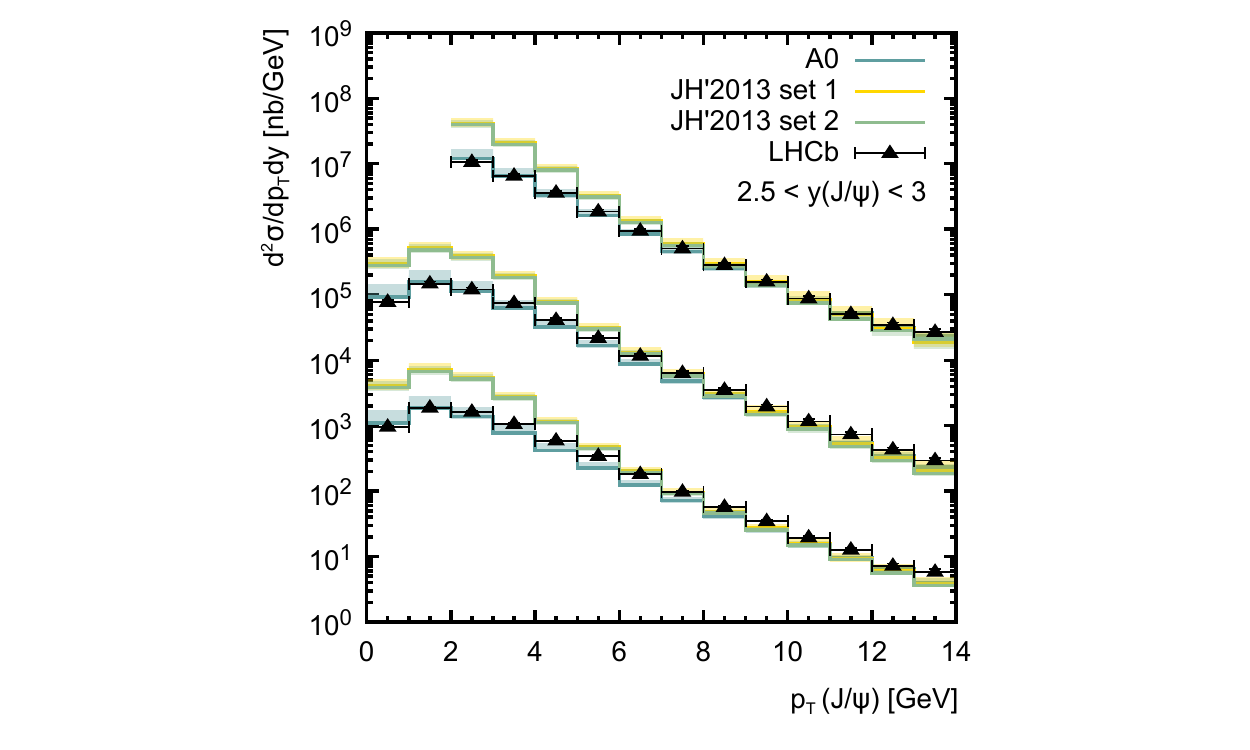}
\includegraphics[width=8.1cm]{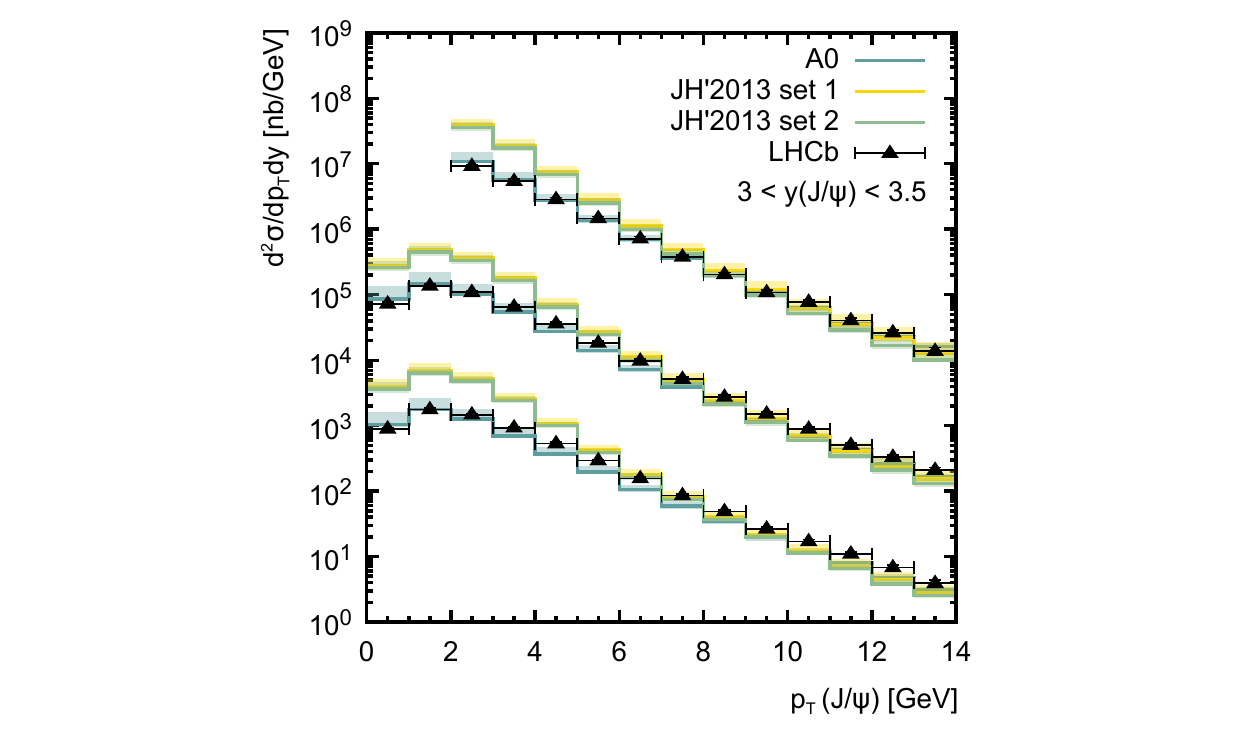}
\includegraphics[width=8.1cm]{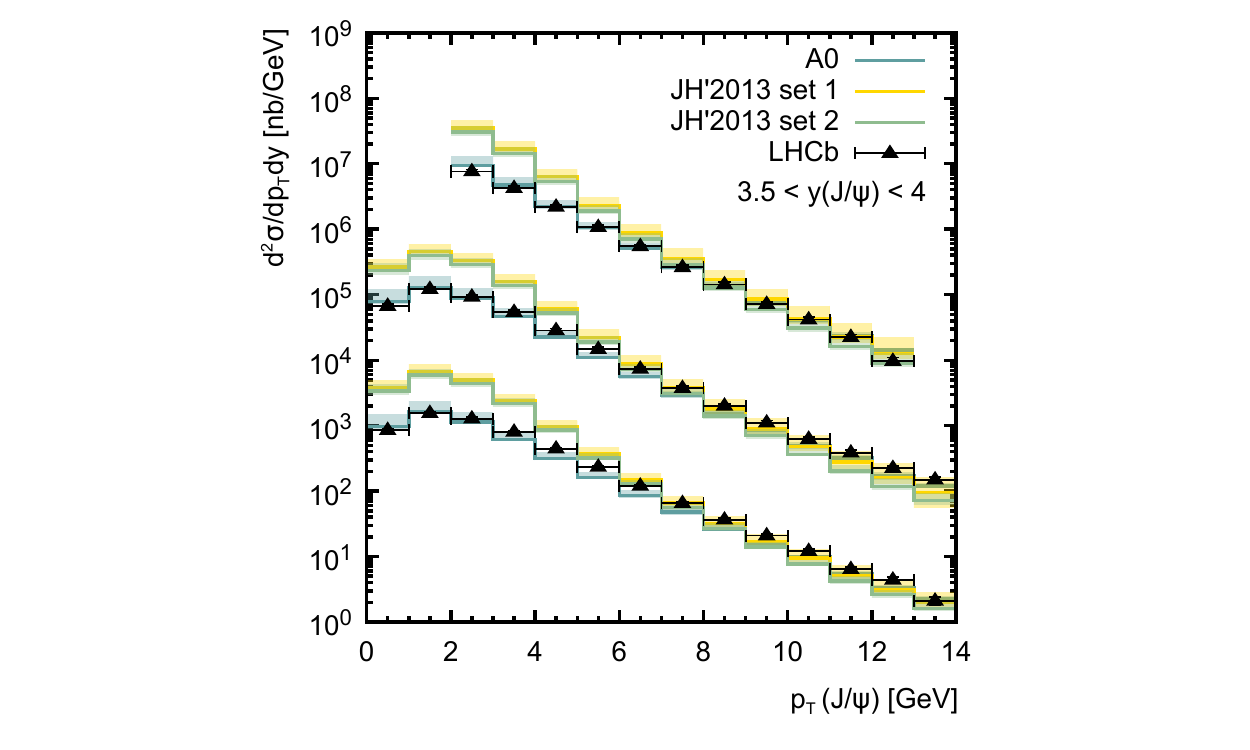}
\includegraphics[width=8.1cm]{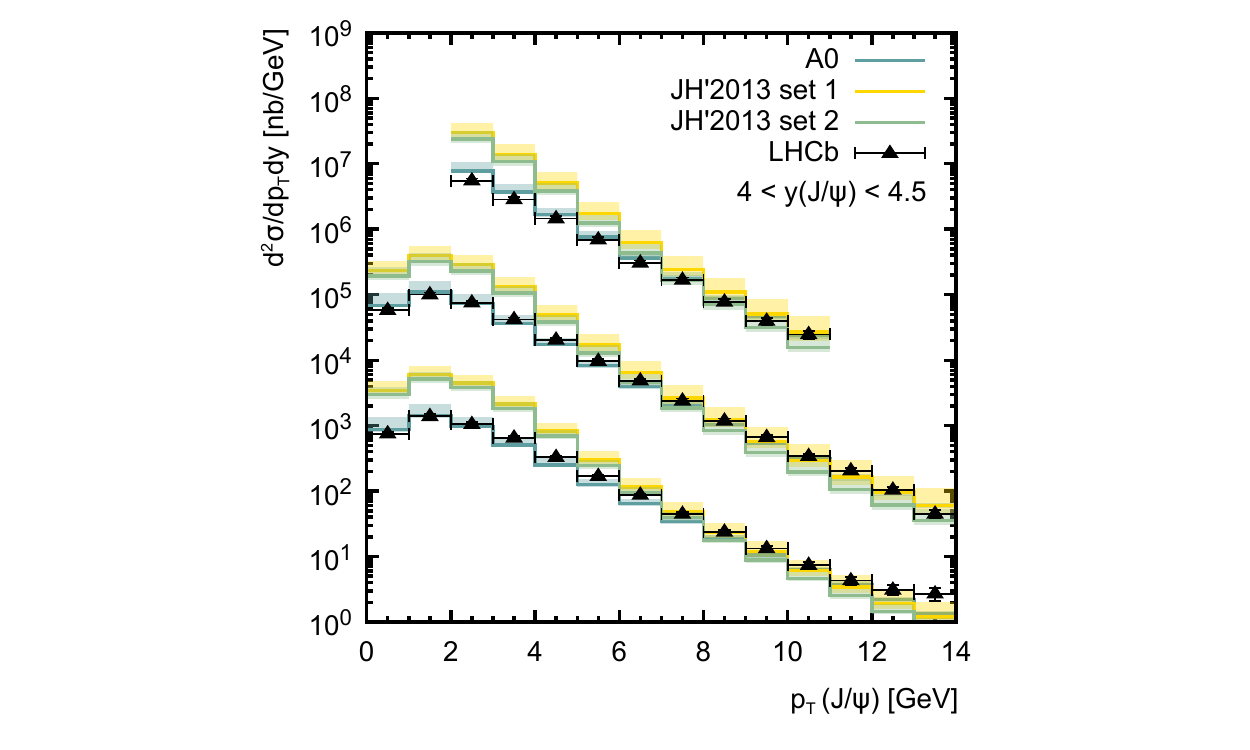}
\caption{Transverse momentum distribution of prompt $J/\psi$ mesons 
produced in $pp$ collisions at $\sqrt s = 7$~TeV (upper histograms, multiplied by $10^4$), 
$\sqrt s = 8$~TeV (middle histograms, multiplied by $10^2$) and
$\sqrt s = 13$~TeV (lower histograms) at different rapidities. Notation of all histograms 
is the same as in Fig.~1. The experimental data are from LHCb\cite{47,48,49}.}
\label{fig9}
\end{center}
\end{figure}

\begin{figure}
\begin{center}
\includegraphics[width=8.1cm]{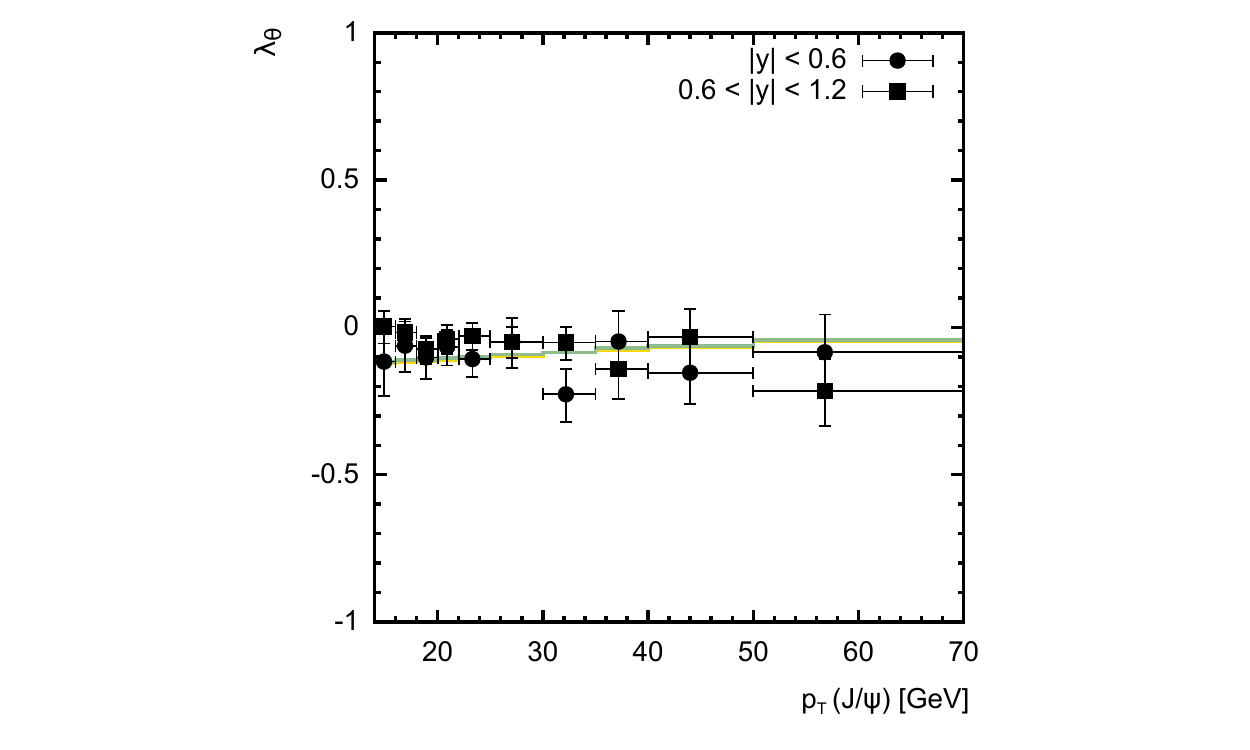}
\includegraphics[width=8.1cm]{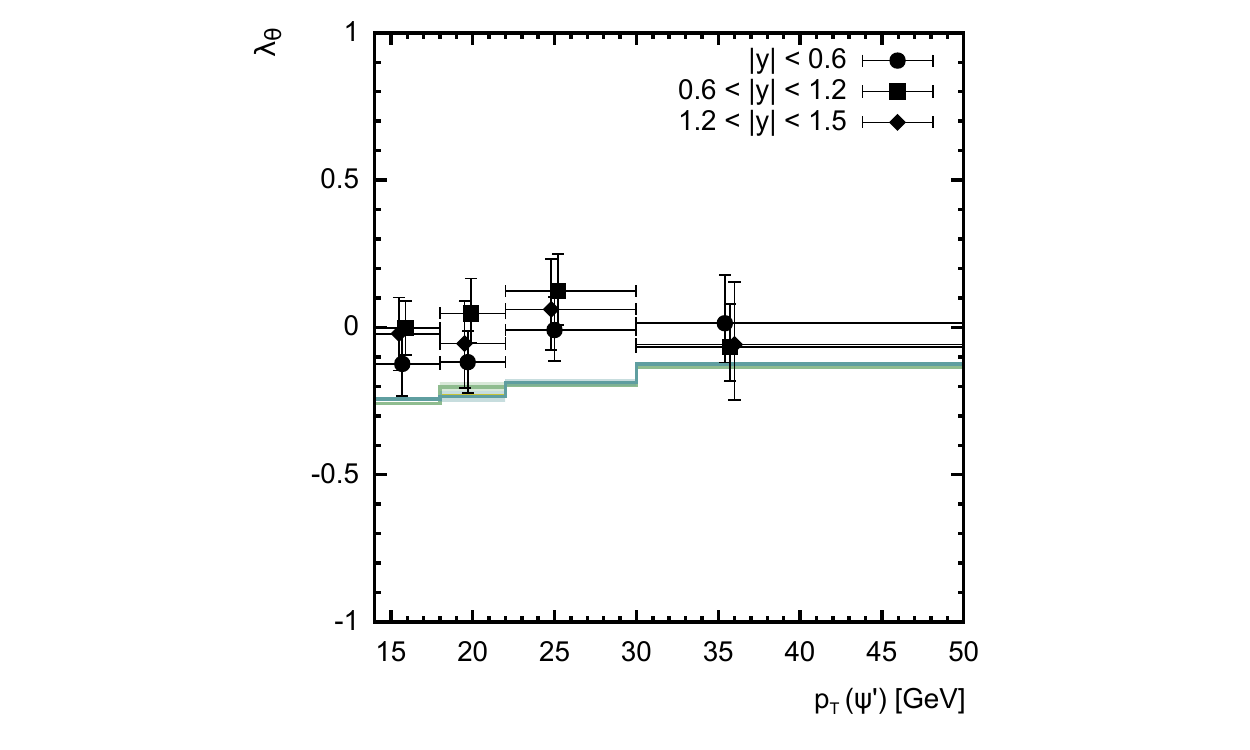}
\includegraphics[width=8.1cm]{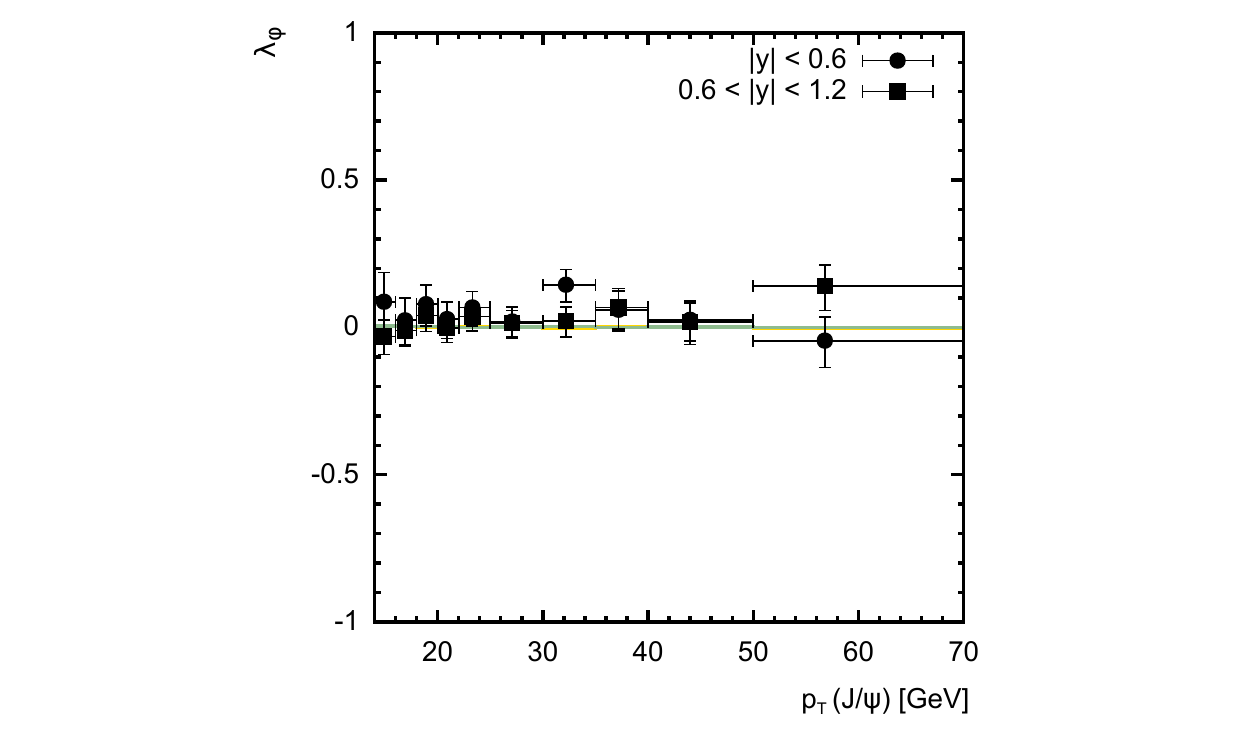}
\includegraphics[width=8.1cm]{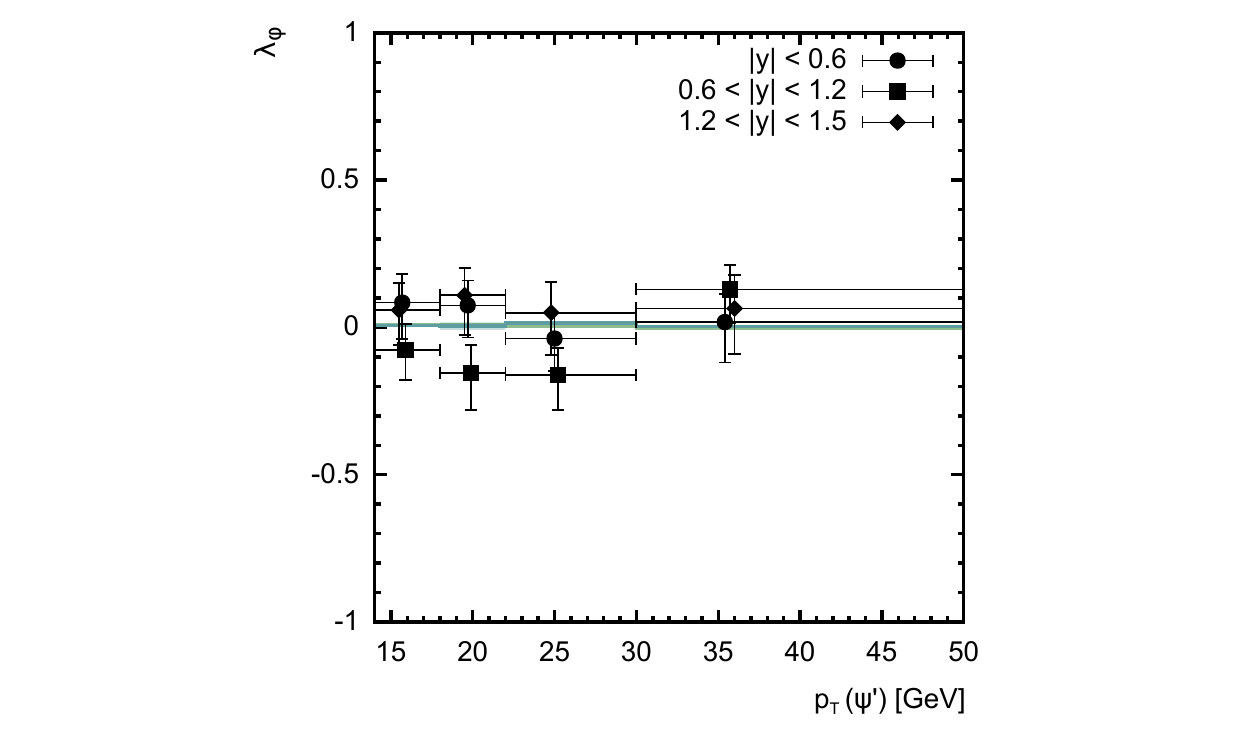}
\includegraphics[width=8.1cm]{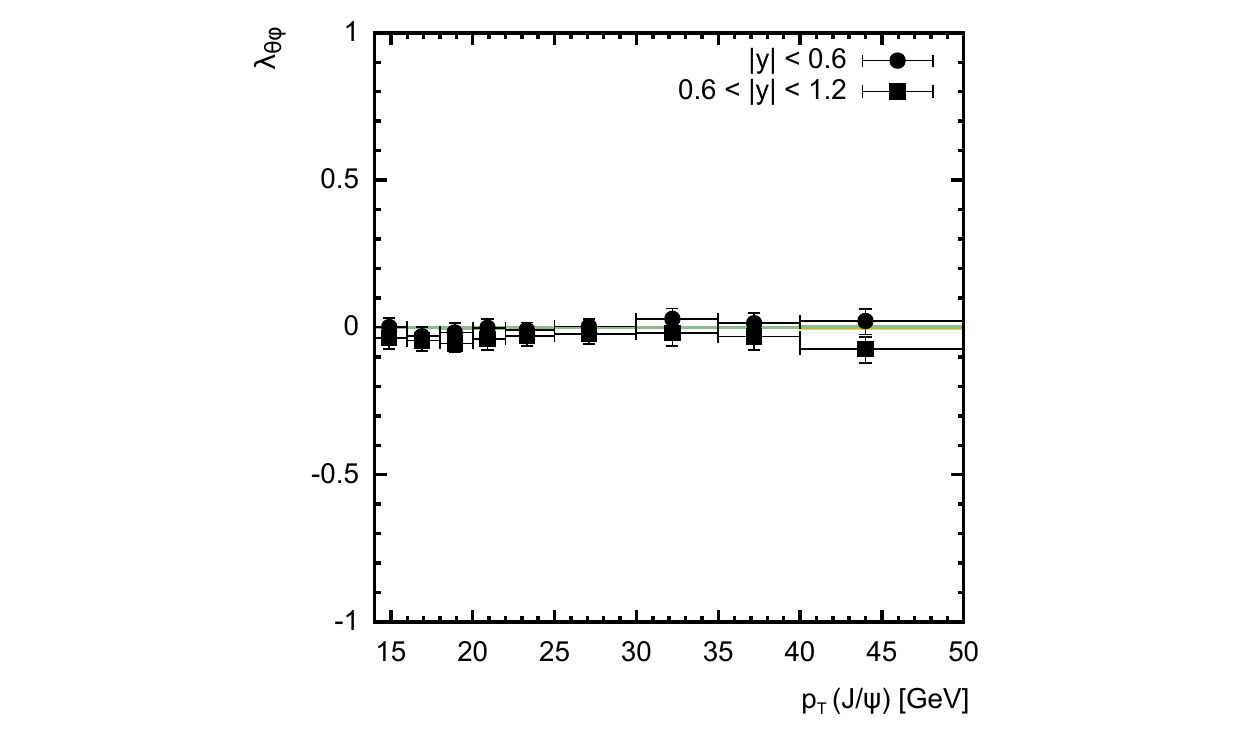}
\includegraphics[width=8.1cm]{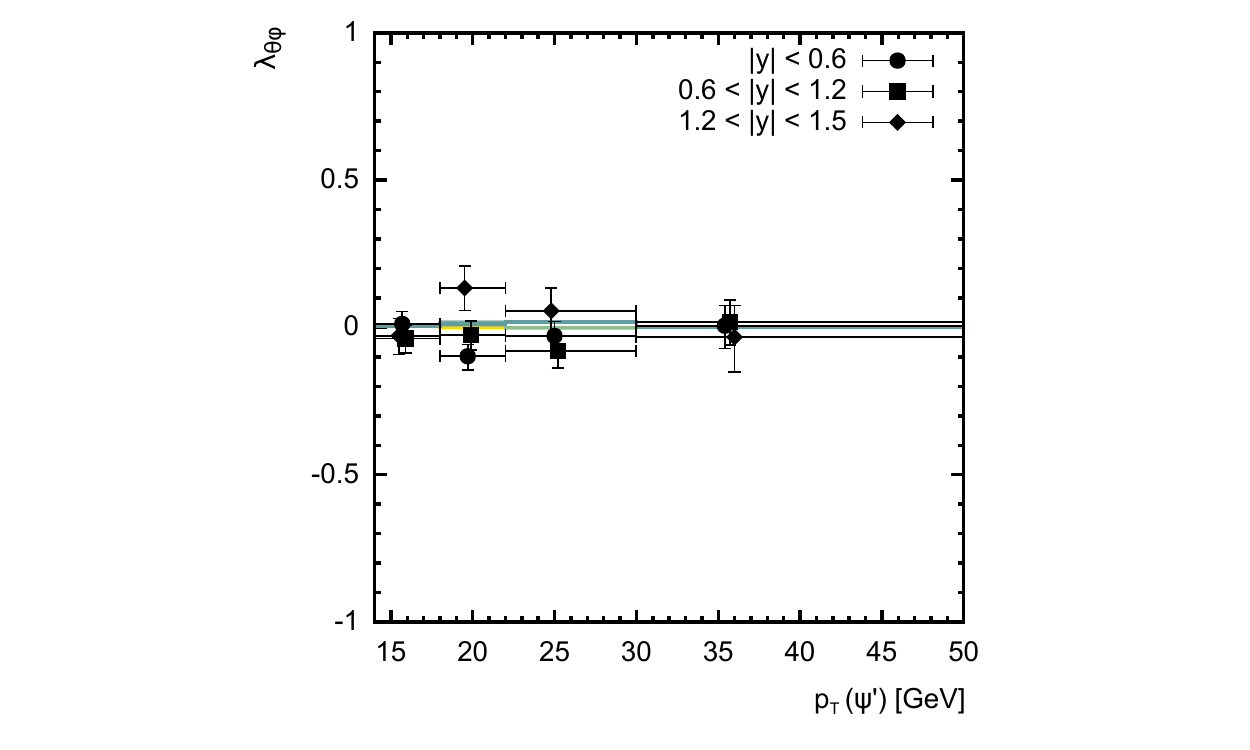}
\includegraphics[width=8.1cm]{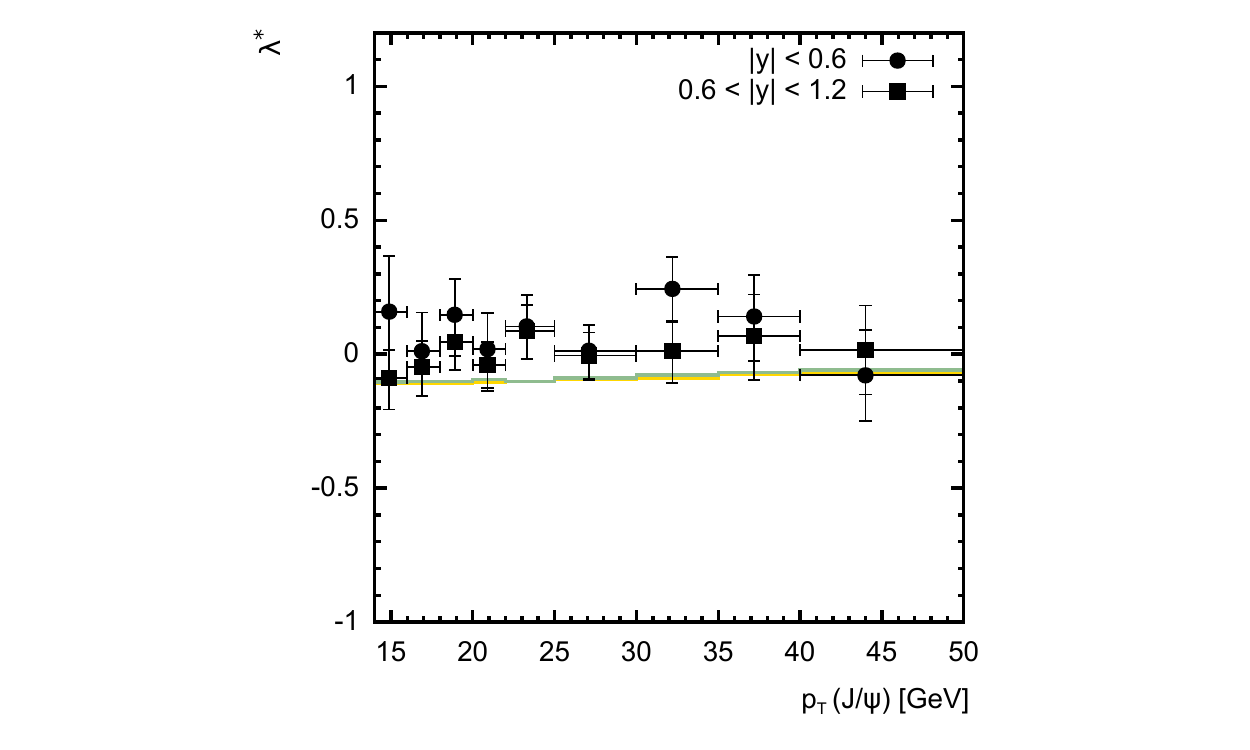}
\includegraphics[width=8.1cm]{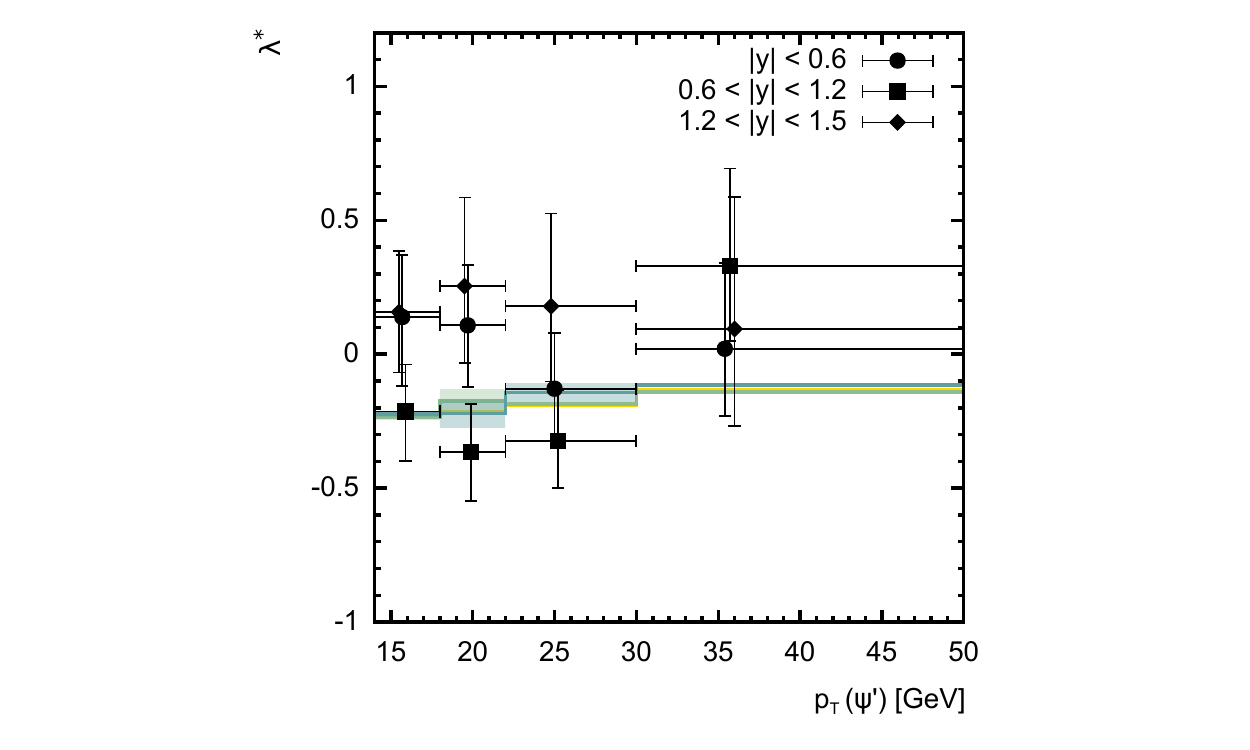}
\caption{Polarization parameters $\lambda_\theta$, $\lambda_\phi$, $\lambda_{\theta \phi}$
and $\lambda^*$ of prompt $J/\psi$ (left panels) and $\psi^\prime$ (right panels) mesons calculated as a function of
transverse momentum in the Collins-Soper frame. The yellow, blue and green
histograms correspond to the predictions obtained at $|y| < 0.6$, $0.6 < |y| < 1.2$ 
and $1.2 < |y| < 1.5$. The JH'2013 set 2 gluon density is used. The experimental data are from CMS\cite{50}.}
\label{fig10}
\end{center}
\end{figure}

\begin{figure}
\begin{center}
\includegraphics[width=8.1cm]{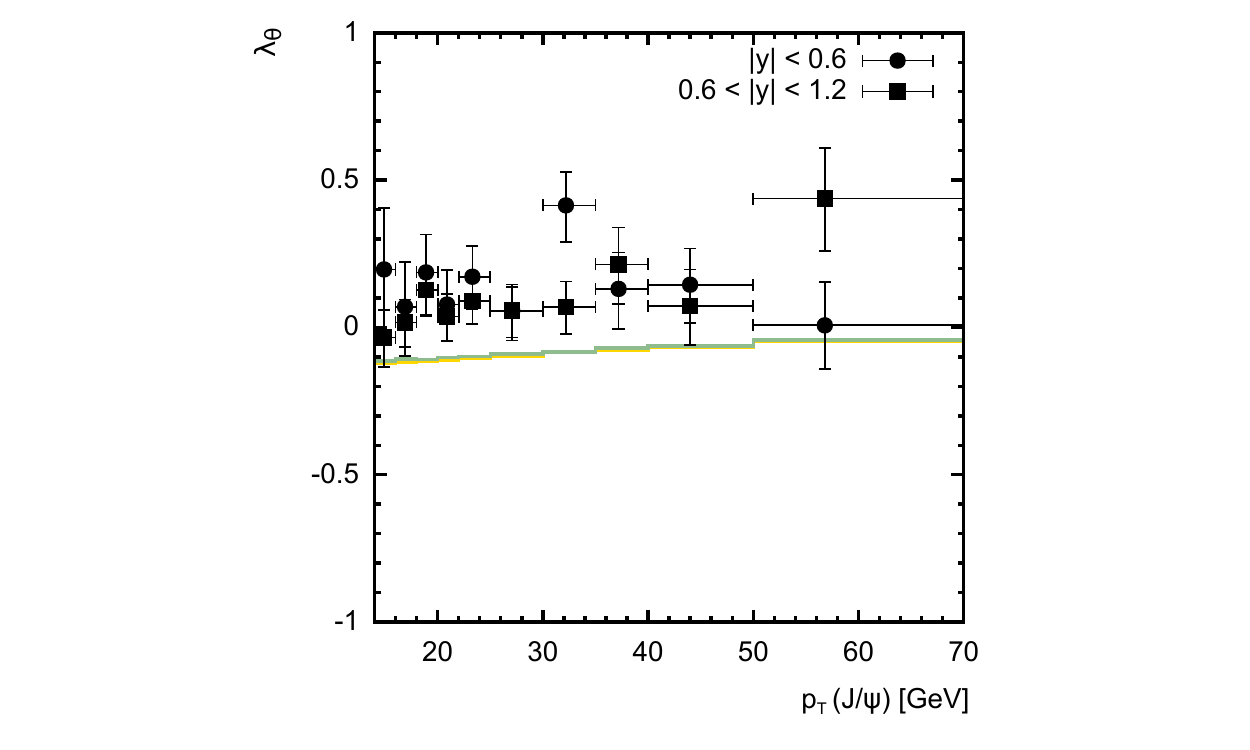}
\includegraphics[width=8.1cm]{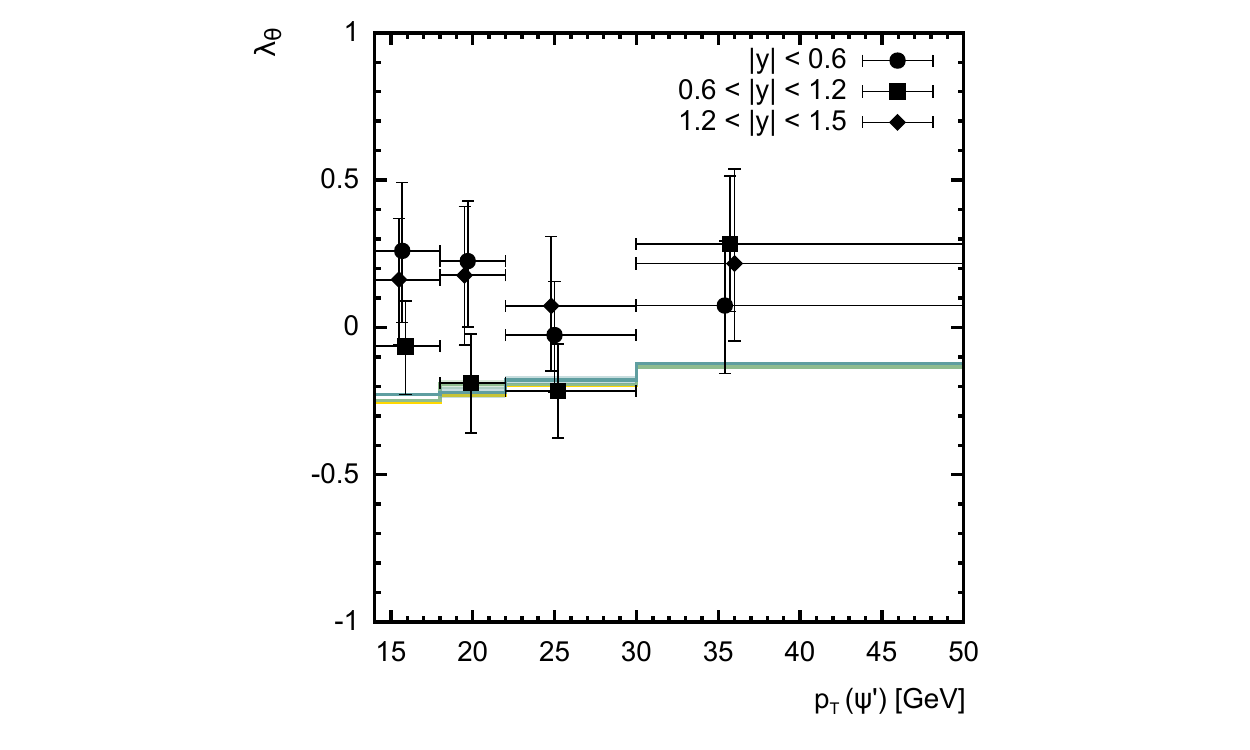}
\includegraphics[width=8.1cm]{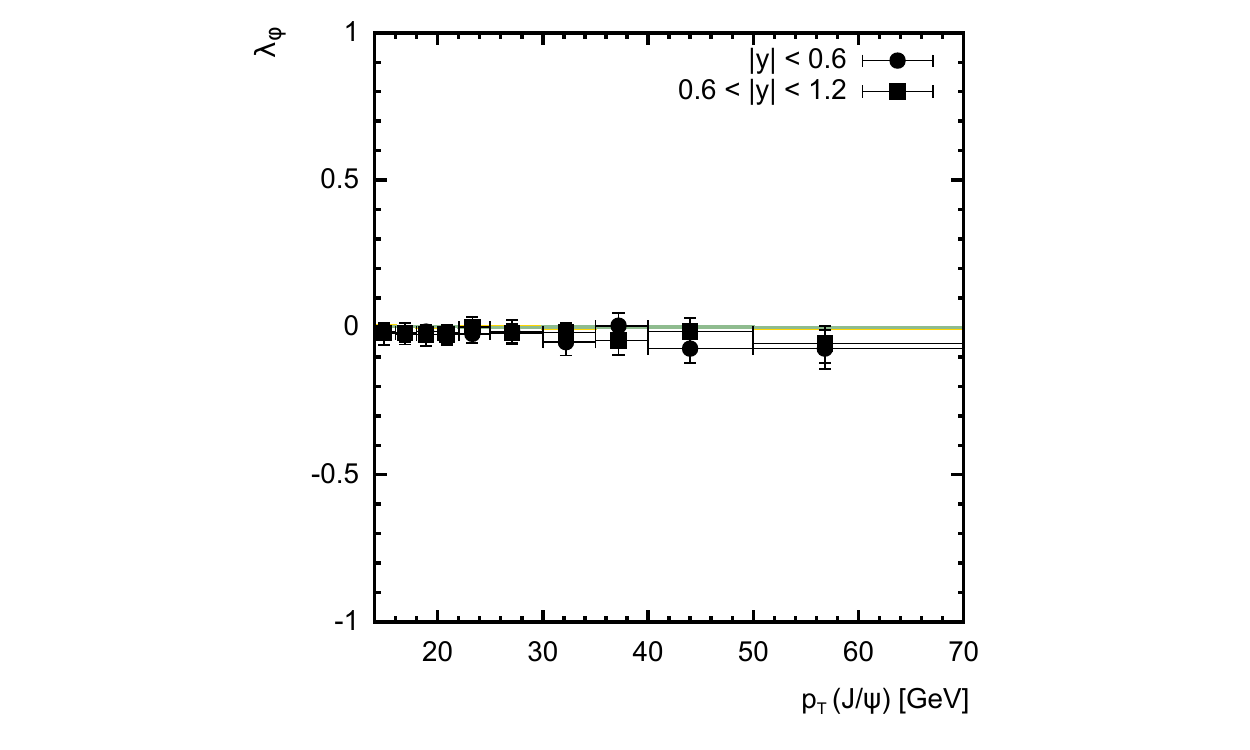}
\includegraphics[width=8.1cm]{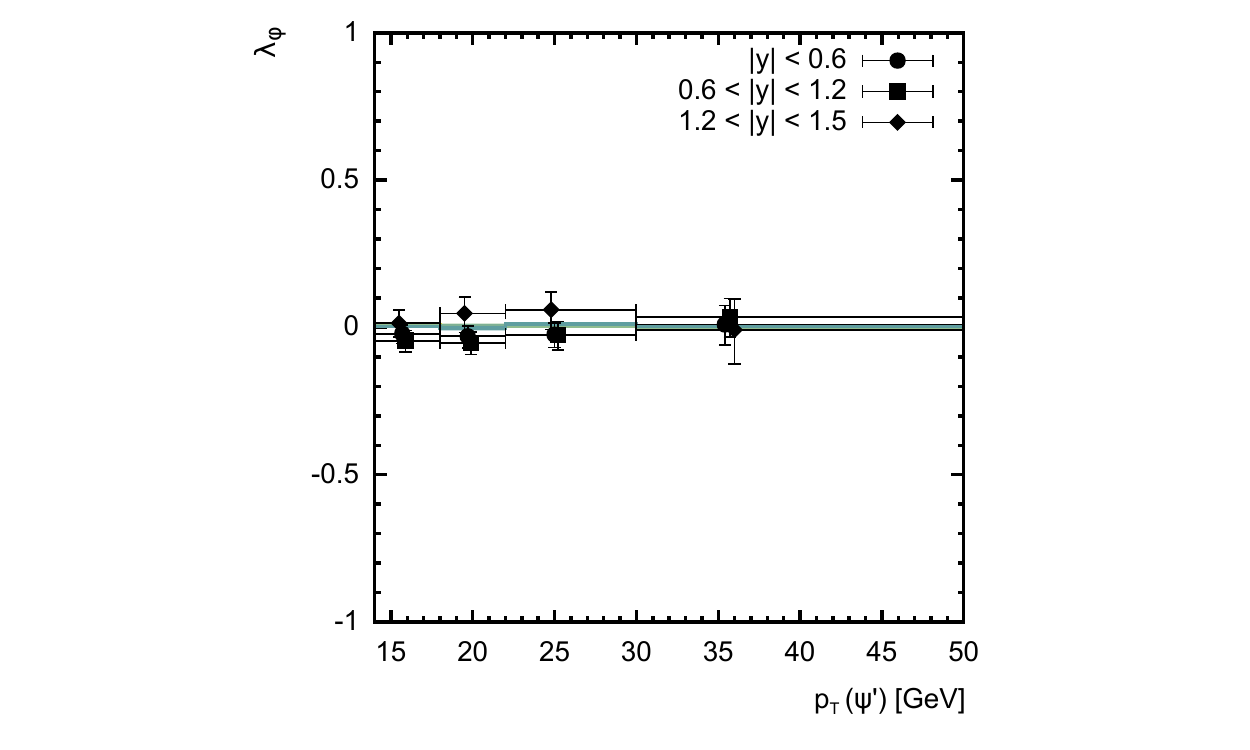}
\includegraphics[width=8.1cm]{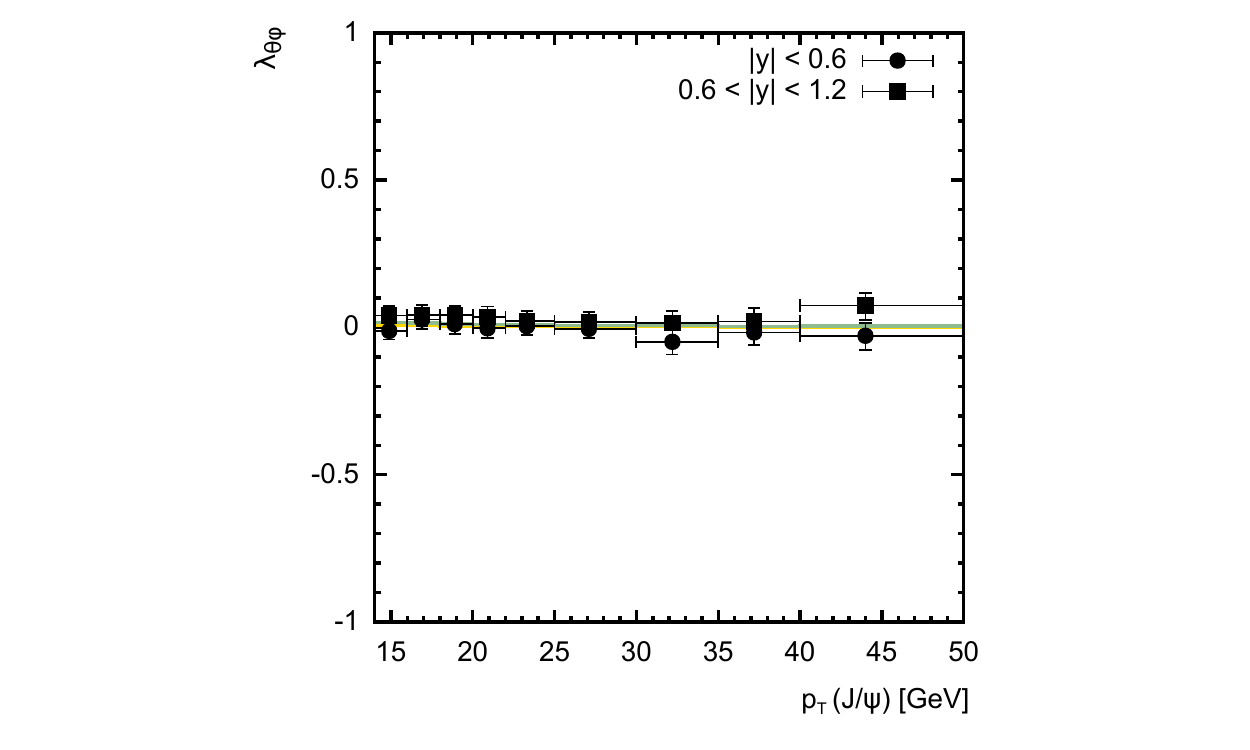}
\includegraphics[width=8.1cm]{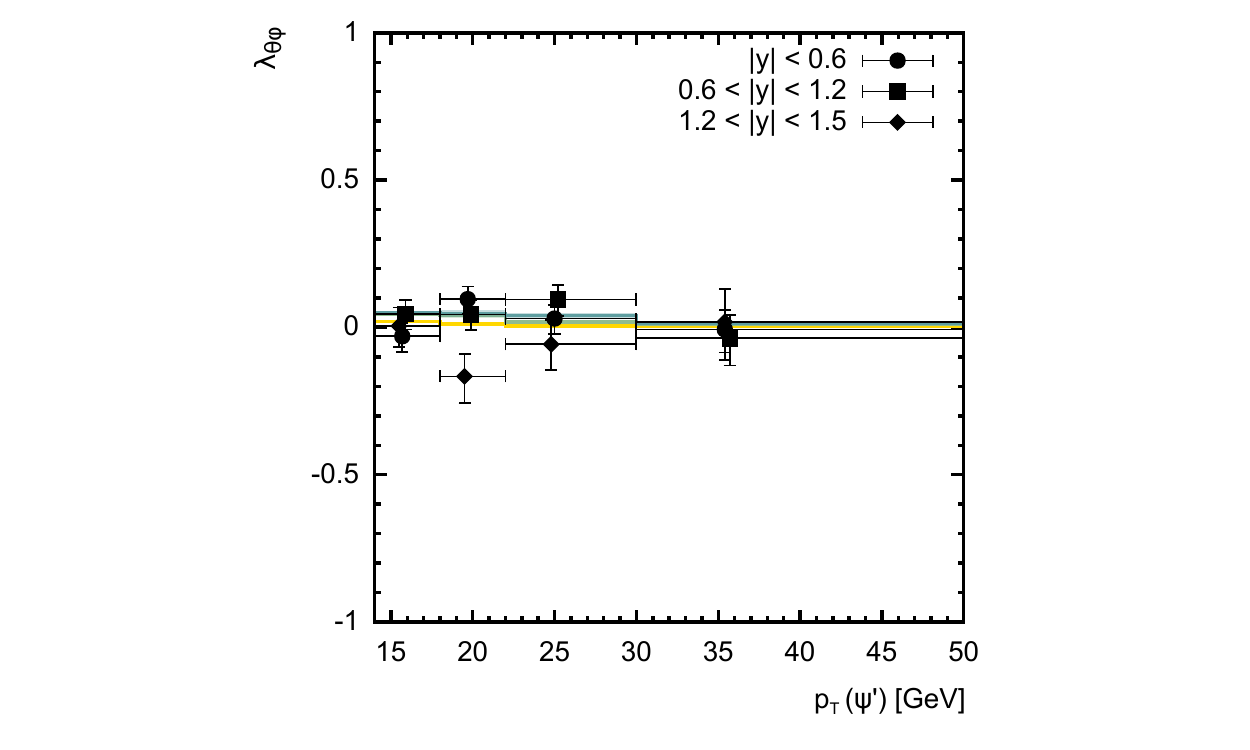}
\includegraphics[width=8.1cm]{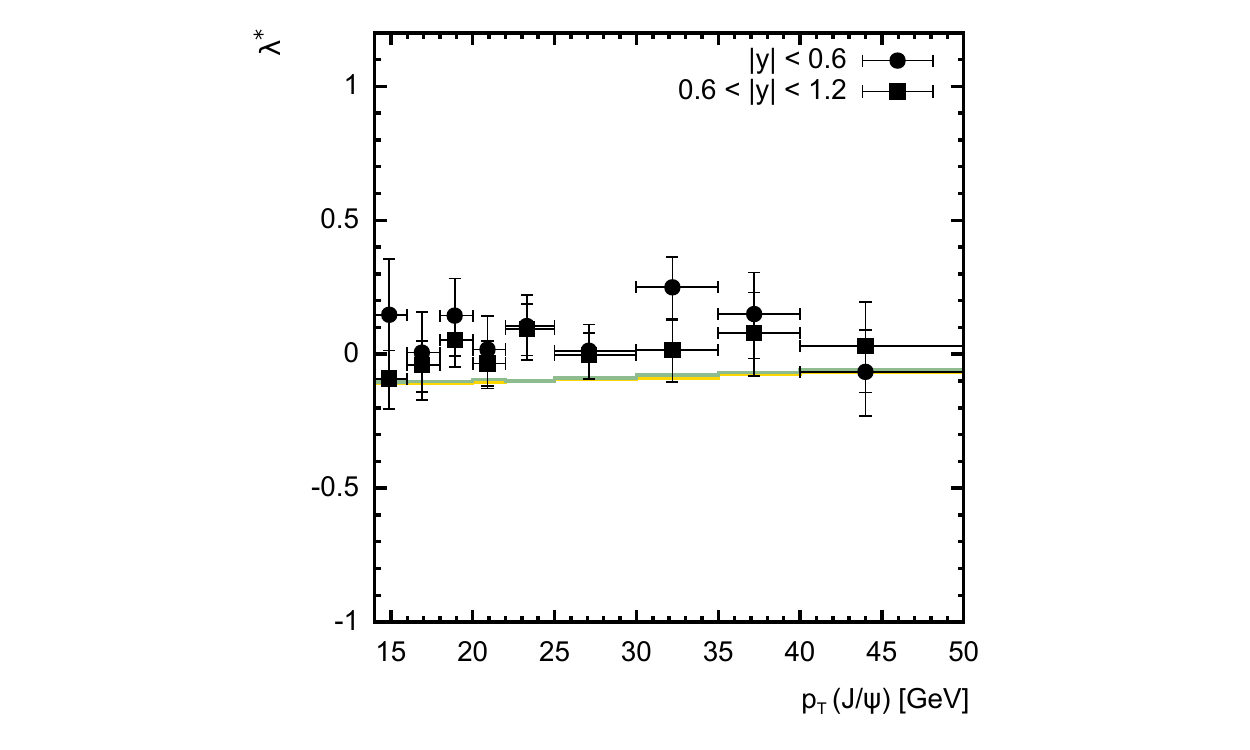}
\includegraphics[width=8.1cm]{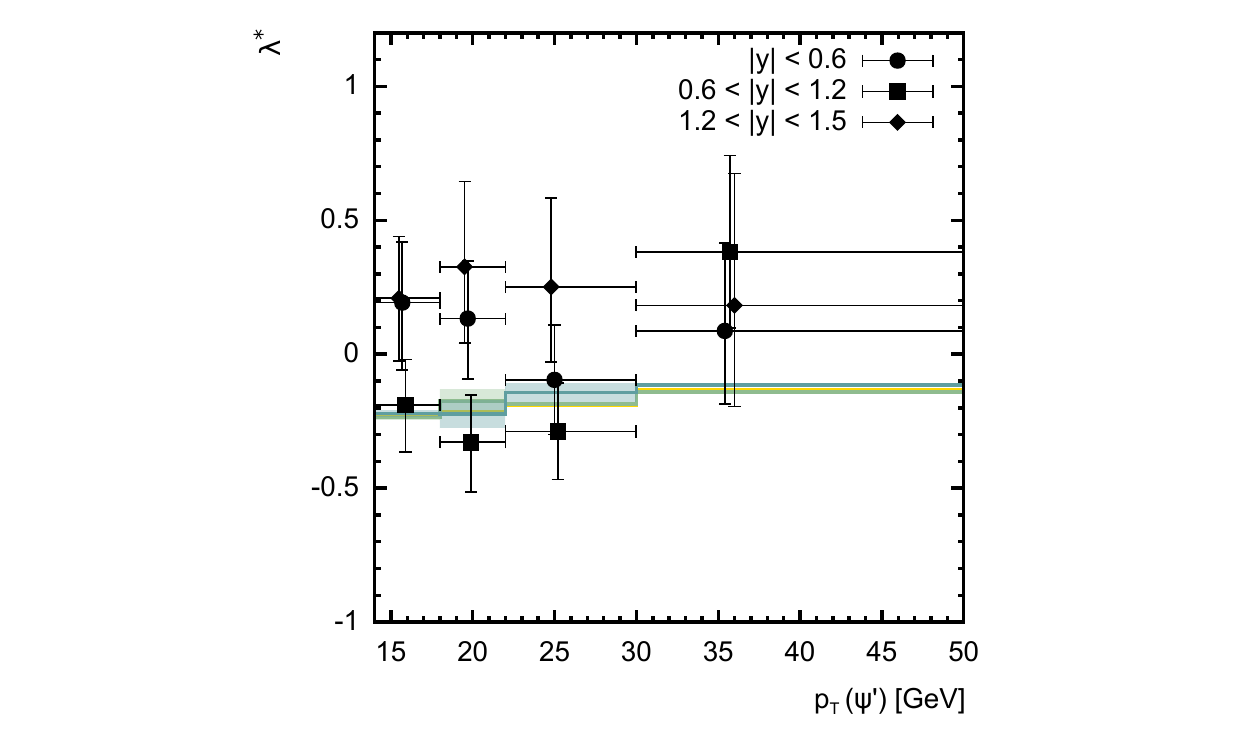}
\caption{Polarization parameters $\lambda_\theta$, $\lambda_\phi$, $\lambda_{\theta \phi}$
and $\lambda^*$ of prompt $J/\psi$ (left panels) and $\psi^\prime$ (right panels) mesons calculated as a function of
transverse momentum in the helicity frame. Notation of all histograms is the same as in Fig.~8.
The experimental data are from CMS\cite{50}.}
\label{fig11}
\end{center}
\end{figure}

\begin{figure}
\begin{center}
\includegraphics[width=8.1cm]{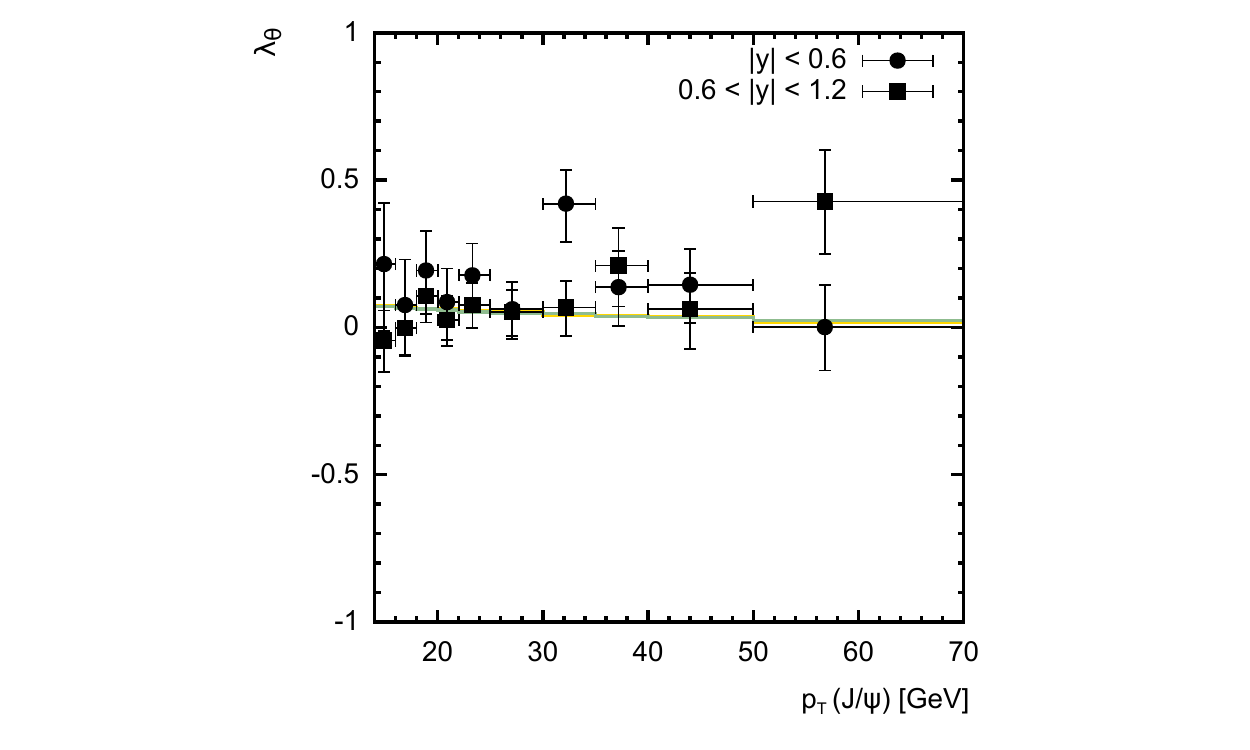}
\includegraphics[width=8.1cm]{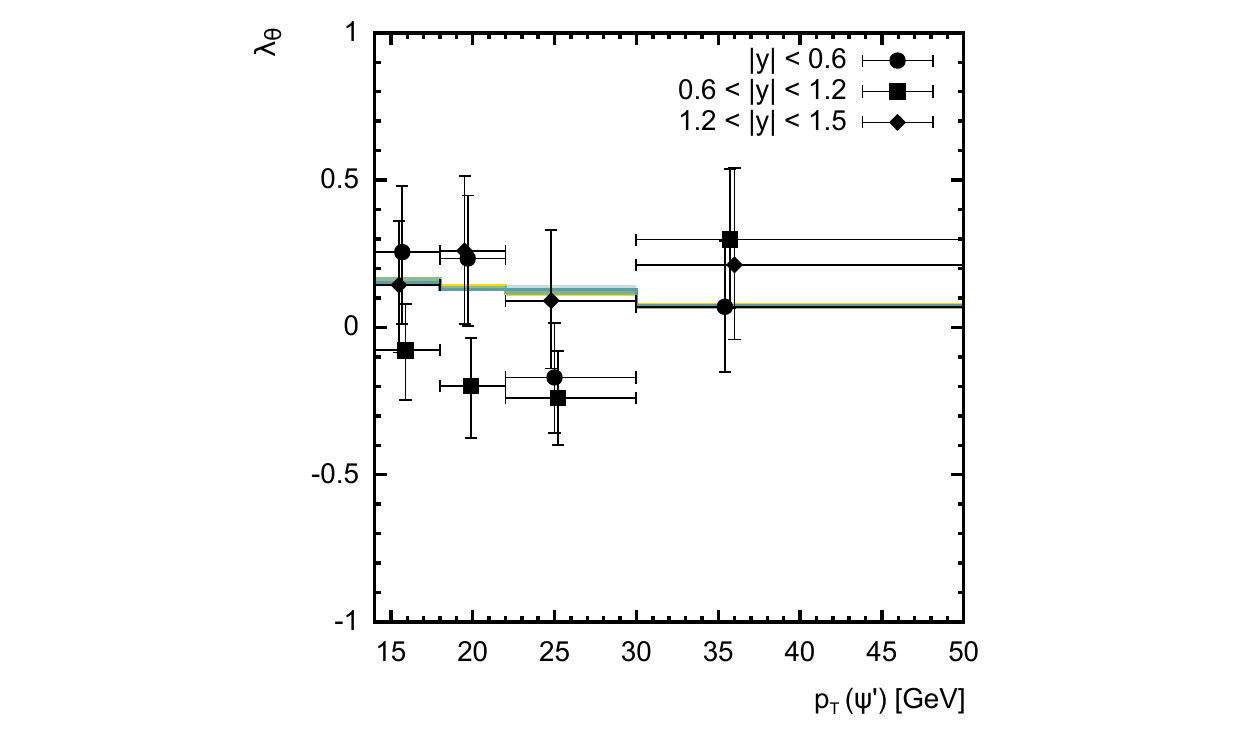}
\includegraphics[width=8.1cm]{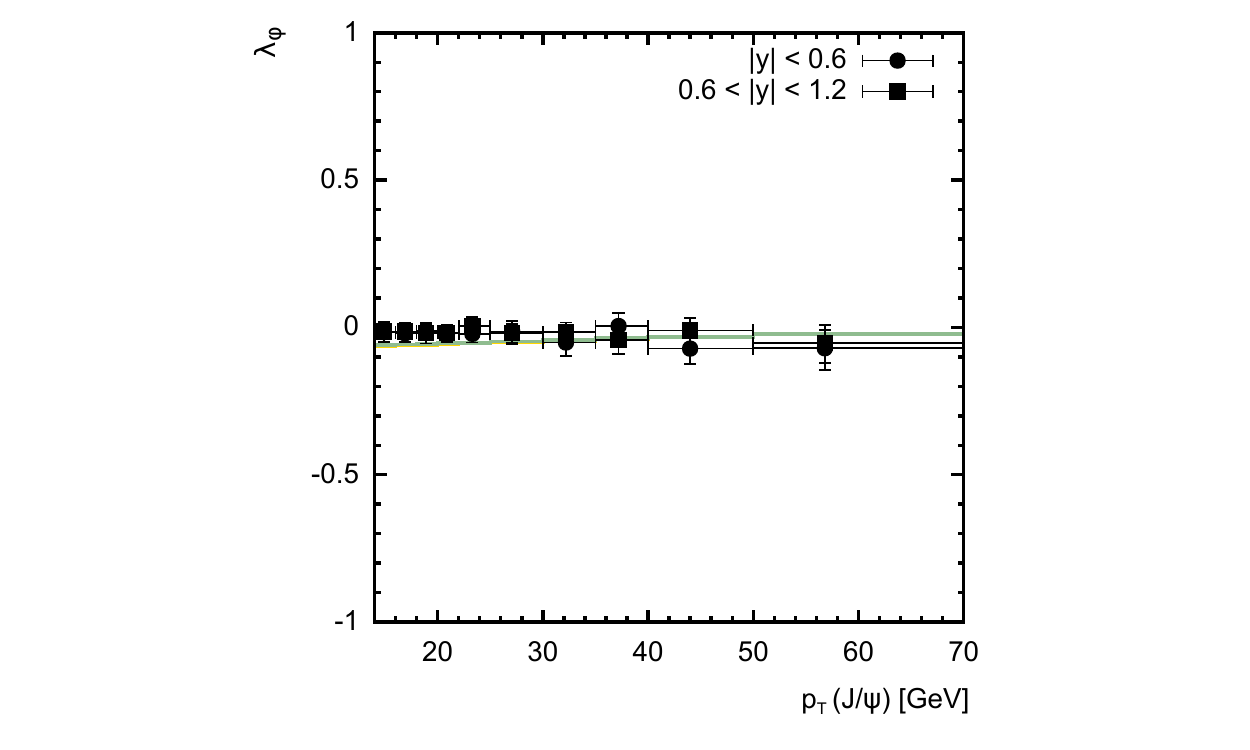}
\includegraphics[width=8.1cm]{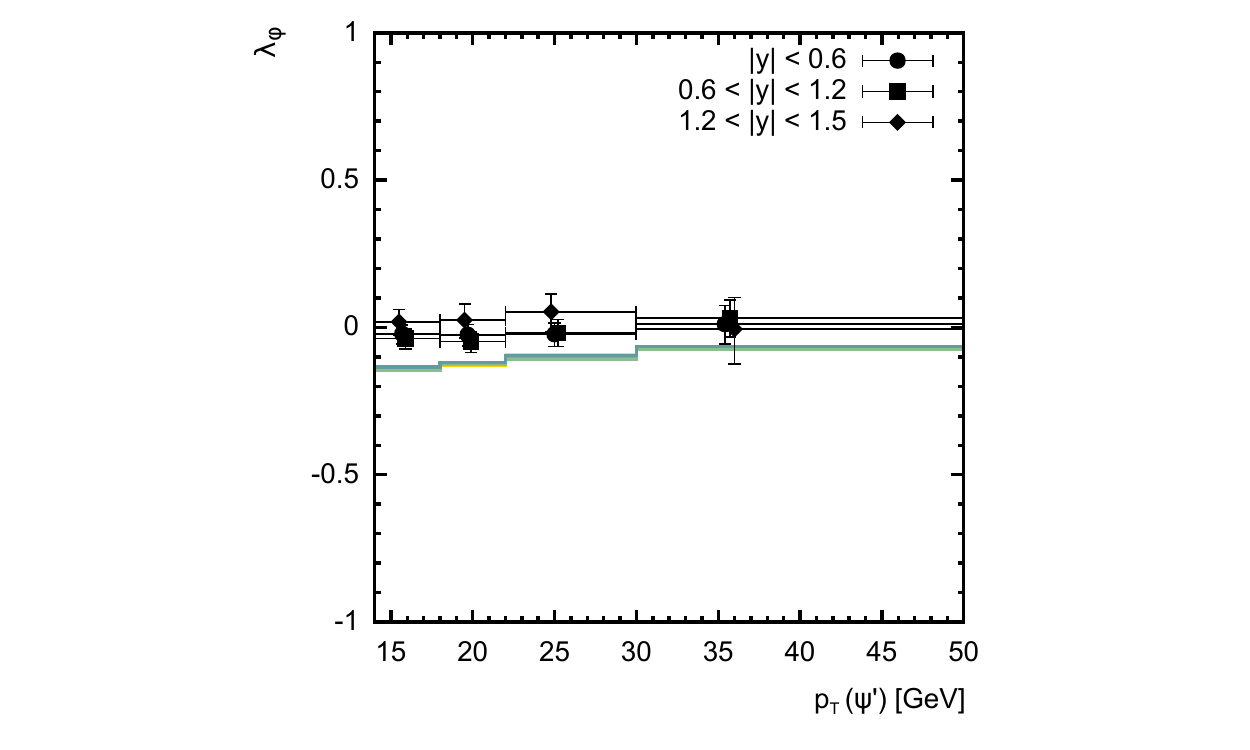}
\includegraphics[width=8.1cm]{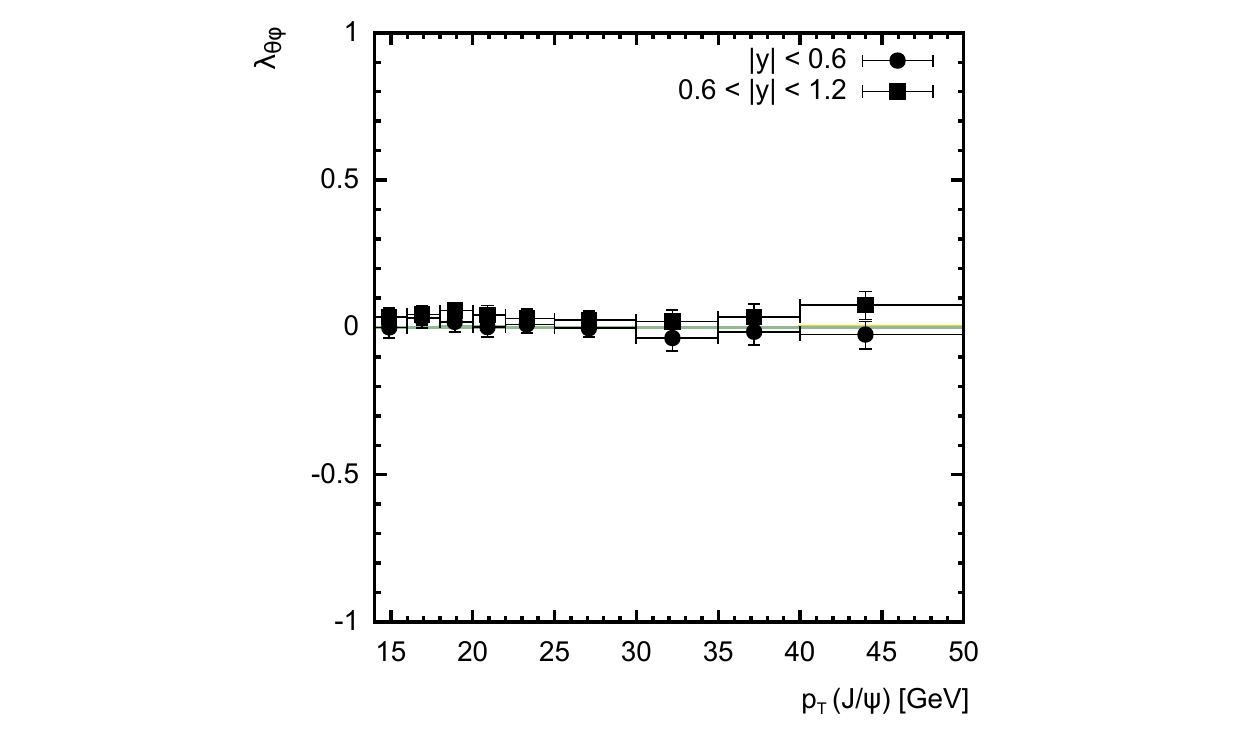}
\includegraphics[width=8.1cm]{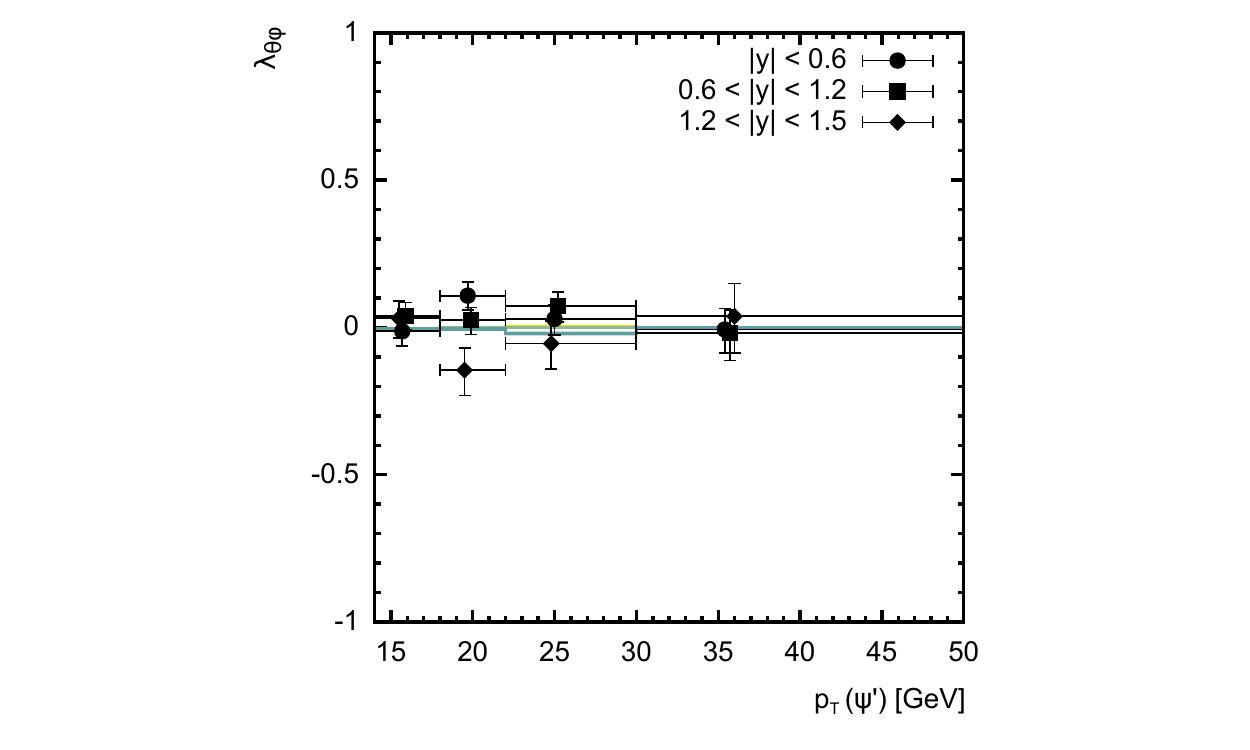}
\includegraphics[width=8.1cm]{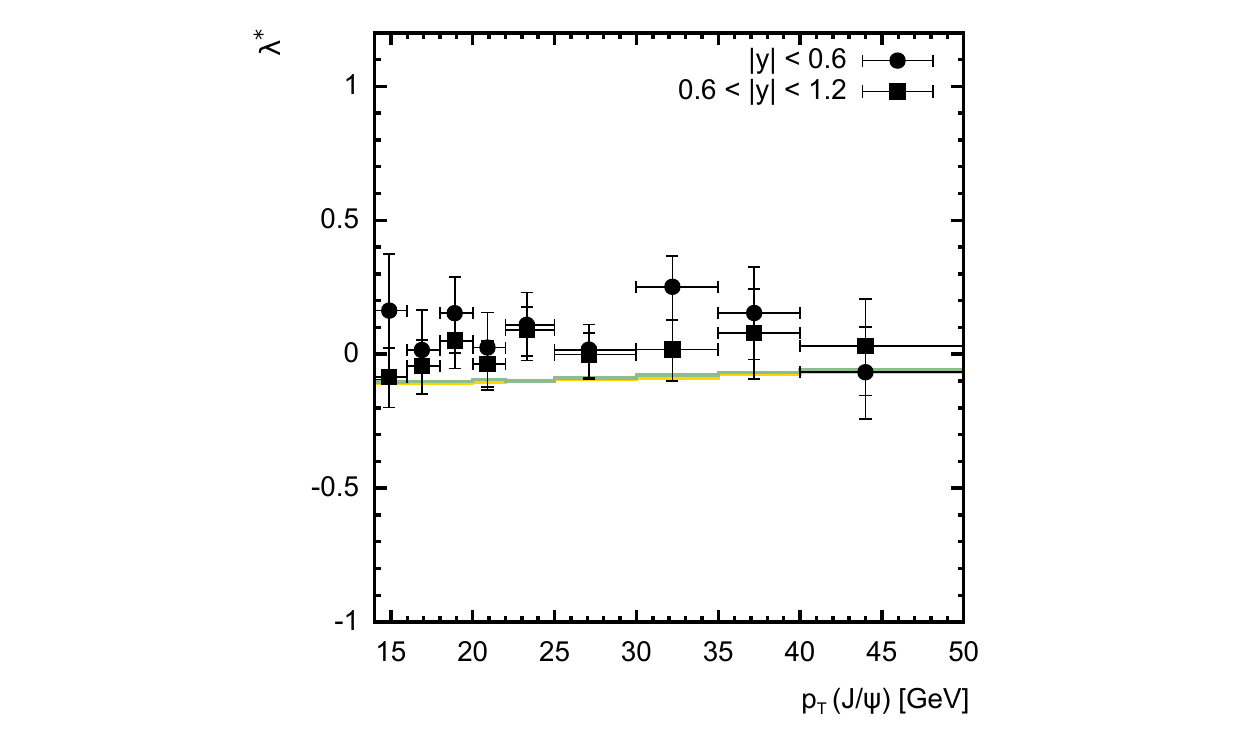}
\includegraphics[width=8.1cm]{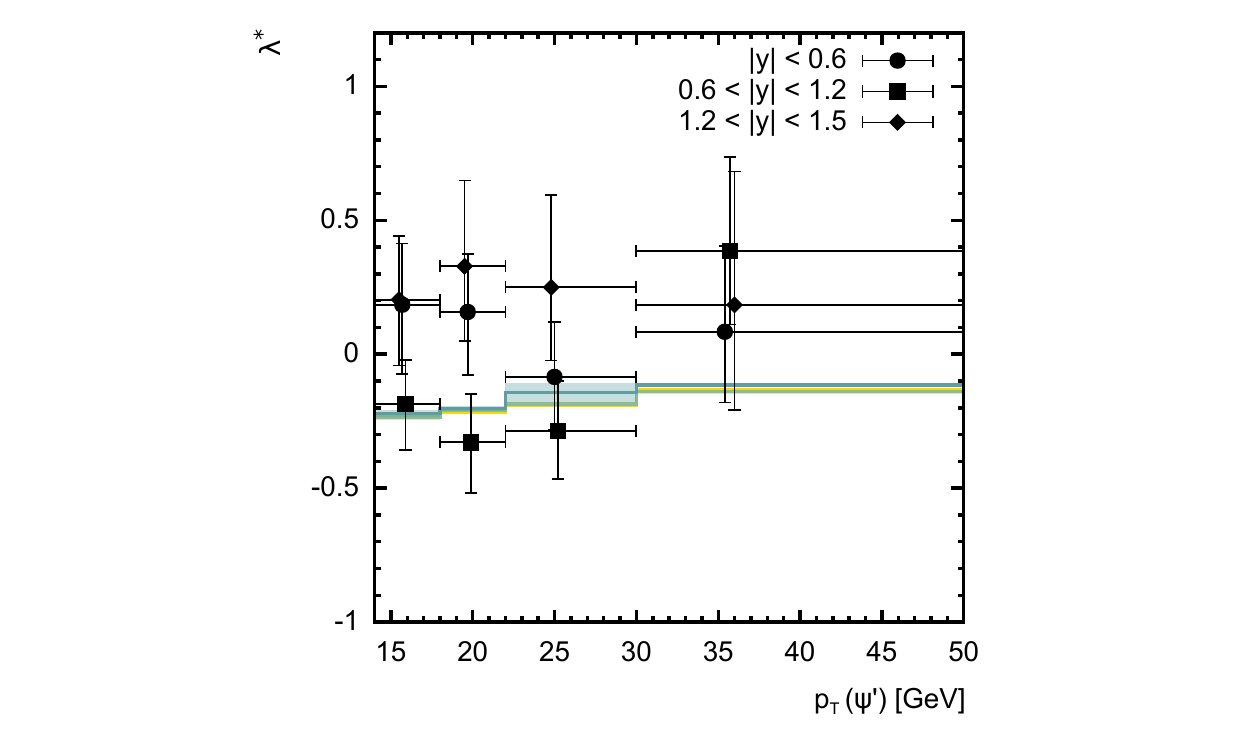}
\caption{Polarization parameters $\lambda_\theta$, $\lambda_\phi$, $\lambda_{\theta \phi}$
and $\lambda^*$ of prompt $J/\psi$ (left panels) and $\psi^\prime$ (right panels) mesons calculated as a function of
transverse momentum in the perpendicular helicity frame. Notation of all histograms is the same as in Fig.~8.
The experimental data are from CMS\cite{50}.}
\label{fig12}
\end{center}
\end{figure}

\end{document}